\documentclass[iop]{emulateapj}
\usepackage{graphicx}
\usepackage{savesym}
\usepackage{listings}
\usepackage[intlimits]{amsmath}
\savesymbol{iint}
\usepackage{txfonts}
\usepackage{bm}
\restoresymbol{TXF}{iint}
\usepackage{amssymb}
\usepackage{natbib}
\usepackage{color}

\def\be{\begin{equation}}
\def\ee{\end{equation}}
\def\ba{\begin{eqnarray}}
\def\ea{\end{eqnarray}}
\def\go{\mathrel{\raise.3ex\hbox{$>$}\mkern-14mu
             \lower0.6ex\hbox{$\sim$}}}
\def\lo{\mathrel{\raise.3ex\hbox{$<$}\mkern-14mu
             \lower0.6ex\hbox{$\sim$}}}

\begin{document}
\title{Magnetic Axis Drift and Magnetic  Spot Formation in Neutron Stars with Toroidal Fields}

\author{Konstantinos N. Gourgouliatos}
\affiliation{Department of Mathematical Sciences, Durham University, Durham,  DH1 3LE, UK}
\affiliation{Department of Applied Mathematics, University of Leeds, Leeds LS2 9JT , UK,}
\email{Konstantinos.Gourgouliatos@durham.ac.uk}
\author{Rainer Hollerbach}
\affiliation{Department of Applied Mathematics, University of Leeds, Leeds LS2 9JT , UK}

\begin{abstract}
We explore magnetic field configurations that lead to the formation of magnetic spots on the surface of neutron stars, and to the displacement of the magnetic dipole axis. We find that a toroidally dominated magnetic field is essential for the generation of a single spot with a strong magnetic field. 
Once a spot forms, it survives for several million years, even after the total magnetic field has decayed significantly. We find that the dipole axis is not stationary with respect to the neutron star's surface and does not in general coincide with the location of the magnetic spot. This is due to non-axisymmetric instabilities of the toroidal field that displace the poloidal dipole axis at rates that may reach $0.4^{\circ}$ per century. A misaligned poloidal dipole axis with the toroidal field leads to more significant displacement of the dipole axis than the fully aligned case. Finally we discuss the evolution of neutron stars with such magnetic fields on the $P-\dot{P}$ diagram and the observational implications. We find that neutron stars spend a very short time before they cross the Death-Line of the $P-\dot{P}$ diagram, compared to their characteristic ages. Moreover, the maximum intensity of their surface magnetic field is substantially higher than the dipole component of the field. We argue that SGR 0418+5729 could be an example of this type of behaviour, having a weak dipole field, yet hosting a magnetic spot responsible for its magnetar behaviour. The evolution on the pulse profile and braking index of the Crab pulsar, which are attributed to an increase of its obliquity, are compatible with the anticipated drift of the magnetic axis.

\end{abstract}

\maketitle

\section{Introduction}

Even though half a century has elapsed since the discovery of pulsars \citep{Hewish:1968}, their emission mechanism is still a matter of debate \citep{Petri:2016}. The pioneering work of \cite{Goldreich:1969} set the framework for a plasma-filled magnetosphere, where radio emission is associated with the magnetic polar region \citep{Radhakrishnan:1969}, through curvature radiation \citep{Sturrock:1971, Ruderman:1975}. Based on these ideas, \cite{Arons:1979} proposed the slot-gap acceleration model, where the radiation originates at the boundary region between the open and closed field lines, with several follow-up elaborations \citep{Arons:1983, Muslimov:2003, Dyks:2003, Gil:2003}. The slot-gap model and its variants, however, require polar magnetic fields significantly higher than the dipole, with the latter being inferred by pulsar timing and determination of the rotation period $P$ and period derivative $\dot{P}$. According to these measurements, the vast majority of radio emitting pulsars have magnetic fields too weak  to power radio emission \citep{Medin:2007}. In this context, it has been argued that appropriate higher multipole components dominate the magnetic field structure near the surface of the star, providing magnetic fields sufficiently strong to power the radiative mechanisms \citep{Gil:2006}. These strong magnetic fields must be located at the polar cap, the region around the magnetic dipole axis, where open magnetic field lines emerge.

Even if such multipoles formed at birth, they would have decayed Ohmically within $100$kyr,  in the absence of any other physical processes sustaining them, given a crust conductivity $\sigma \sim10^{23-25}s^{-1}$ \citep{Potekhin:1999} and a typical size of the spot of $1$km. \cite{Goldreich:1992} showed that the Hall effect operates in the crust and it is possible to generate small-scale structure through secular evolution \citep{Pons:2009, Gourgouliatos:2014a}, instabilities \citep{Pons:2010,Rheinhardt:2002,Rheinhardt:2004, Gourgouliatos:2016b} and turbulent cascade \citep{Wareing:2009b, Wareing:2010}. These structures can actually survive for several Ohmic timescales due to the self-organisation effect of Hall evolution \citep{Gourgouliatos:2014b}.

Motivated by the question of formation and survival of magnetic spots, \cite{Geppert:2014} performed axially symmetric magnetothermal simulations in the crusts of neutron stars, exploring a broad range of initial conditions. They found that a magnetic field in the form of a twisted torus \citep{Lander:2009, Ciolfi:2013, Fujisawa:2014} where practically all the energy is contained in the toroidal component, can  push the poloidal magnetic field lines towards one of the magnetic poles. Consequently, the magnetic field near the pole becomes strong enough to activate the slot-gap mechanism. 

Theoretical and observational works have explored the presence and the role of crustal toroidal fields in neutron stars. These fields were motivated by the star's radiative properties \citep{Ho:2011, Shabaltas:2012, Storch:2014}, theoretical requirements related to stability \citep{Braithwaite:2009, Reisenegger:2009, Marchant:2011, Ciolfi:2012} or as outcomes of dynamo and differential rotation \citep{Braithwaite:2006, Spruit:2009, Mastrano:2011, Ferrario:2015, Mosta:2015}. If such fields are indeed present in the crusts of neutron stars it would be useful to understand their implications for the intermediate and long term magnetic evolution of neutron stars. 

Apart from rotation powered radio pulsars, magnetars are also believed to host spots of strong magnetic field. Their x-ray radiation can be modelled as a black-body radiation emerging from a $\sim 1$km sized hot spot \citep{Rea:2012, Rea:2013,  Guillot:2015} where the local magnetic field can be about two orders of magnitude higher than the dipole component \citep{Tiengo:2013}. These spots, unlike the ones speculated in the slot-gap model, do not need to be colocated with the magnetic dipole axis, as thermal emission from the surface does not depend on the position of the open magnetic field lines.

Therefore, the aim of this paper is to explore the 3-D evolution of the crustal magnetic field structure in neutron stars containing strong toroidal field. We focus on the effect of this evolution on physical parameters that can be extracted from observations, in particular the intensity of the magnetic dipole and its direction. Moreover, we identify the conditions favouring the formation of magnetic spots and finally the evolution of a pulsar with the simulated magnetic field on the $P-\dot{P}$ diagram.  

The plan of the paper is as follows. We set the mathematical framework in section 2. We present the results of the simulations for the models explored in section 3. We discuss the results of the simulation in section 4. We discuss the implications of these simulation to neutron stars  in section 5. We conclude in section 6.

\section{Mathematical Setup and Initial Conditions}

The crust of a neutron star can be approximated to good accuracy by an ion Coulomb lattice, where only electrons have the freedom to move \citep{Cumming:2004}. Therefore, the electric current will be carried solely by electrons so that $\bm{j} = -e n_{e} \bm{v}_{e}$, where $\bm{j}$ is the electric current density, $e$ the elementary electron charge, $n_{e}$ the electron number density and $\bm{v}_{e}$ the electron velocity. We write Amp\`ere's law $\bm{j}=\left(c/4\pi\right) \nabla \times \bm{B}$, where $c$ is the speed of light and $\bm{B}$ the magnetic induction; and Ohm's law $\bm{j}= \sigma \left(\bm{E} +\bm{v}_{e}\times \bm{B}/c\right)$, where $\bm{E}$ is the electric field. Substituting these expressions into Faraday's law yields the following equation: 
\begin{eqnarray}
\frac{\partial B}{\partial t}= -\nabla \times \left(\frac{c}{4 \pi e n_{e}}\left(\nabla \times \bm{B}\right)\times \bm{B} +\frac{c^{2}}{4 \pi \sigma}\nabla \times \bm{B}\right)\,,
\label{HALL}
\end{eqnarray} 
whose first term in the right hand side describes the evolution of the magnetic field under the Hall effect, and the second term the Ohmic decay \citep{Goldreich:1992}. We  then integrate (\ref{HALL}) in time numerically and follow the evolution of the magnetic field using the Meissner condition for the internal boundary \citep{Hollerbach:2002, Hollerbach:2004} and a current-free field for the exterior. We do so using a 3-D code, a suitably modified version of the PARODY code of \cite{Aubert:2008},  that has been utilised in previous studies \citep{Wood:2015, Gourgouliatos:2016a}. The crust and conductivity parameters used are the same as in the above mentioned papers. 

We examine how a toroidal field of geometry $\ell=1$ affects the evolution of a poloidal dipole, a configuration that has the form of a twisted torus. The exact profiles we use to initialise the simulations are those given in equation (5) and (6) of \cite{Gourgouliatos:2016a} with a superimposed small non-axisymmetric perturbation containing less than $10^{-4}$ of the total magnetic energy. We explore states where the initial toroidal field contains $50\%$, $90\%$ and $99\%$ of the total magnetic energy, and cases where the axis of the dipole field is inclined with respect to that of the toroidal field (see Table \ref{Table:1}). We have run the simulations long enough so that we can trace a pulsar's life, up to the point where it crosses the so-called Death Line \citep{Bhattacharya:1992, Chen:1993} in the $P-\dot{P}$ diagram, assuming an initial period of $50$ms.
\begin{table}
\caption{Summary of the simulations performed. The first column indicates the name of the run; for subsequent columns: $B_{d}$ is the initial dipole magnetic field, $B_{max}$  is the maximum value of the magnetic field in the crust, $E_{tot}$ is the total magnetic energy stored in the crust, $E_{tor}/E_{tot}$  is the ratio of the energy of the toroidal magnetic field to the poloidal field, $\alpha$ is the inclination between the magnetic dipole axis to the axis of symmetry of the toroidal field and the last column is $t_{d}$ the time a pulsar will ned to cross the Death Line in the $P-\dot{P}$ diagram.} 
\centering
\begin{tabular}{l l l r l l l} \hline\hline
NAME & $B_{d}$ & $B_{max}$ & $E_{tot}$ & $E_{tor}/E_{tot}$ & $\alpha$& $t_{d}$\\ 
 & $(10^{14}$G) &($10^{15}$G)& $(10^{46}$erg) & & (deg)& (Myr) \\ \hline
\hline
A50-1 &0.50& 0.59&  0.56 &	0.50	&0& 1.7\\
A50-2 &1.0&1.2&2.25&0.50	&0& 1.2\\
A50-3&2.0&2.4&9.00	&0.50	&0&0.8\\
A50-4&4.0&4.7&18.00	&0.50	&0&0.55\\
\hline
A90-1 &0.23& 0.61&  0.56 &	0.90	&0&2.8\\
A90-2&0.46&1.2&	2.25	& 0.90 & 0&2.0\\
A90-4&1.8&4.9&18.00	&0.90	&0&0.9\\
\hline
A99-1&0.071& 0.61&  0.56 &	0.99&	0&4.7\\
A99-2 &0.14&1.3&2.25&0.99	&0&4.3\\
A99-4&0.57&4.9&18.00	&0.99	&0&2.3\\
\hline
B90-2&0.46& 1.5&  2.25 &	0.90&	45&1.8\\
B90-4 &1.8&5.2 &18.00&0.90	&10&0.84\\
B99-4&0.57&5.1&18.00	&0.99	&45&2.0\\
\hline  \hline
\end{tabular}
\label{Table:1}
\end{table}

\section{Results}
\begin{figure*}
\includegraphics[width=0.68\columnwidth]{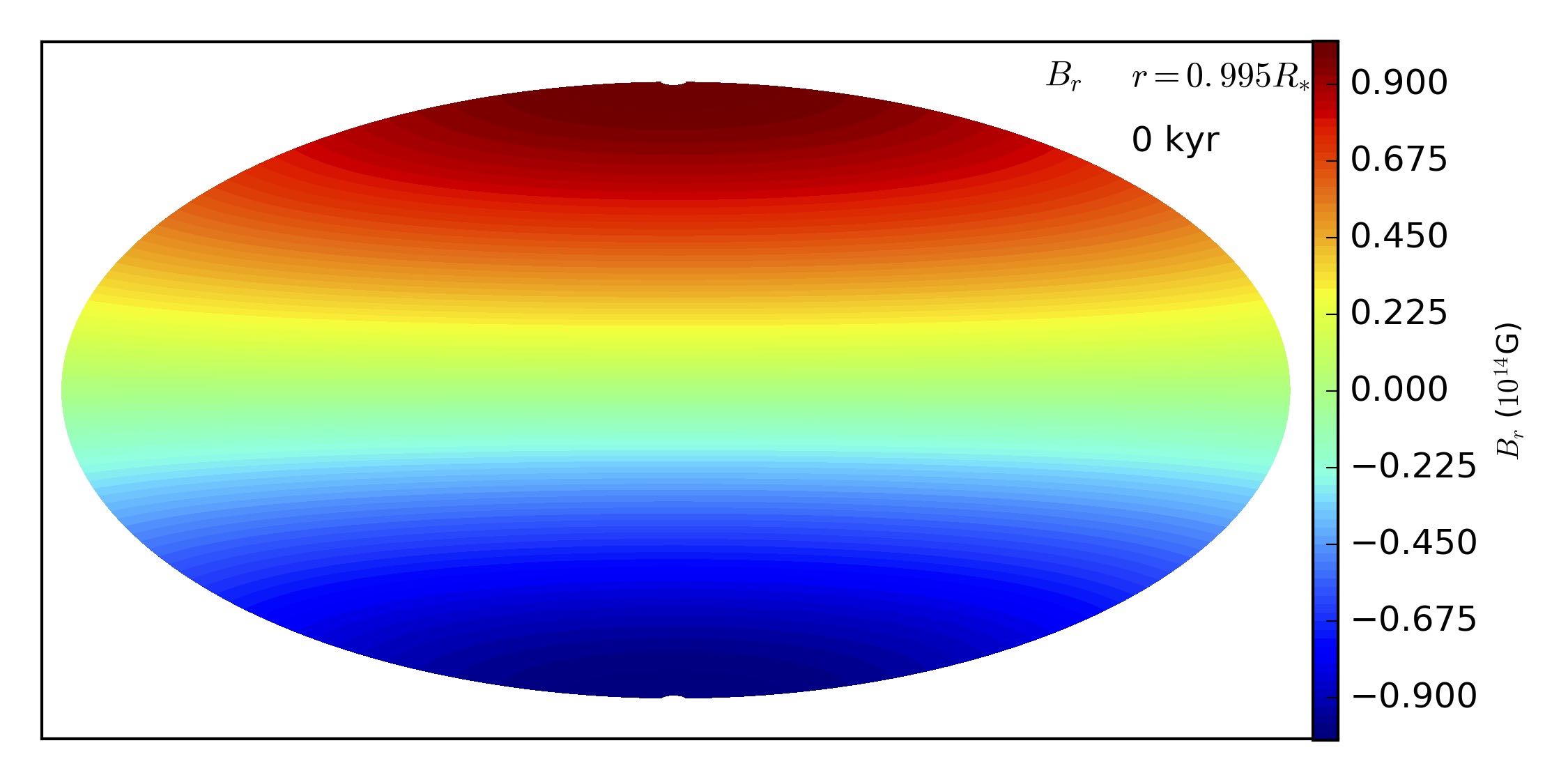}
\includegraphics[width=0.68\columnwidth]{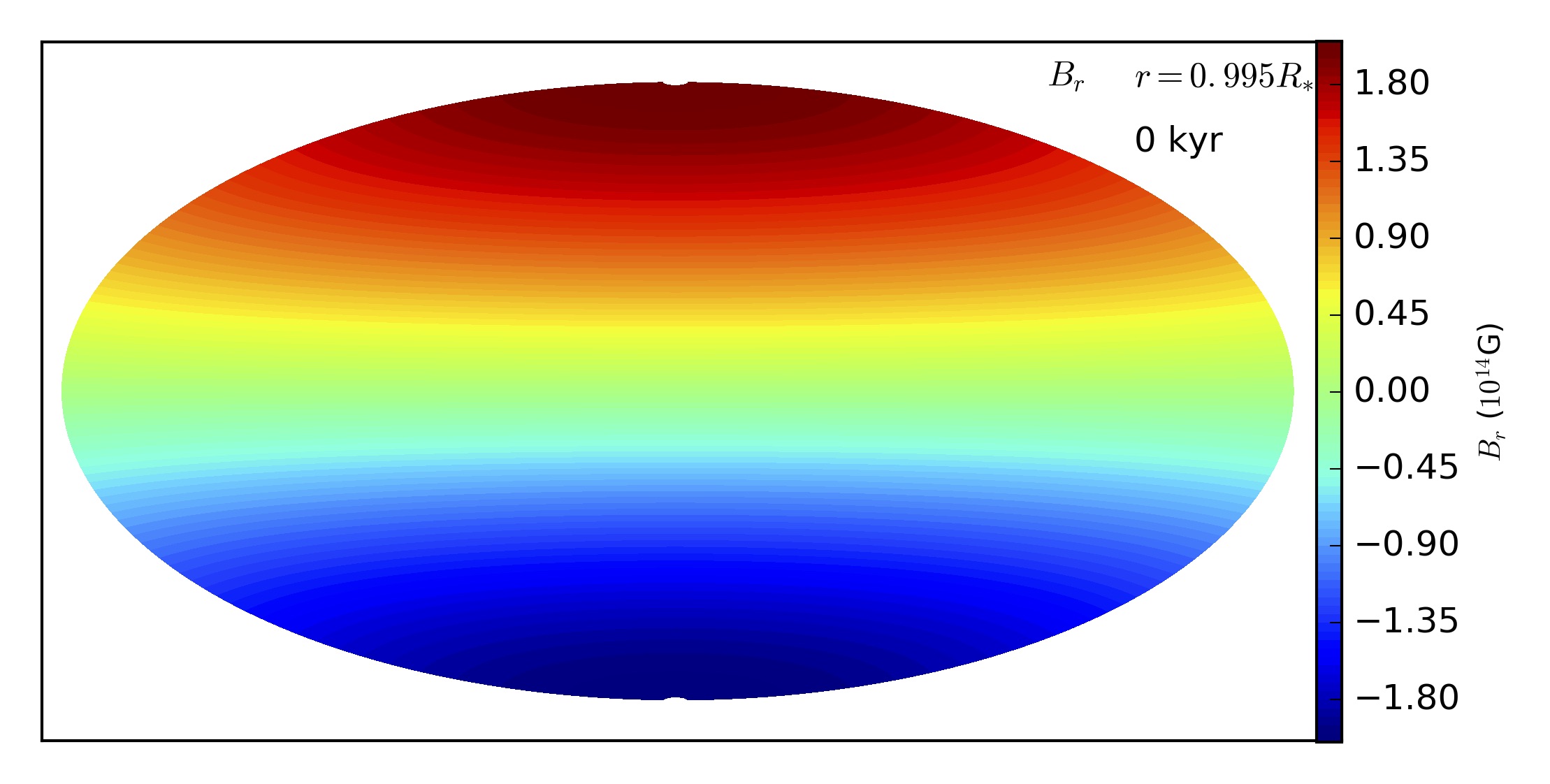}
\includegraphics[width=0.68\columnwidth]{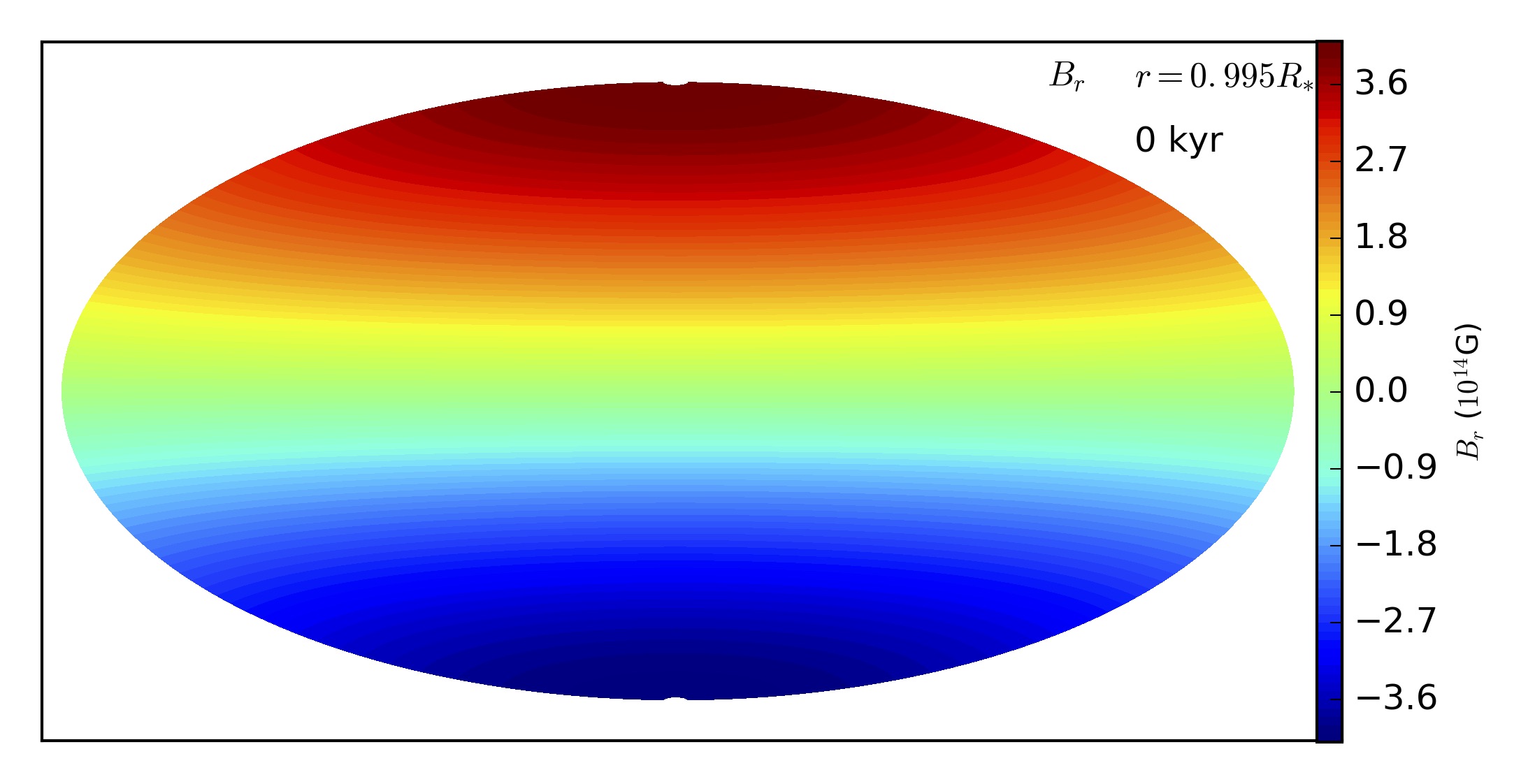}

\includegraphics[width=0.68\columnwidth]{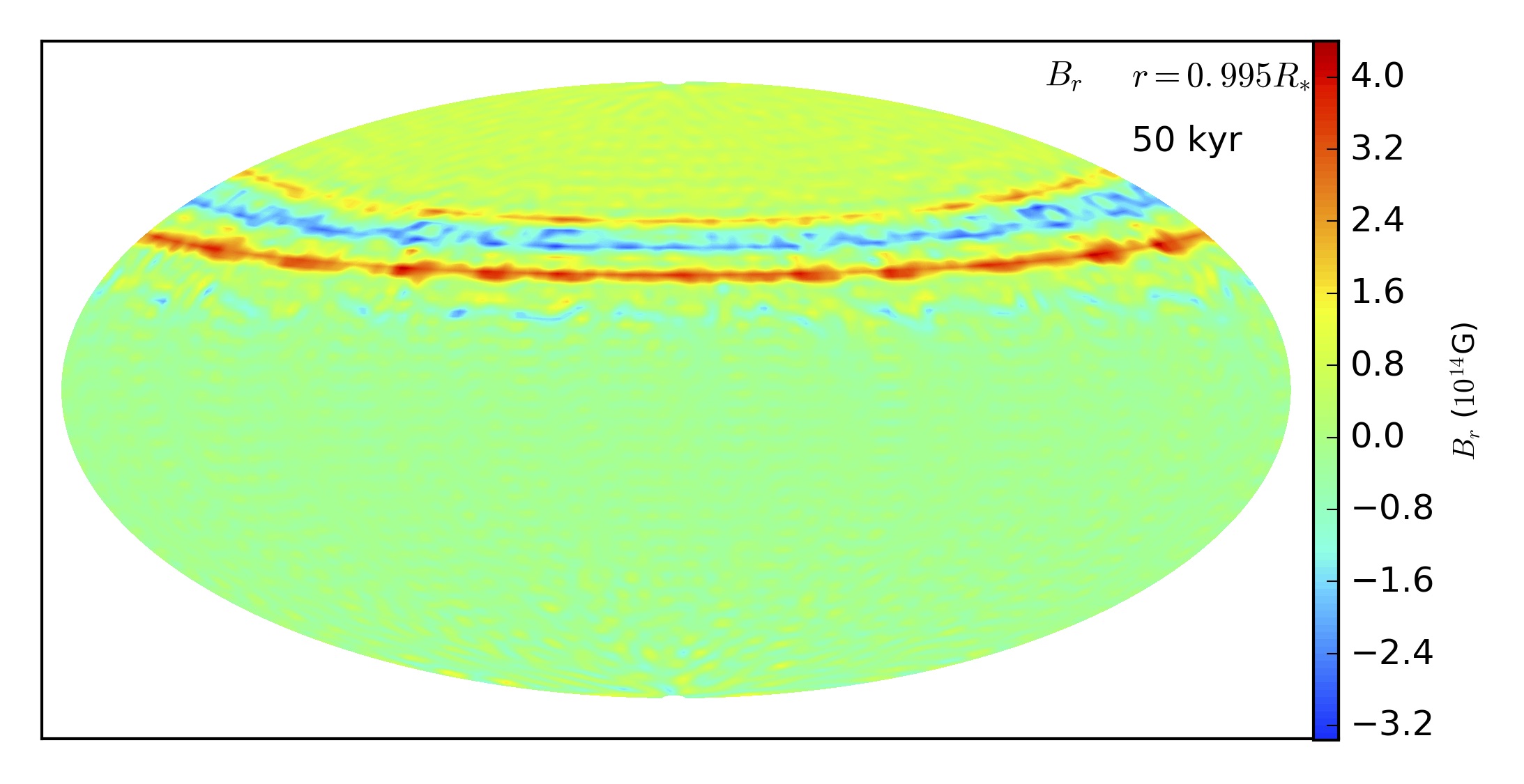}
\includegraphics[width=0.68\columnwidth]{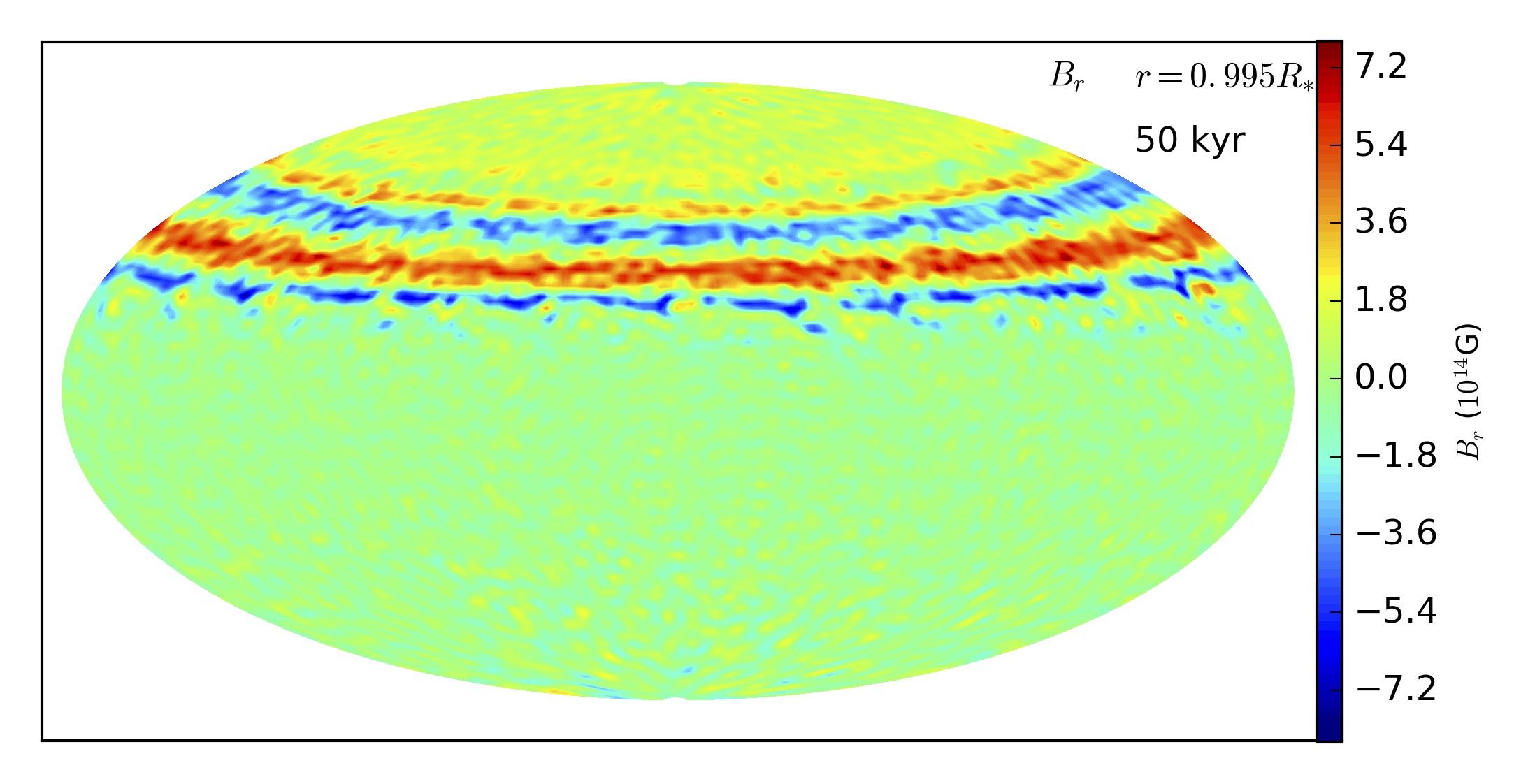}
\includegraphics[width=0.68\columnwidth]{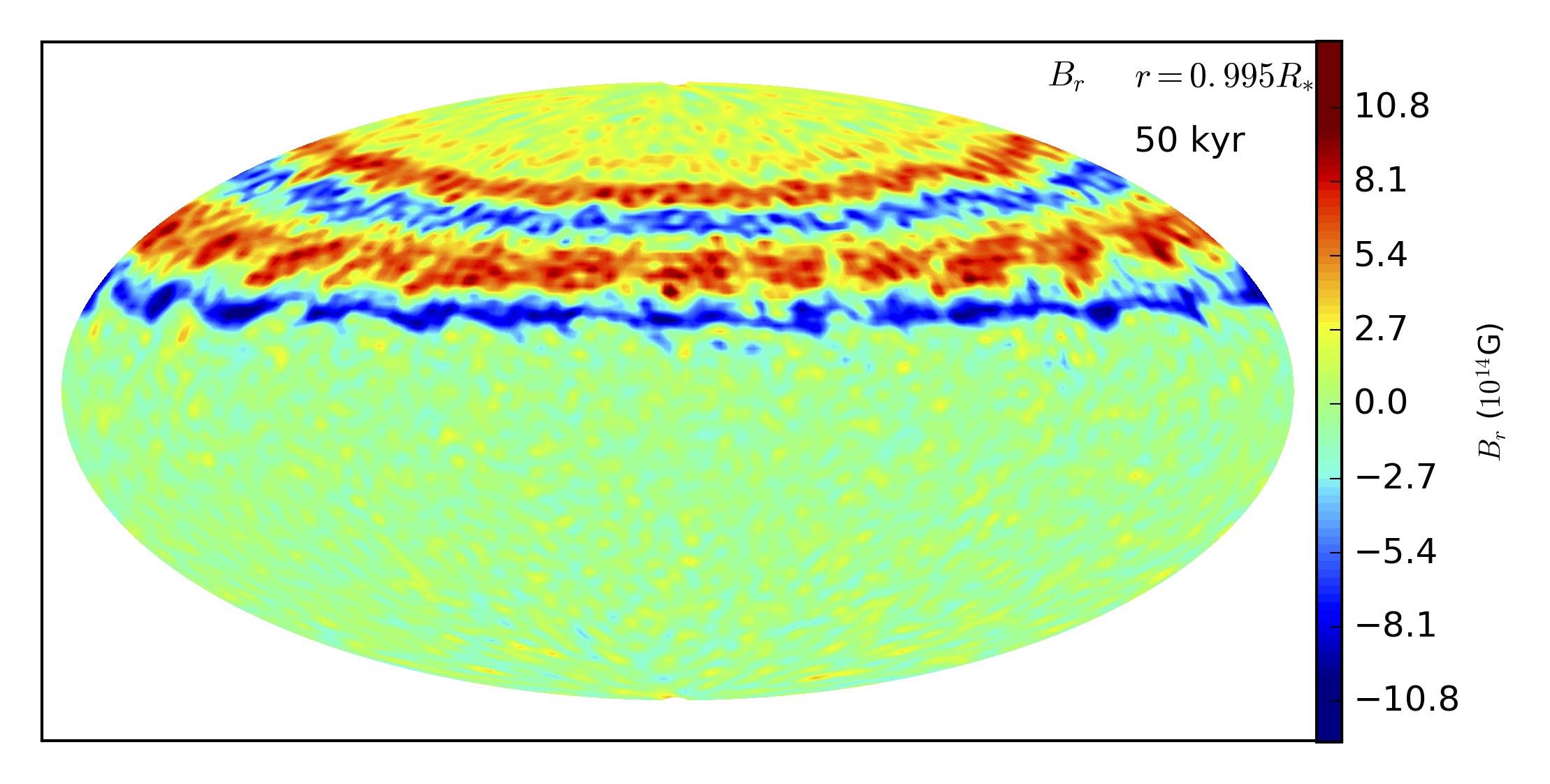}

\includegraphics[width=0.68\columnwidth]{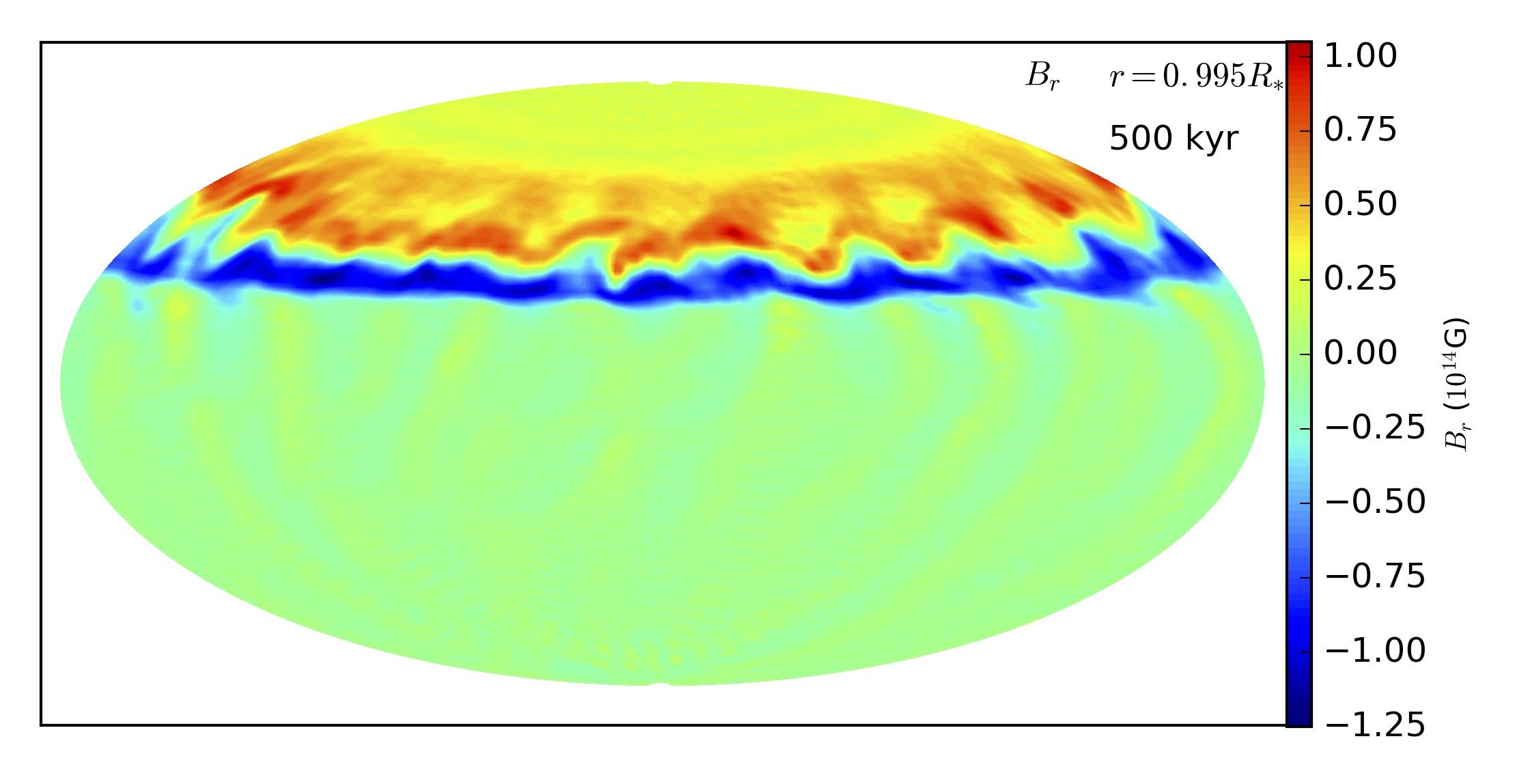}
\includegraphics[width=0.68\columnwidth]{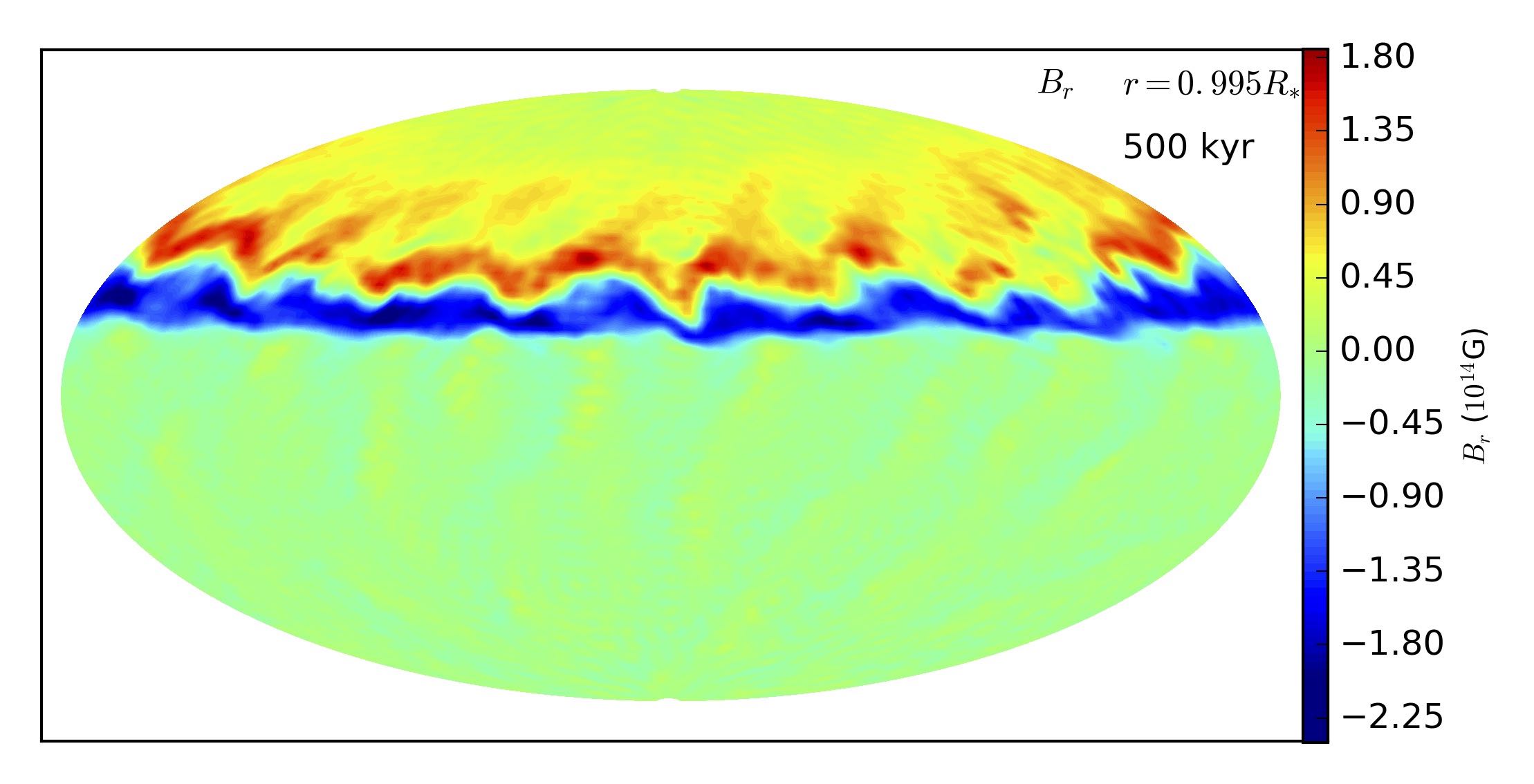}
\includegraphics[width=0.68\columnwidth]{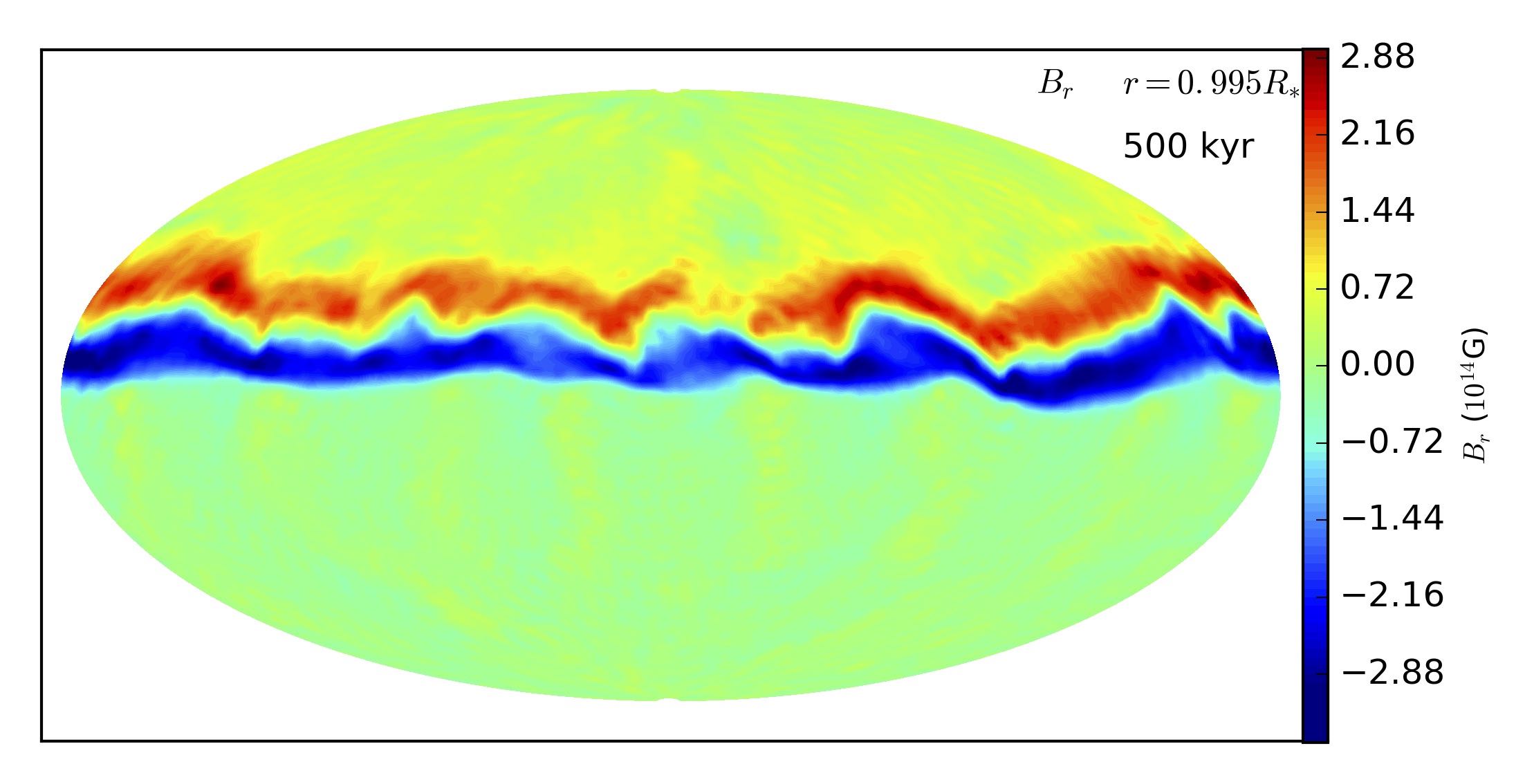}

\includegraphics[width=0.68\columnwidth]{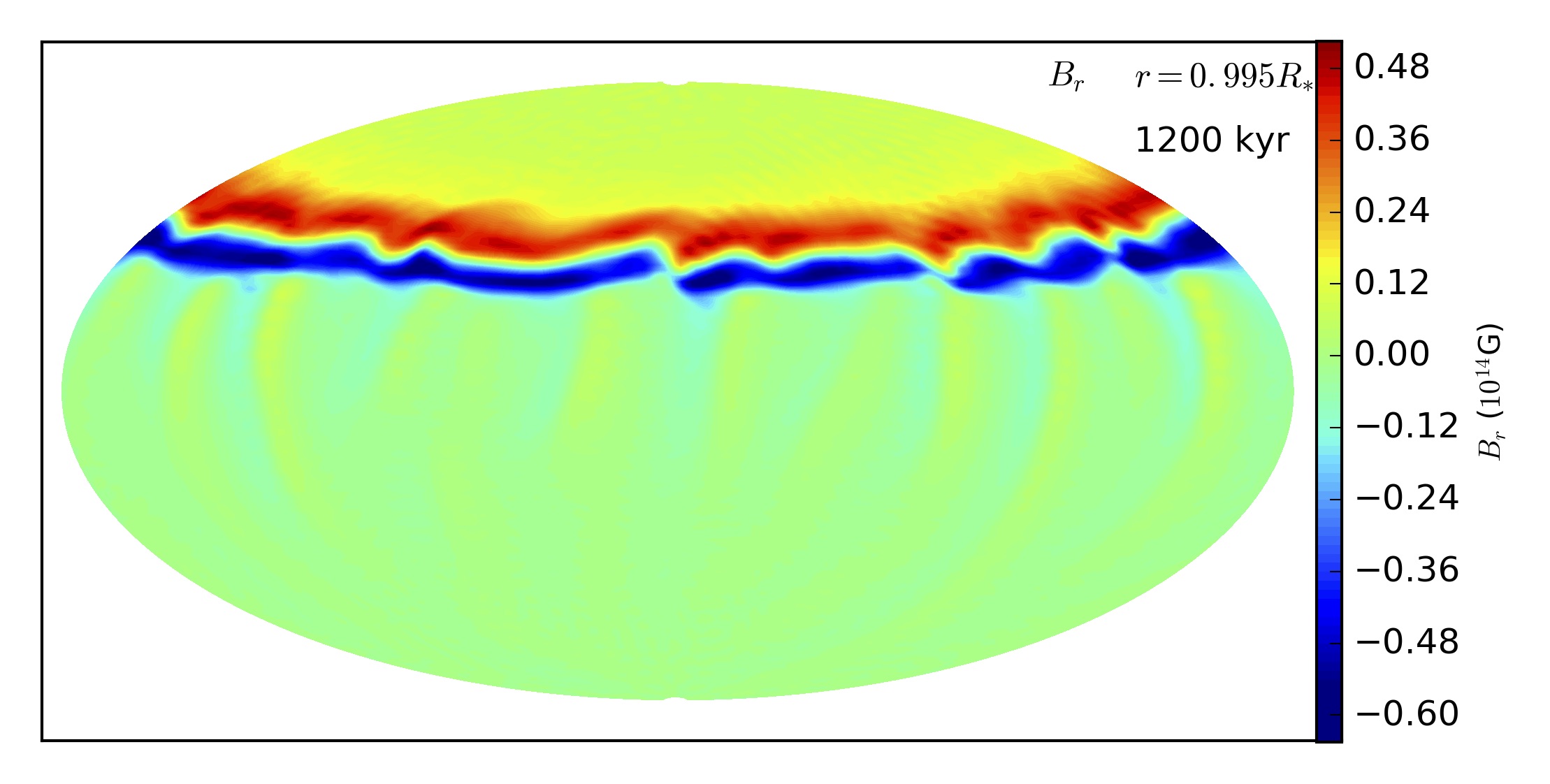}
\includegraphics[width=0.68\columnwidth]{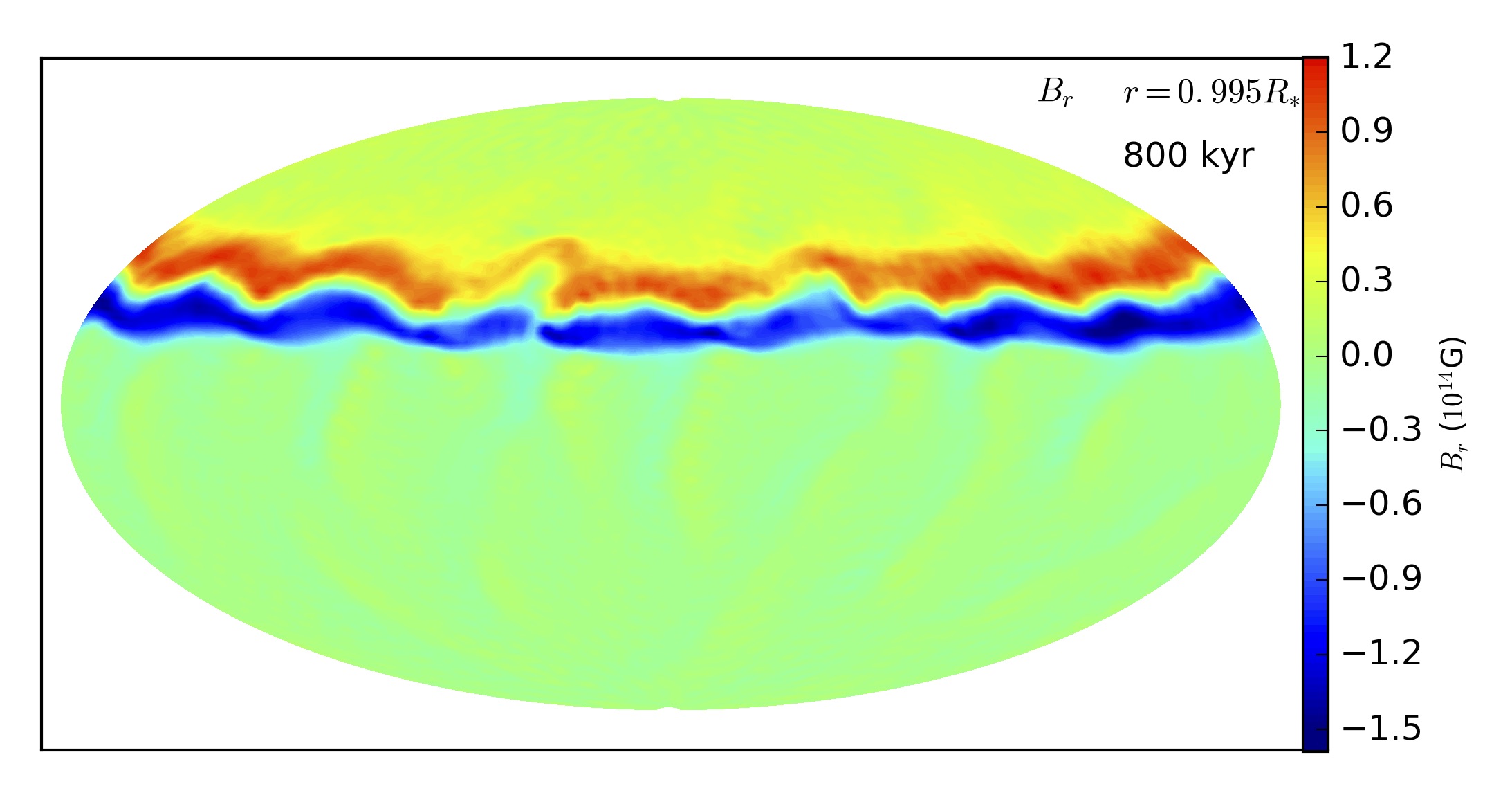}
\includegraphics[width=0.68\columnwidth]{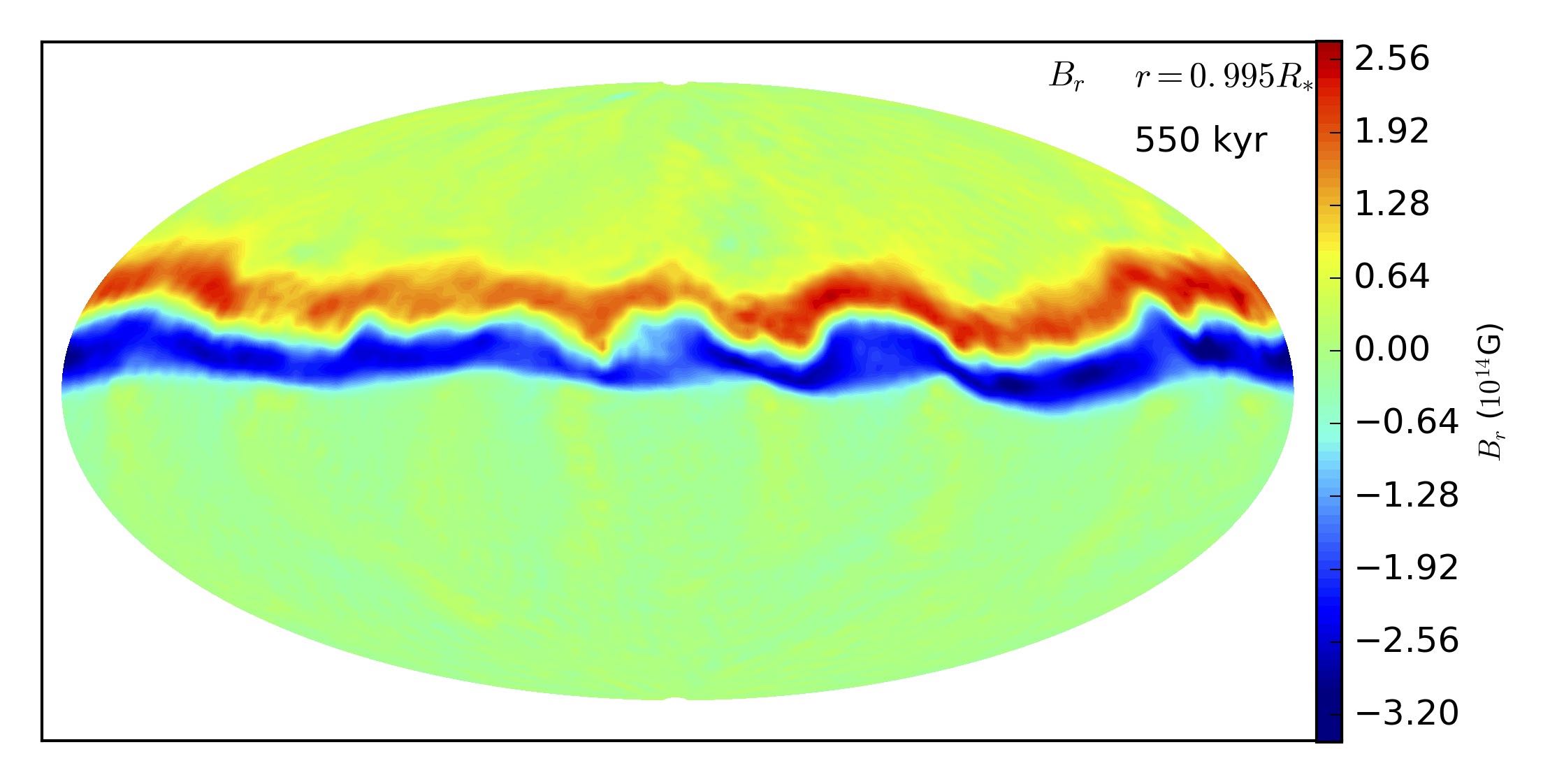}
\caption{Contour plot of the radial component of the magnetic field at $r=0.995R_{*}$, where $R_{*}$ is the neutron star radius,  for models where $50\%$ of the initial energy is in the toroidal field. The plots correspond to times $t=0$ (1st row), $t=50$kyr (2nd row), $t=500$kyr (3rd row), and the time at which a pulsar crosses the Death Line in the $P-\dot{P}$  diagram (fourth row).  The first column shows model A50-2, the second one A50-3 and the third one A50-4.}
\label{Figure:1}
\end{figure*}
\begin{figure*}
\includegraphics[width=0.68\columnwidth]{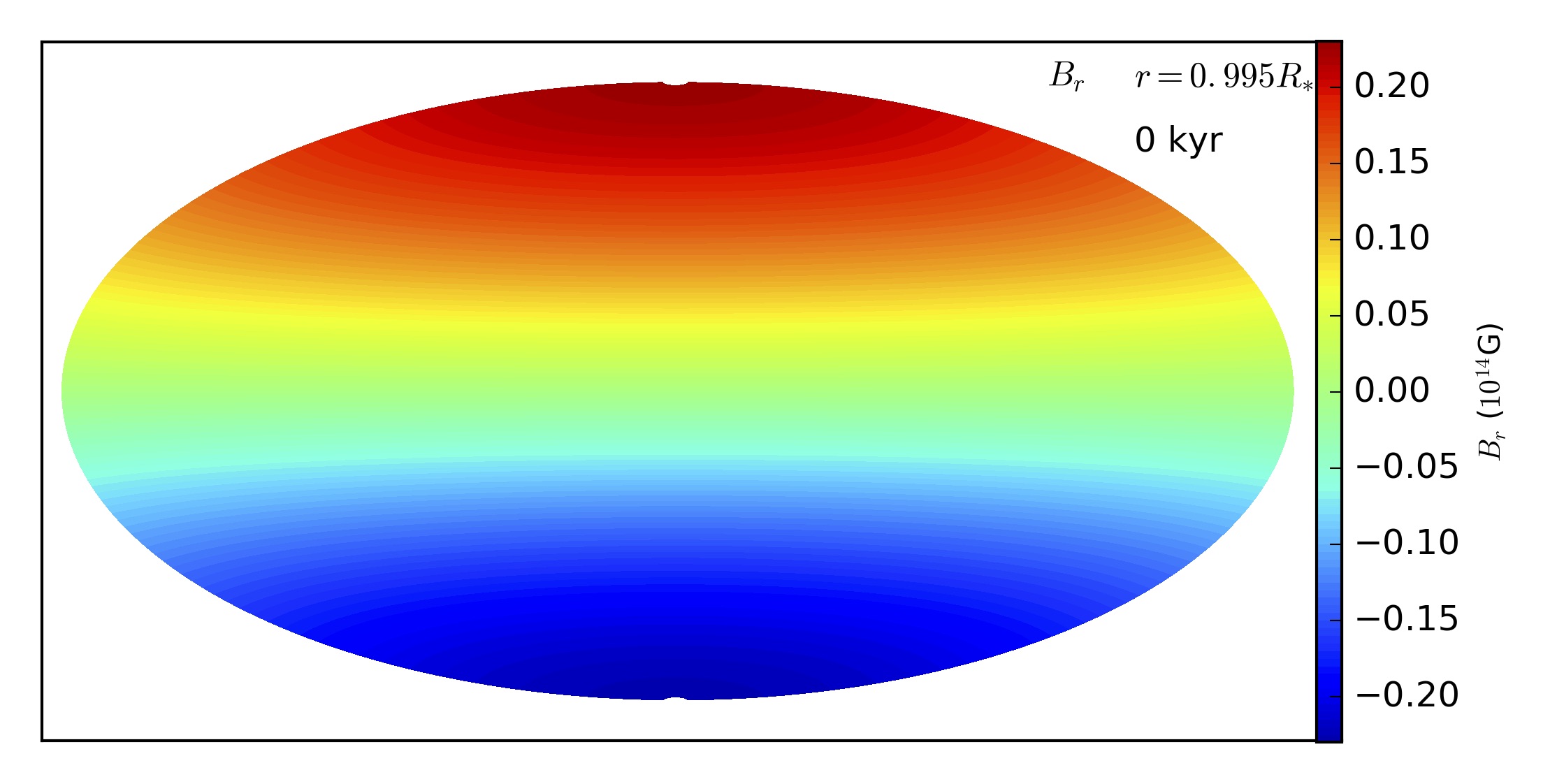}
\includegraphics[width=0.68\columnwidth]{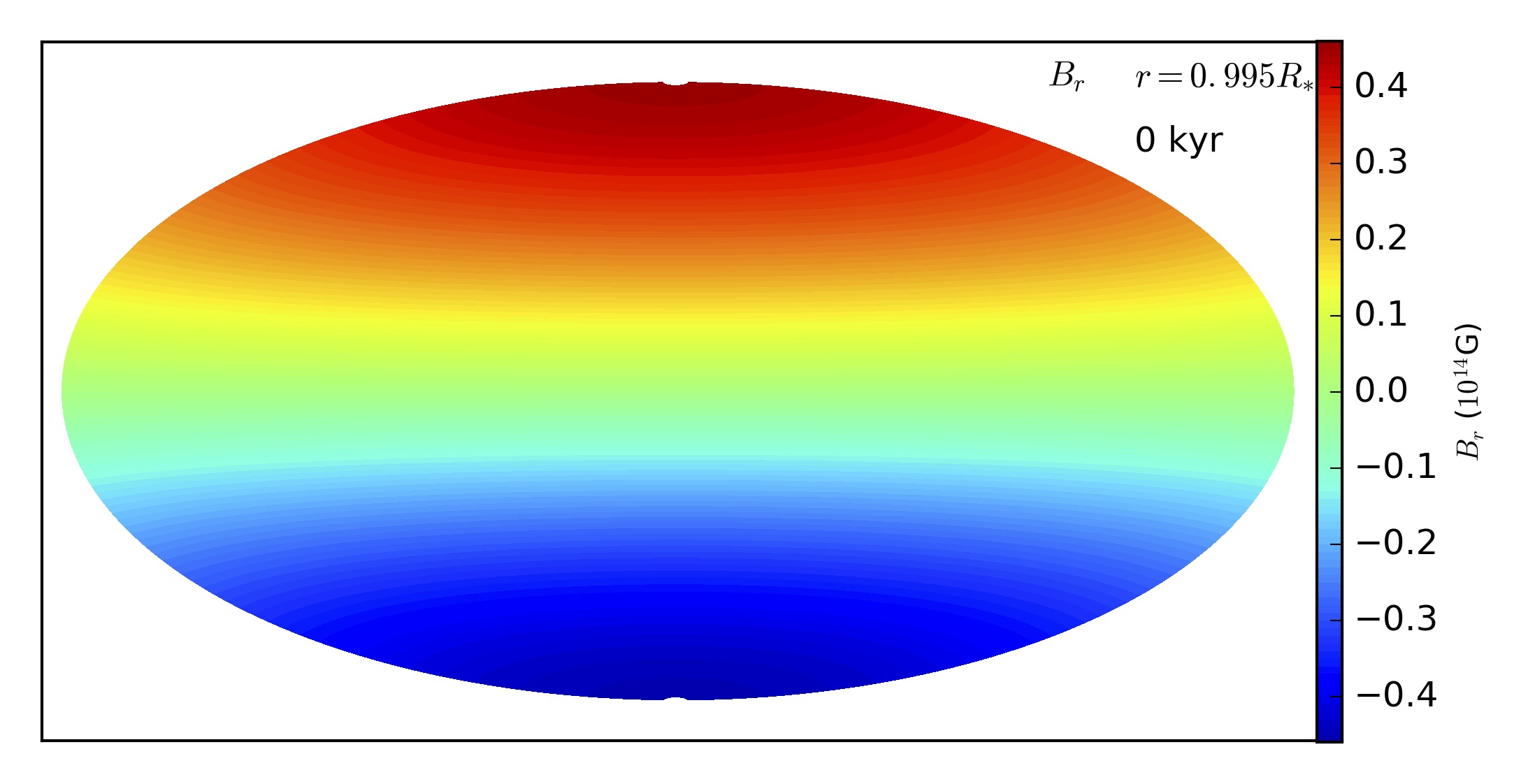}
\includegraphics[width=0.68\columnwidth]{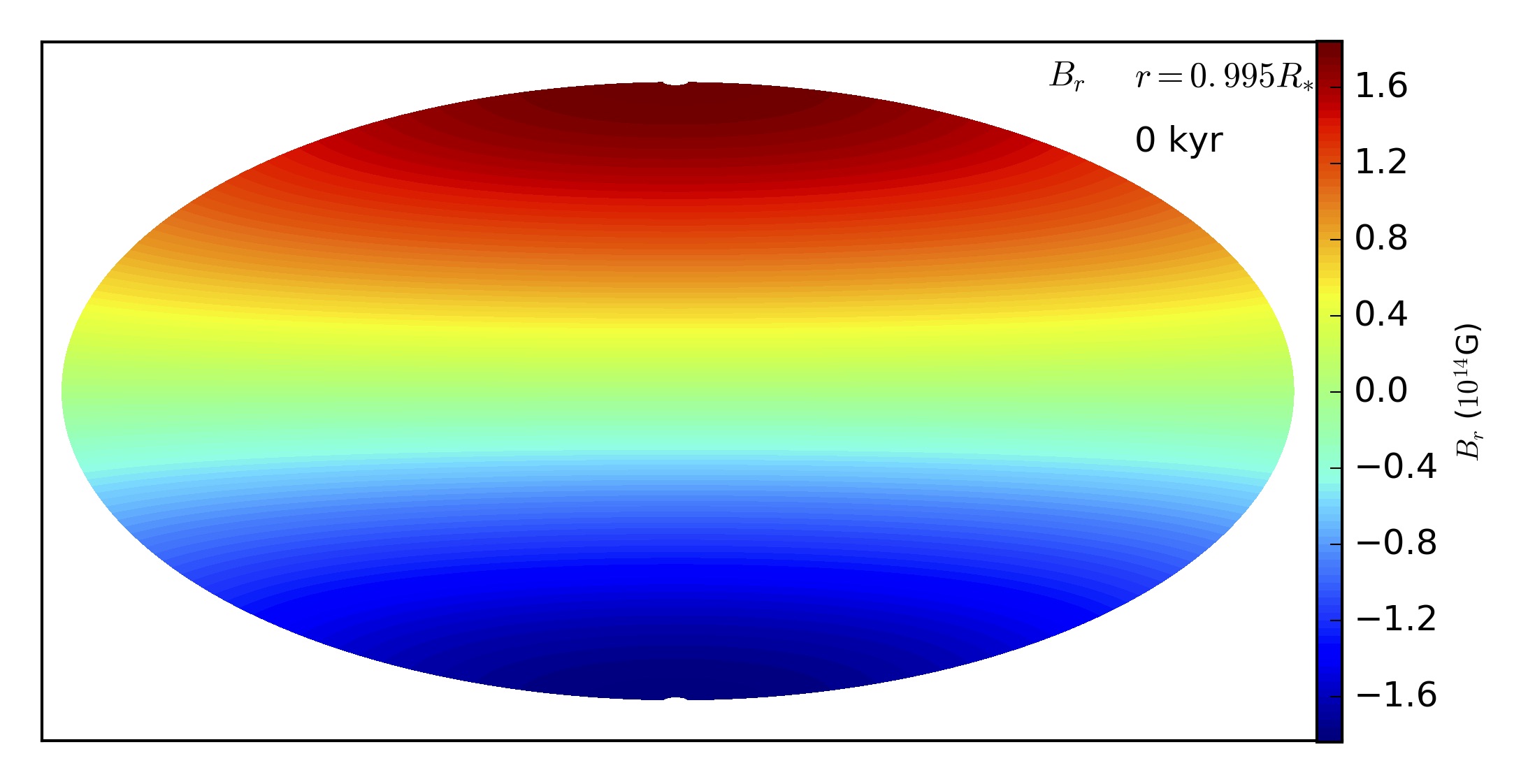}

\includegraphics[width=0.68\columnwidth]{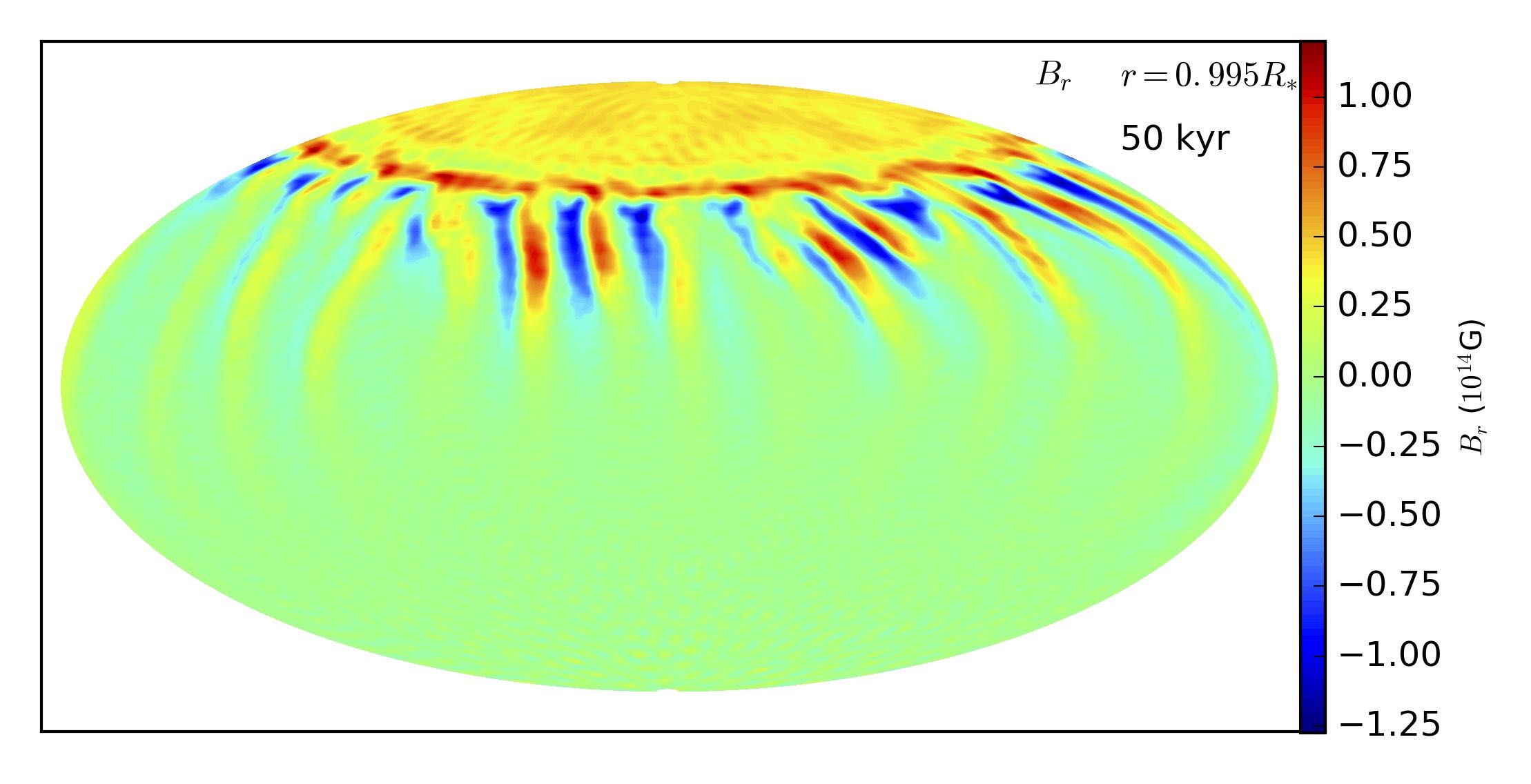}
\includegraphics[width=0.68\columnwidth]{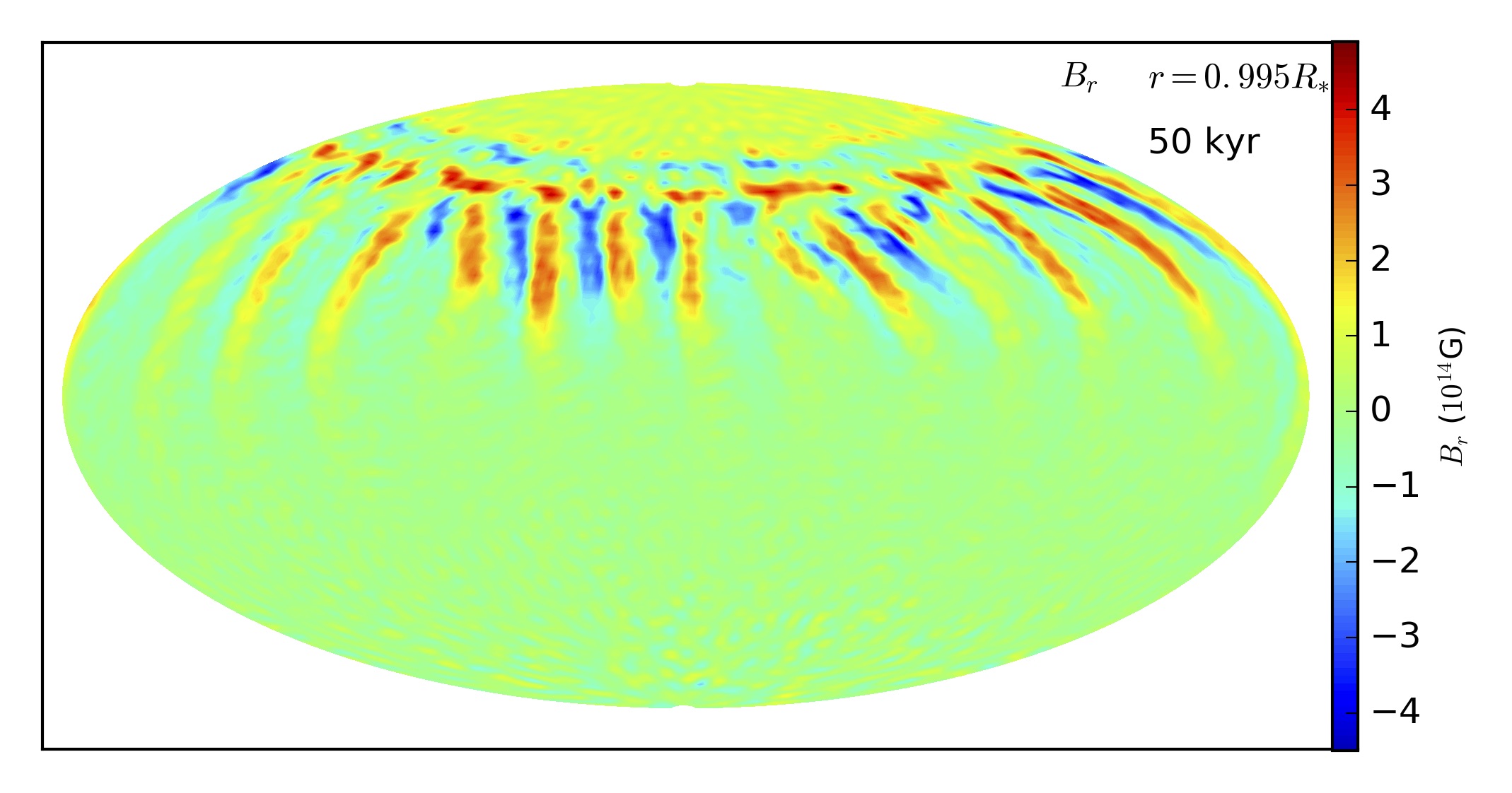}
\includegraphics[width=0.68\columnwidth]{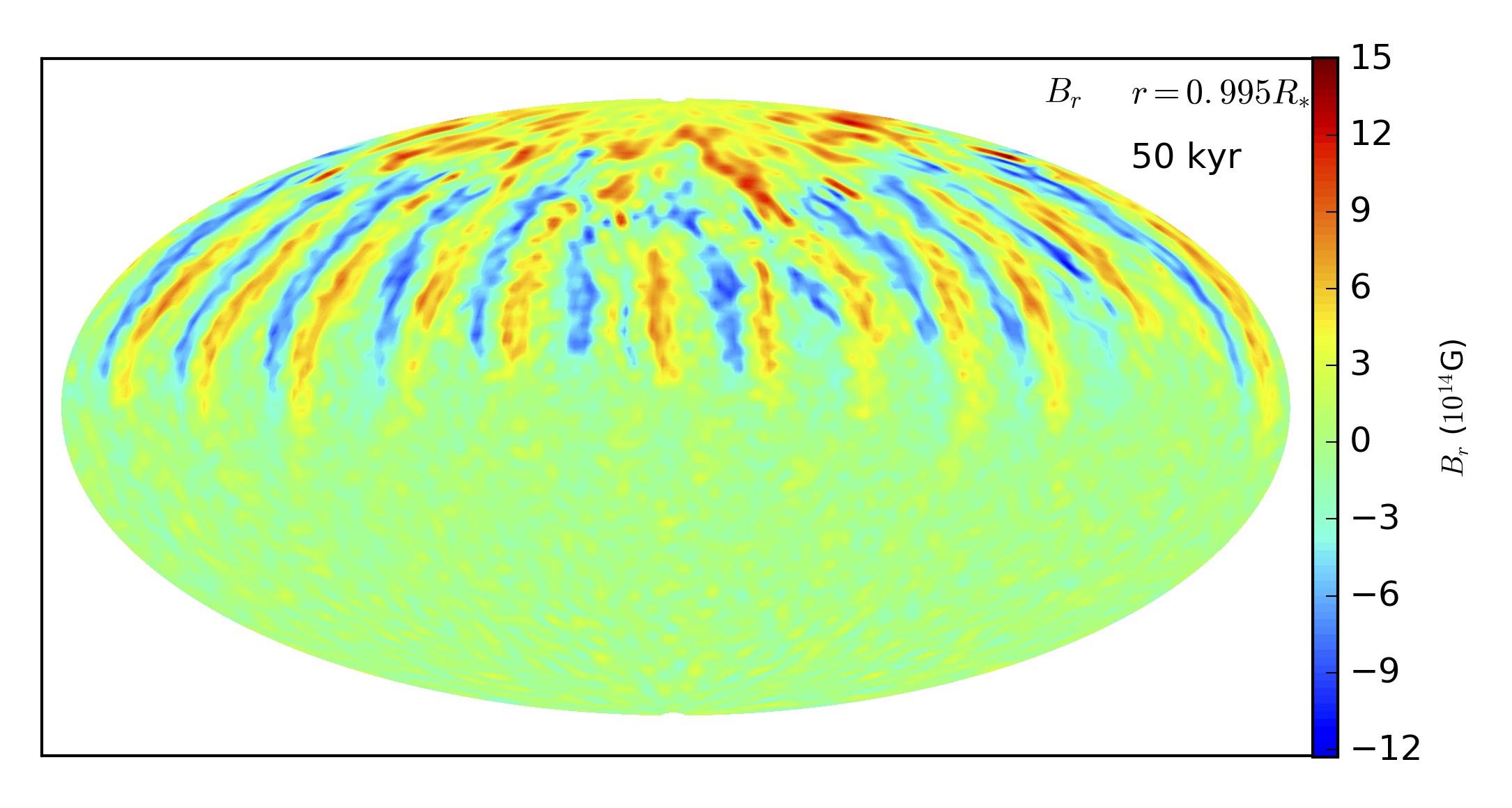}

\includegraphics[width=0.68\columnwidth]{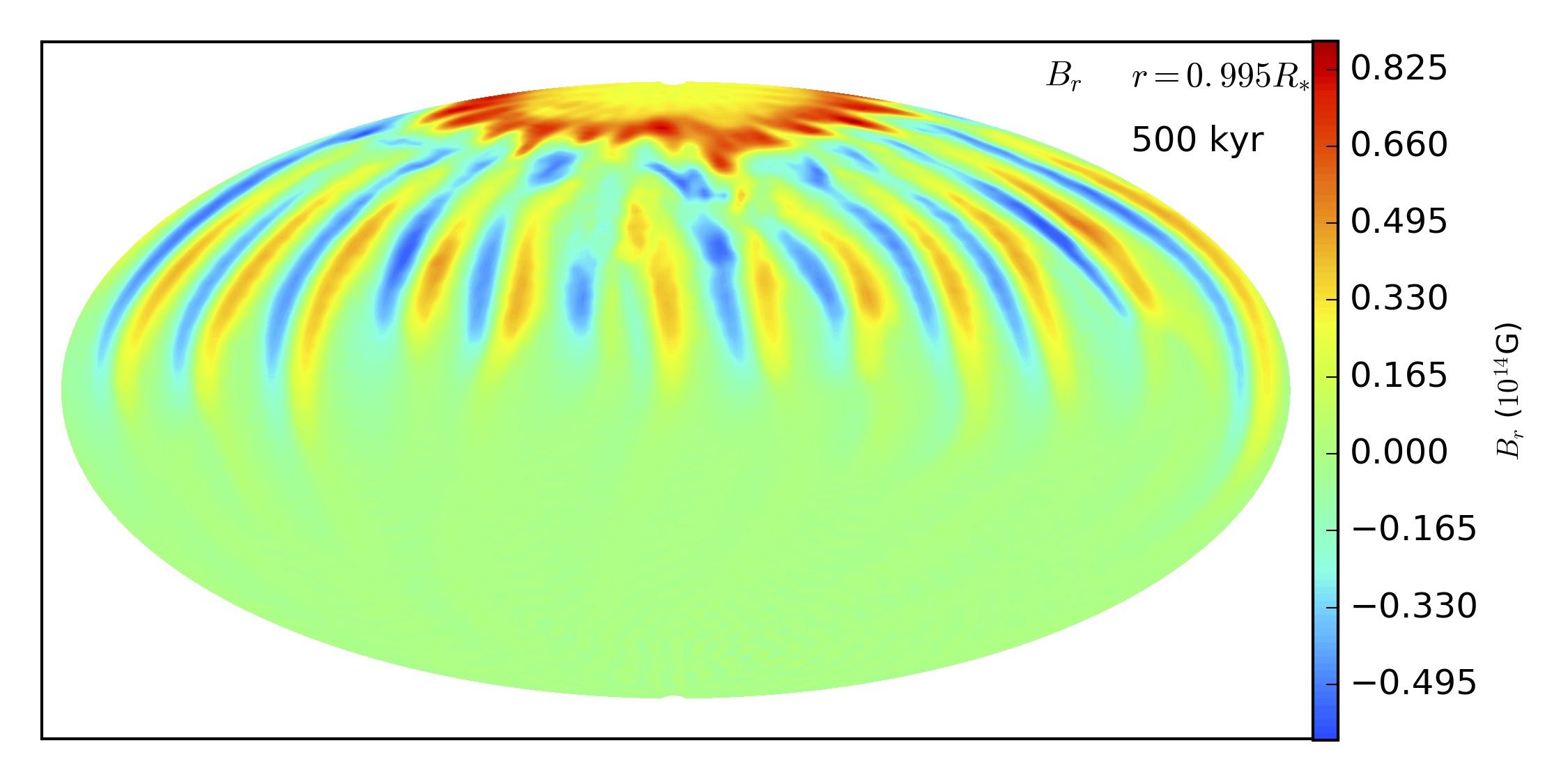}
\includegraphics[width=0.68\columnwidth]{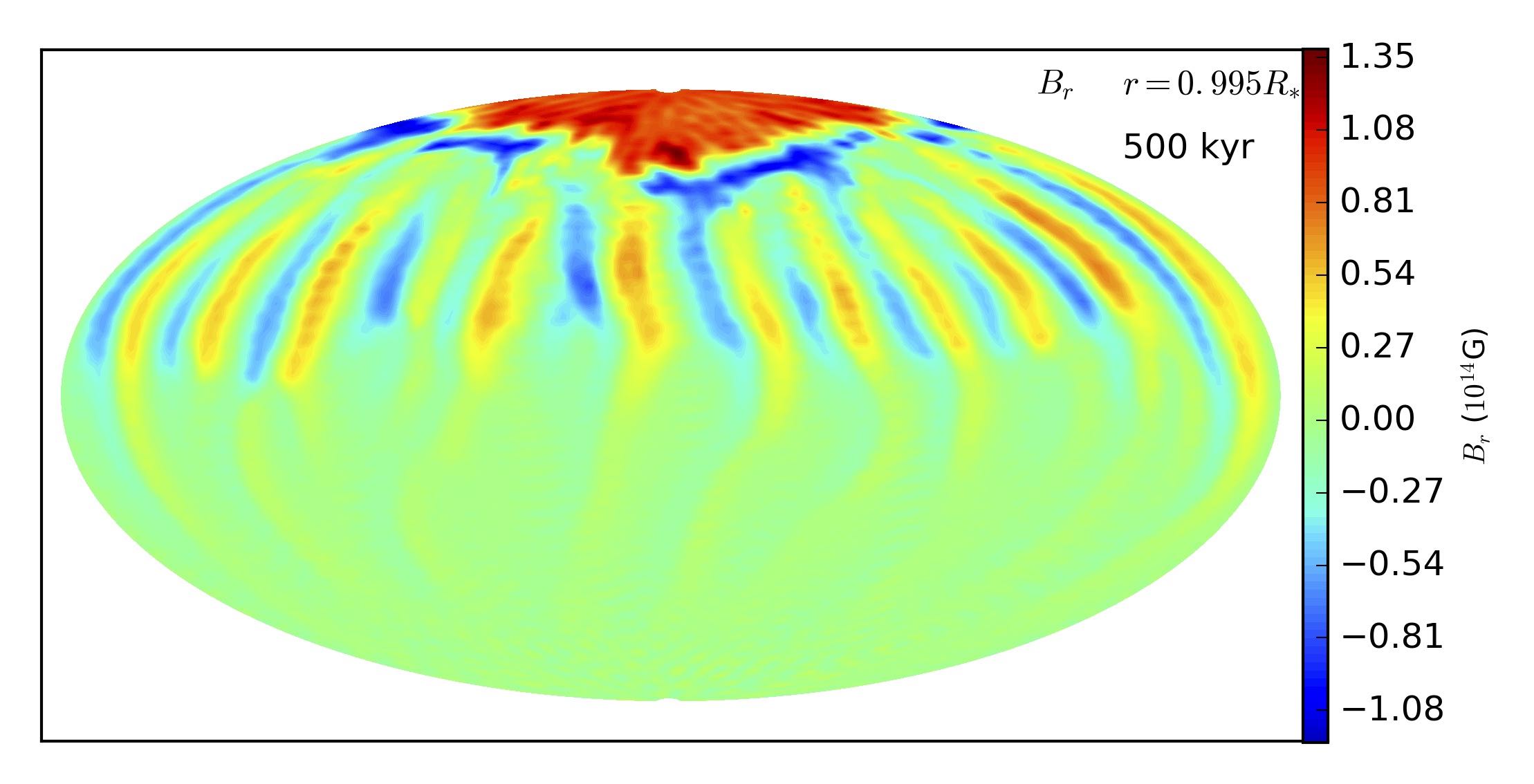}
\includegraphics[width=0.68\columnwidth]{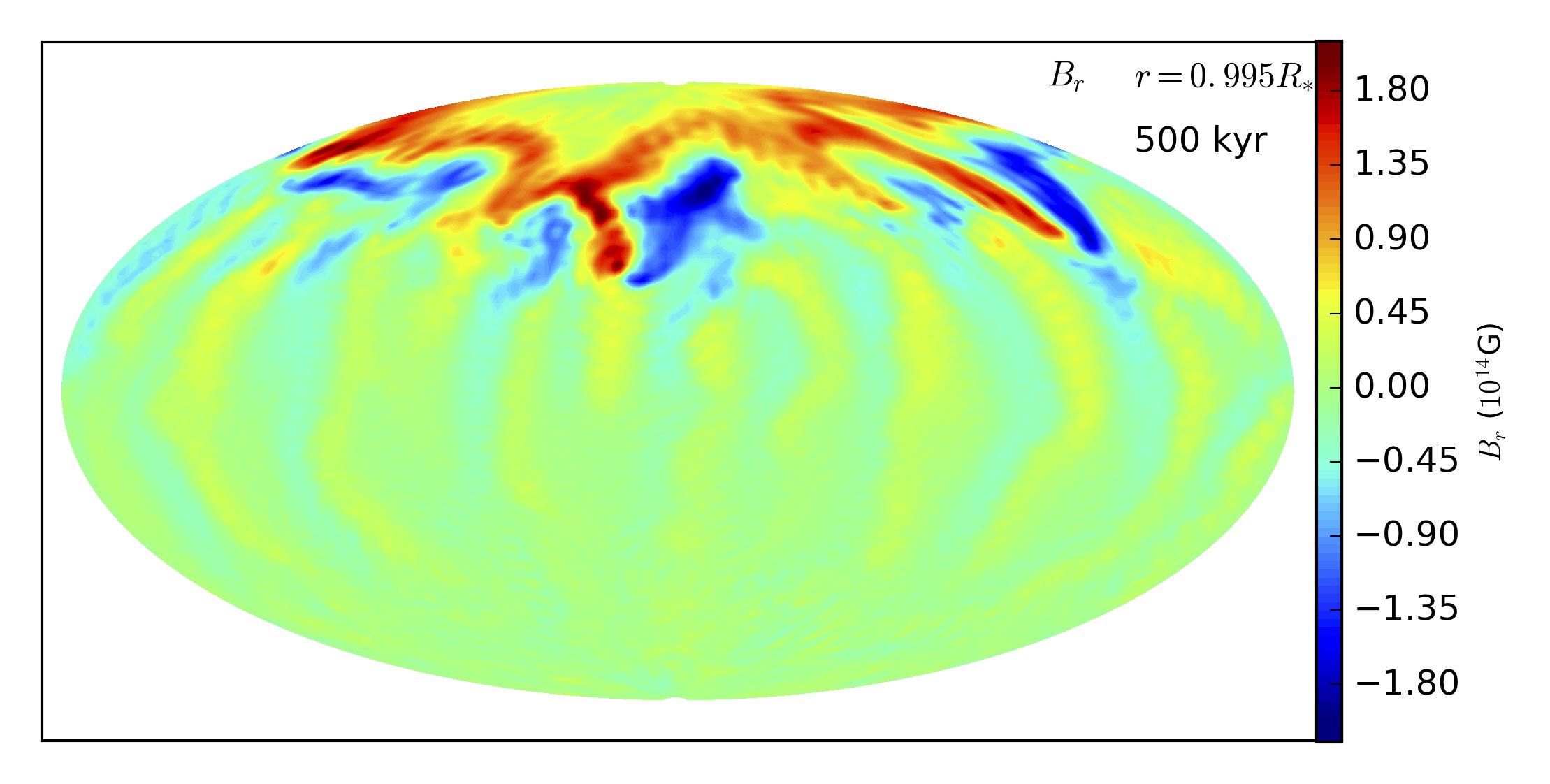}

\includegraphics[width=0.68\columnwidth]{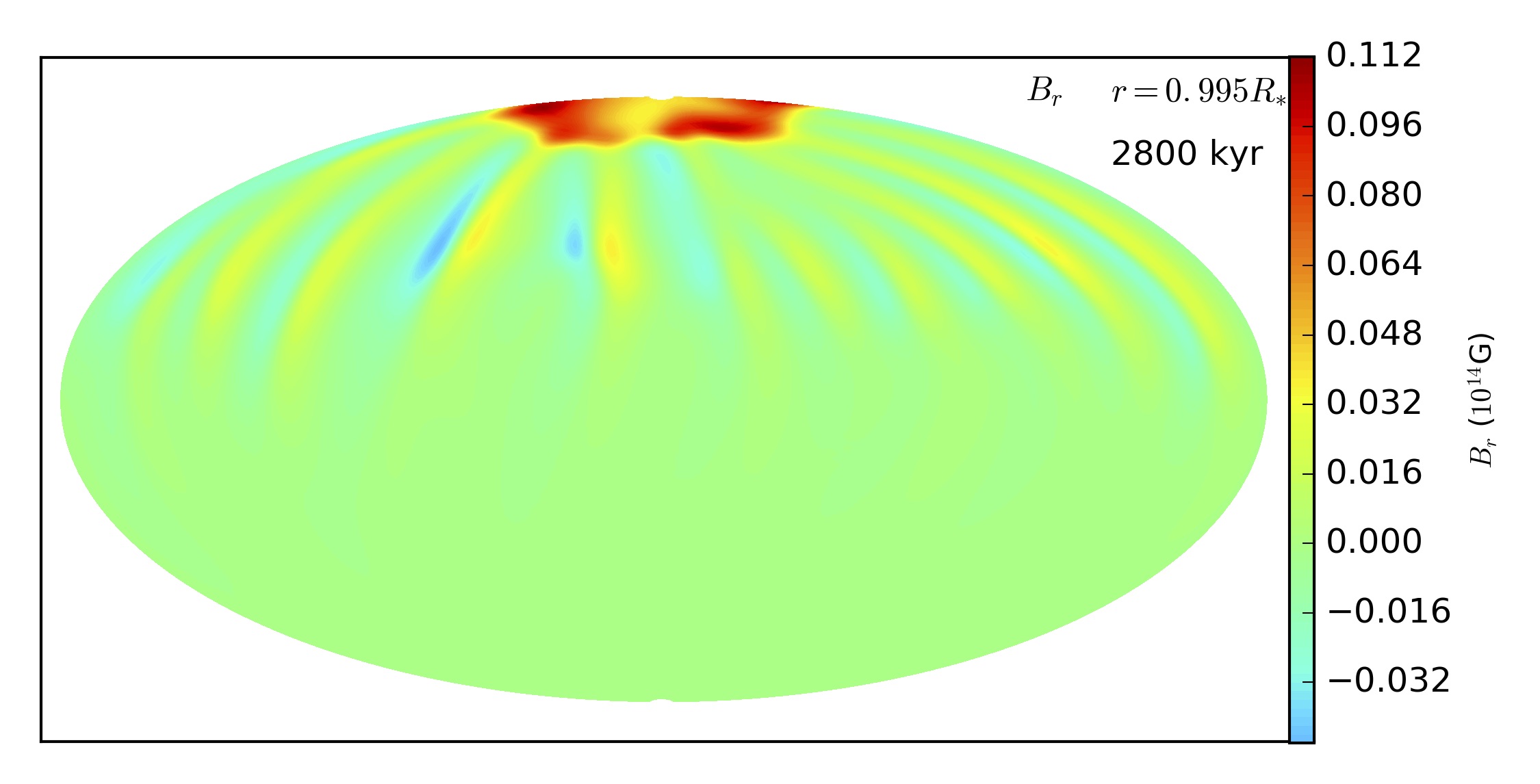}
\includegraphics[width=0.68\columnwidth]{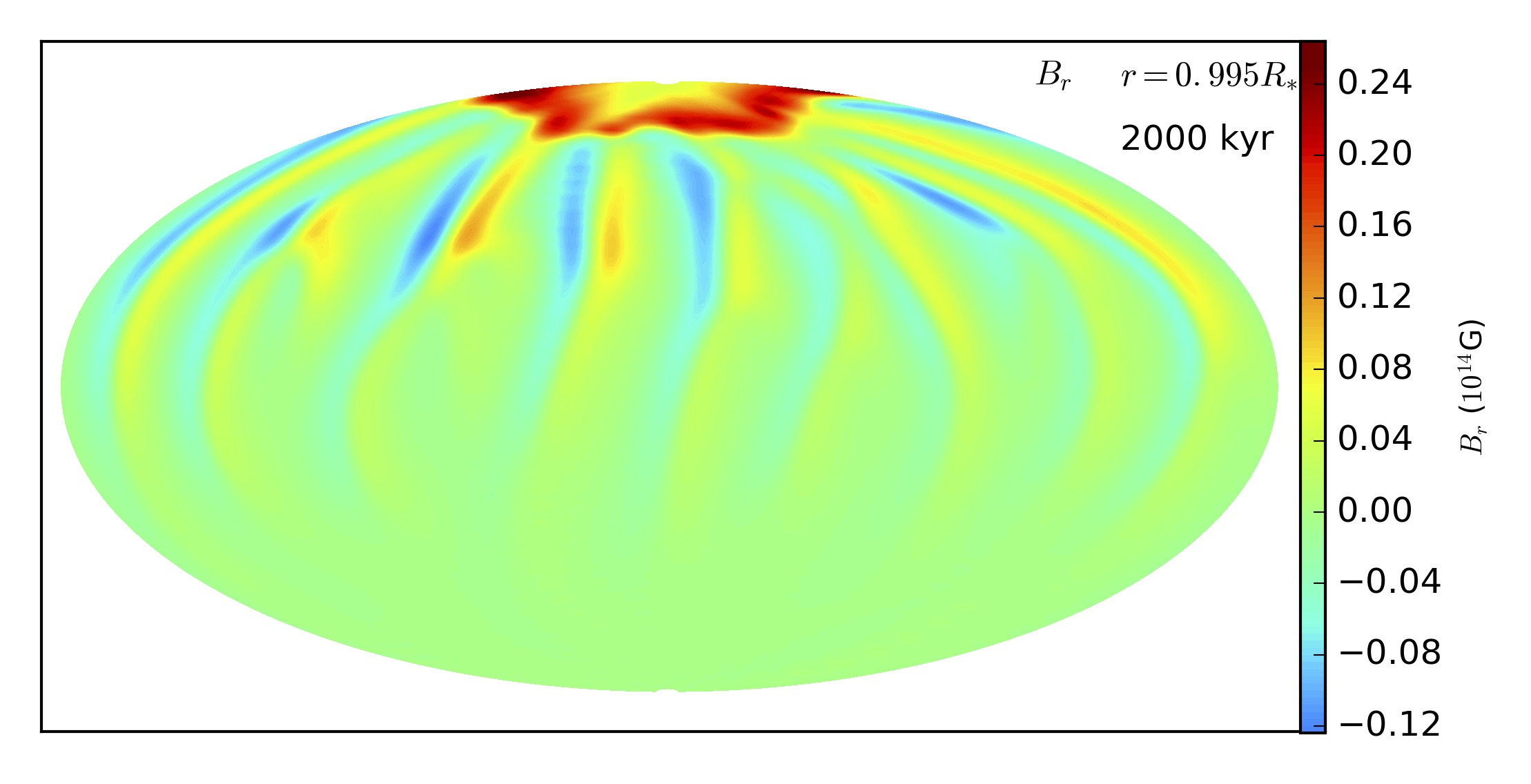}
\includegraphics[width=0.68\columnwidth]{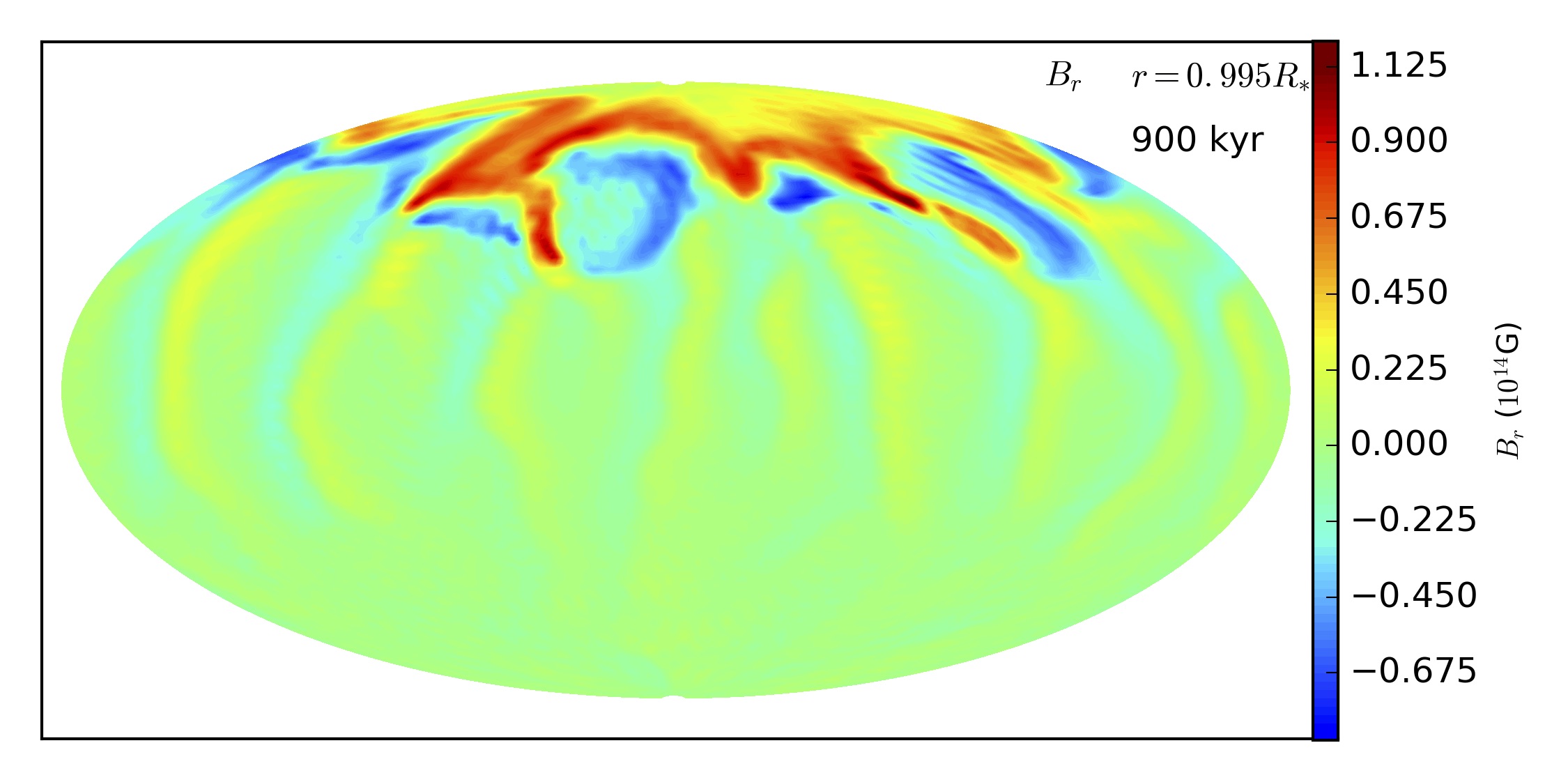}
\caption{Same as Figure \ref{Figure:1}, but for models where $90\%$ of the energy is in the toroidal field. The first column shows model A90-1, the second one A90-2 and the third one  A90-4.}
\label{Figure:2}
\end{figure*} 
\begin{figure*}
\includegraphics[width=0.68\columnwidth]{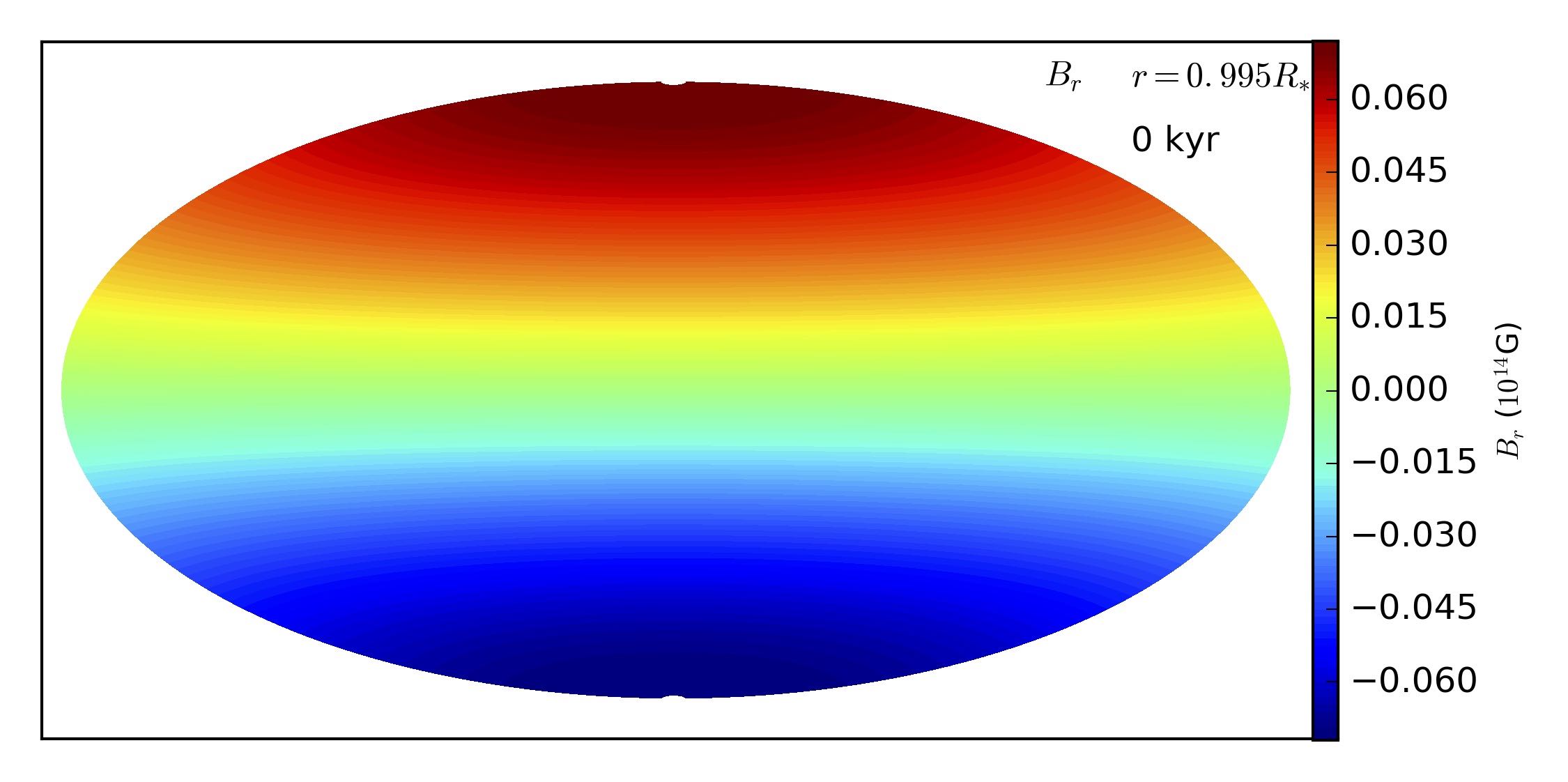} 
\includegraphics[width=0.68\columnwidth]{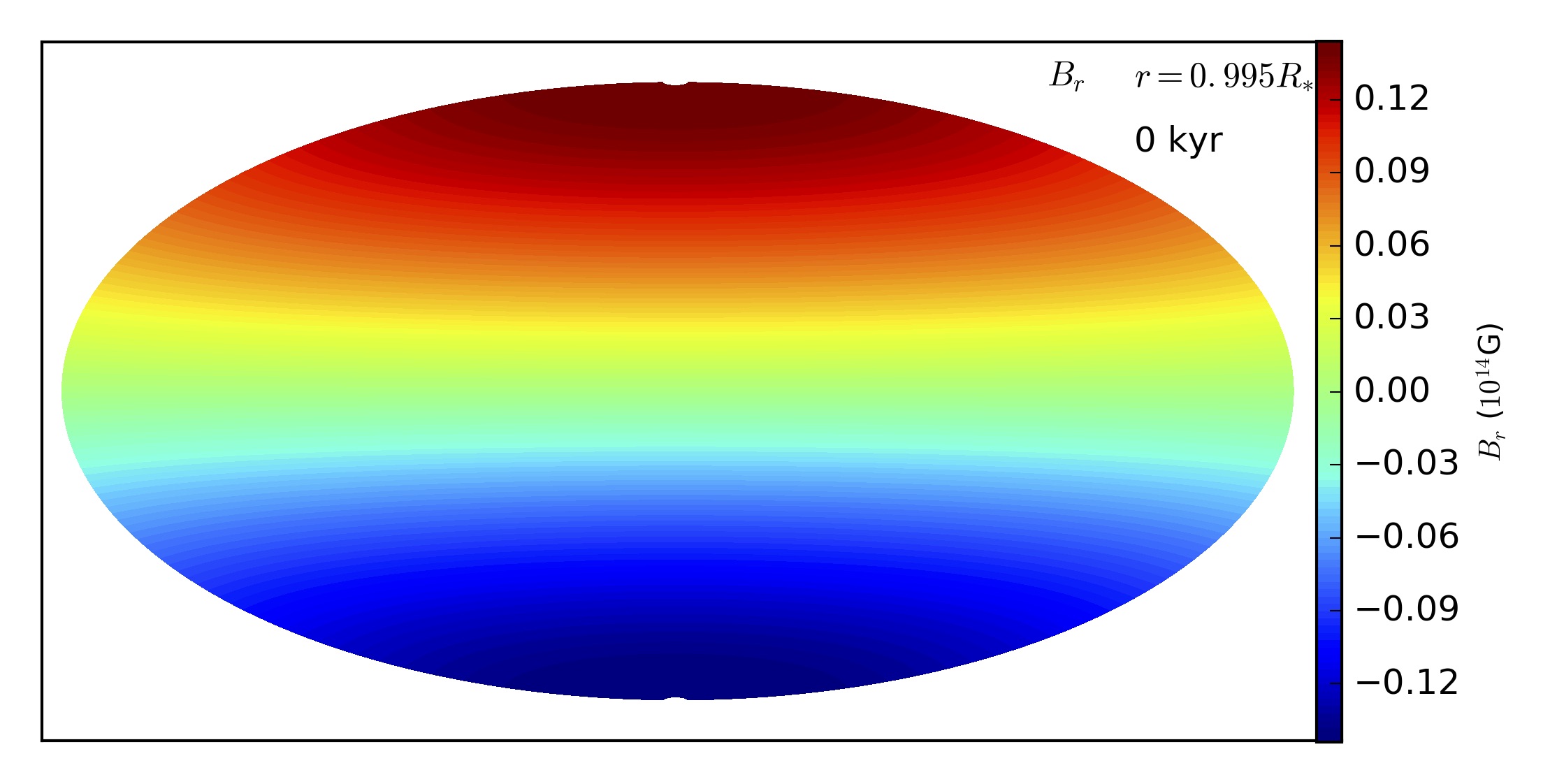}
\includegraphics[width=0.68\columnwidth]{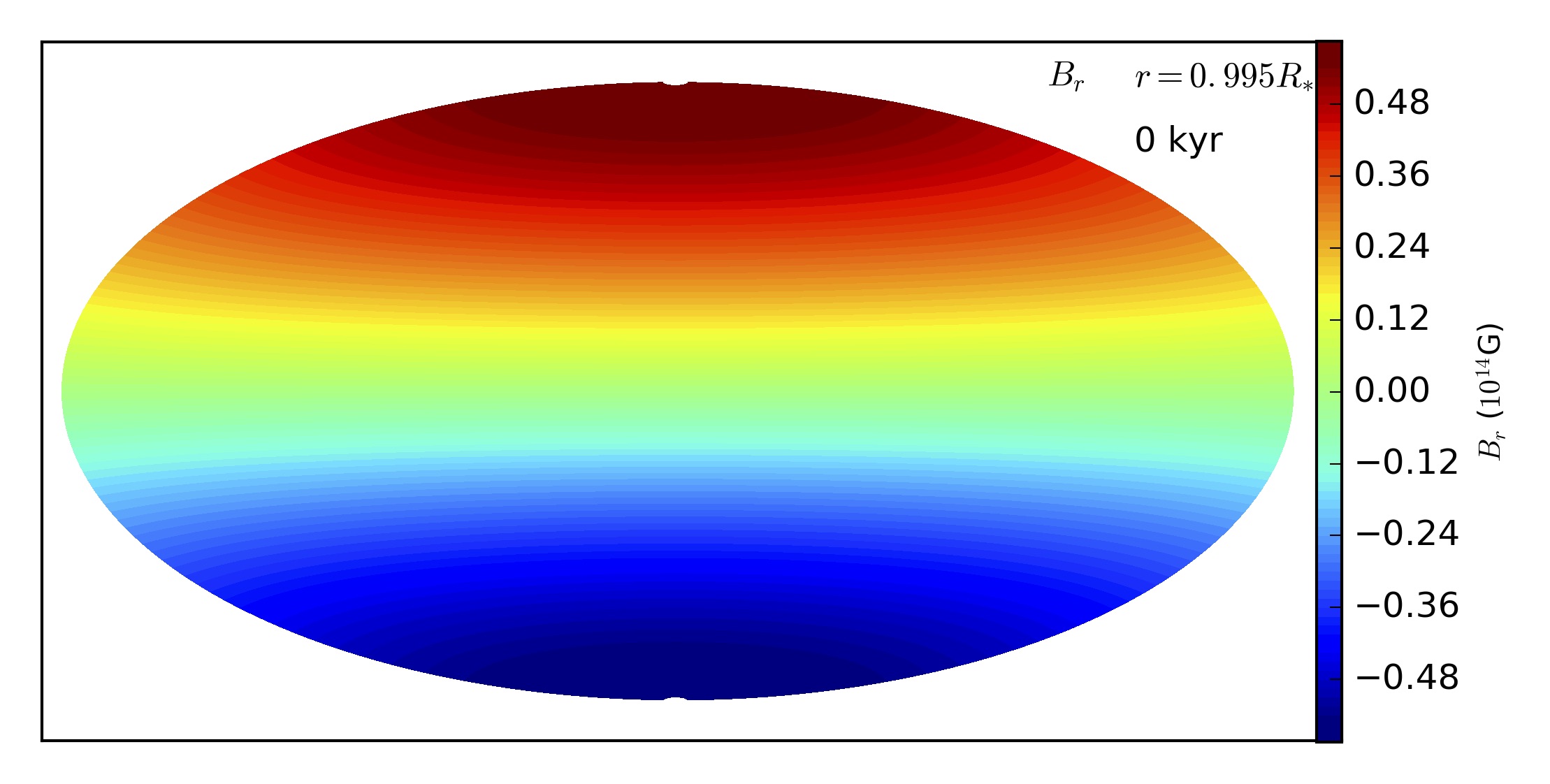}

\includegraphics[width=0.68\columnwidth]{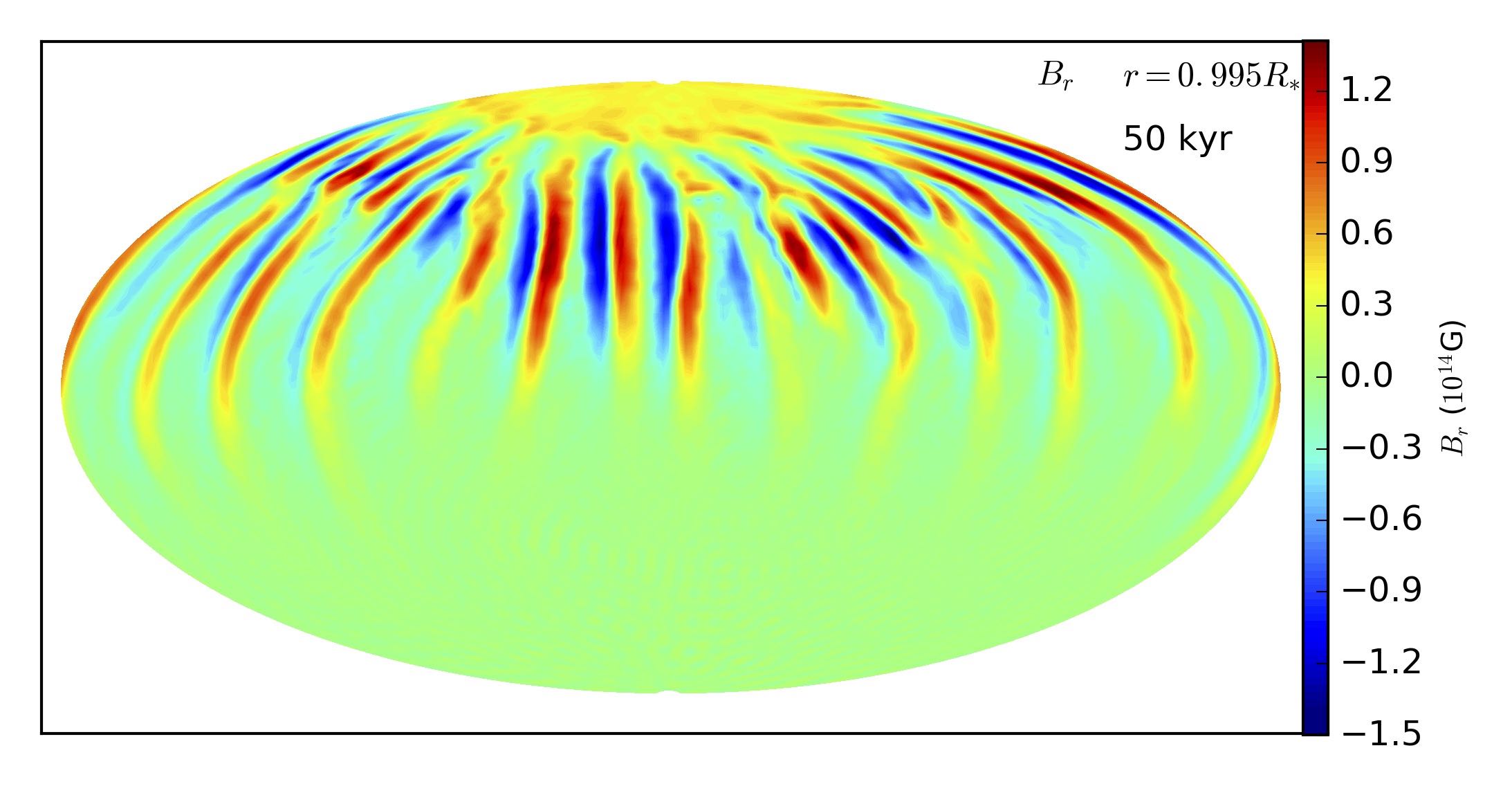}
\includegraphics[width=0.68\columnwidth]{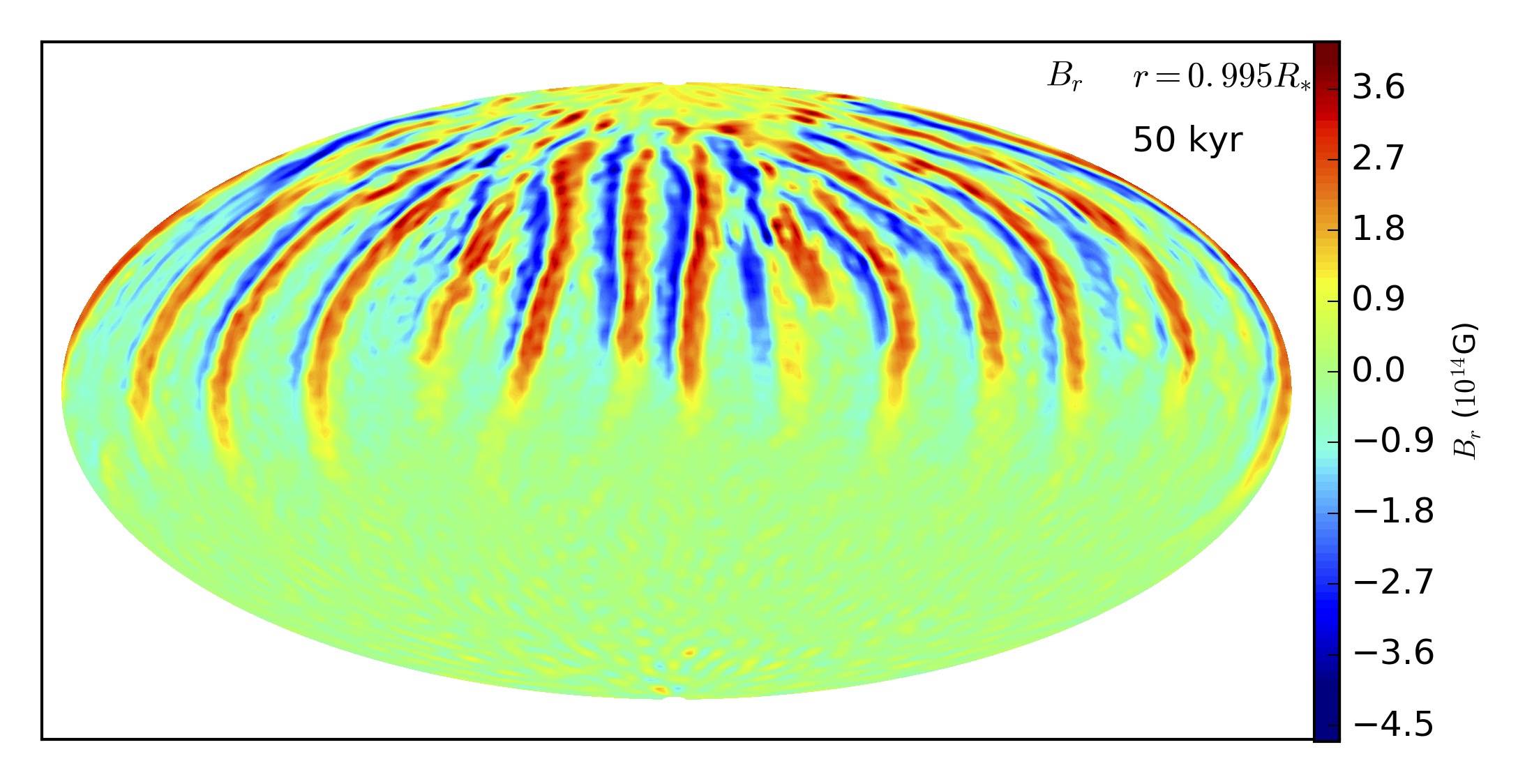}
\includegraphics[width=0.68\columnwidth]{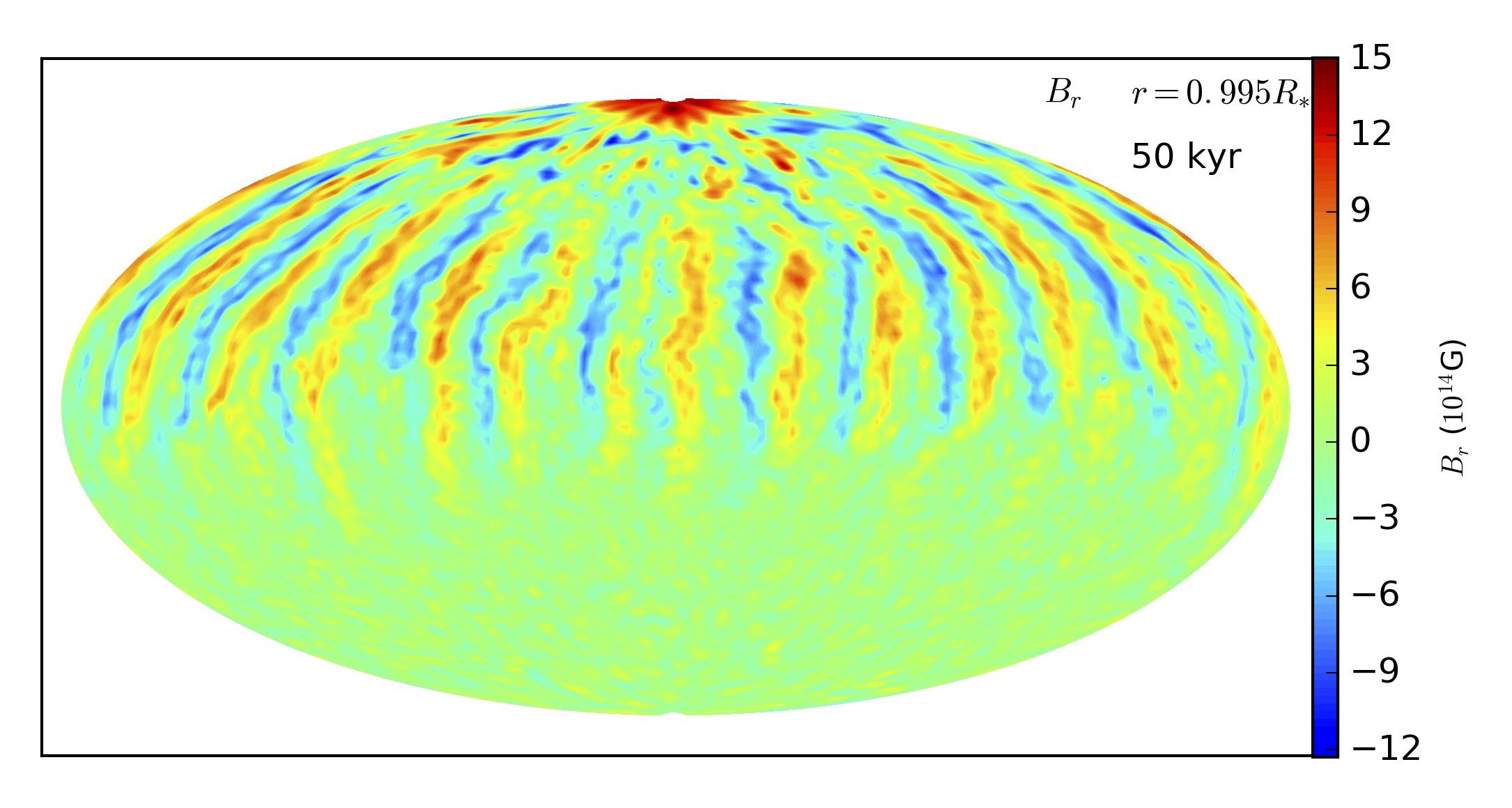}

\includegraphics[width=0.68\columnwidth]{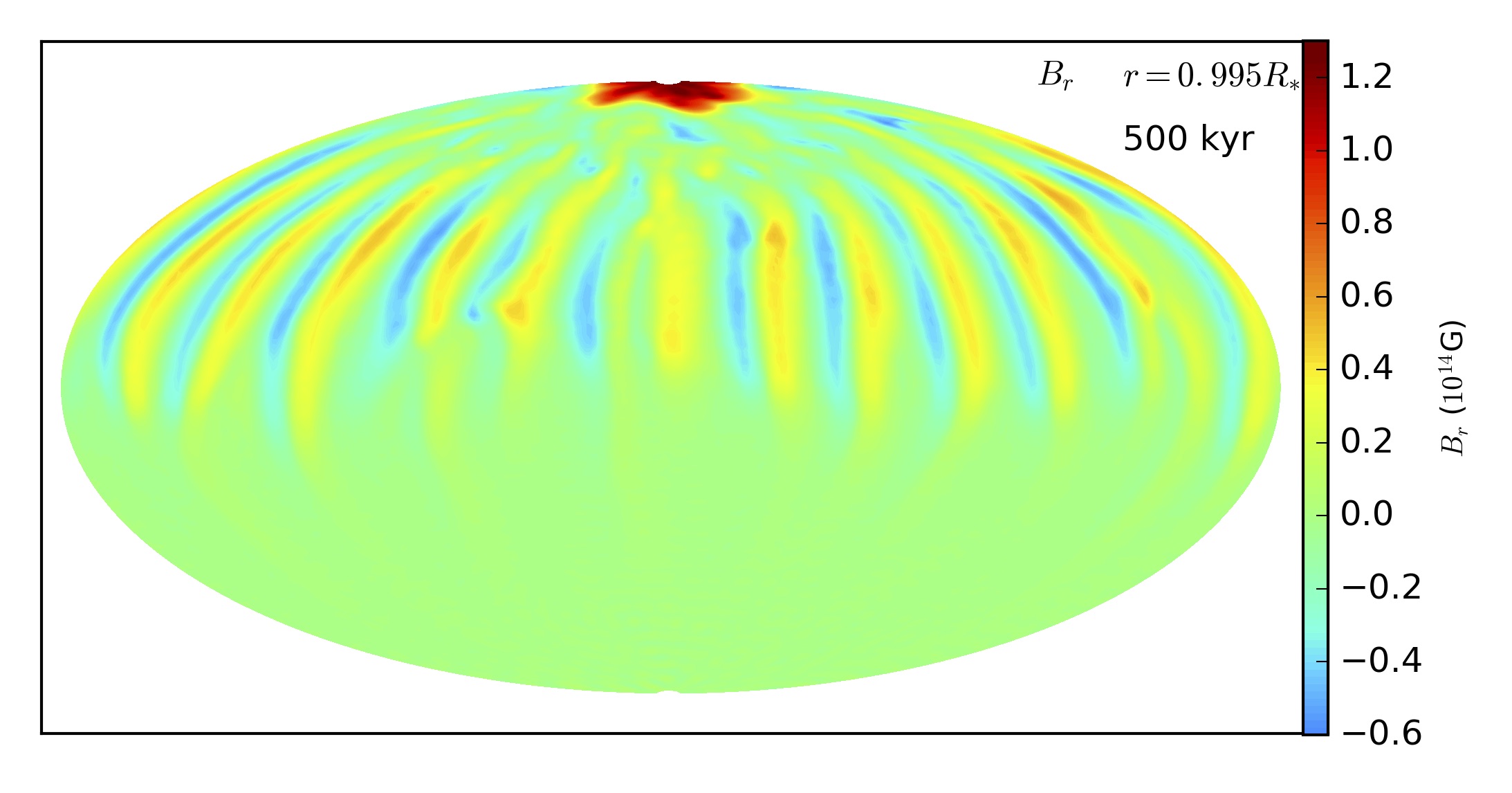}
\includegraphics[width=0.68\columnwidth]{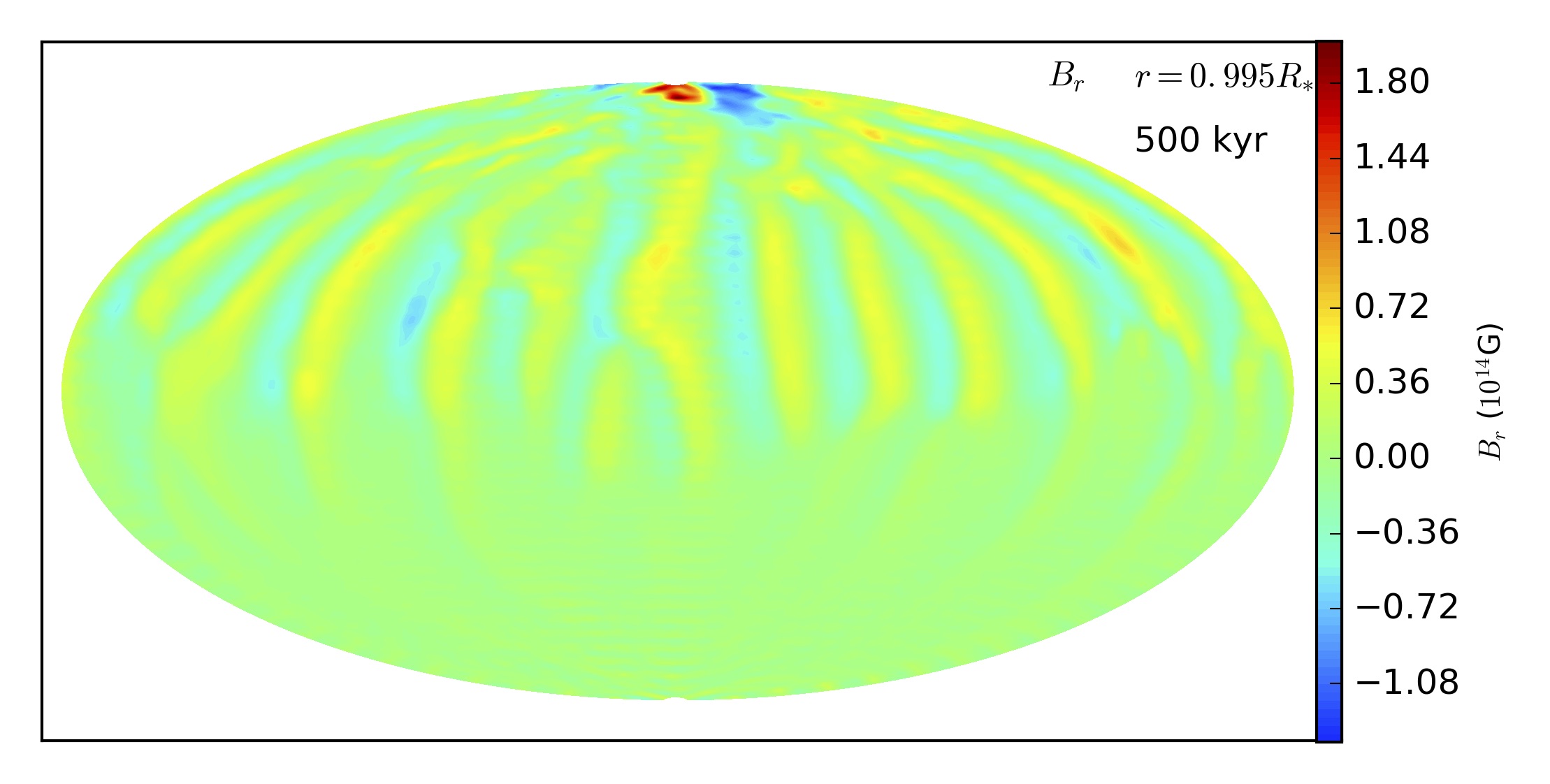}
\includegraphics[width=0.68\columnwidth]{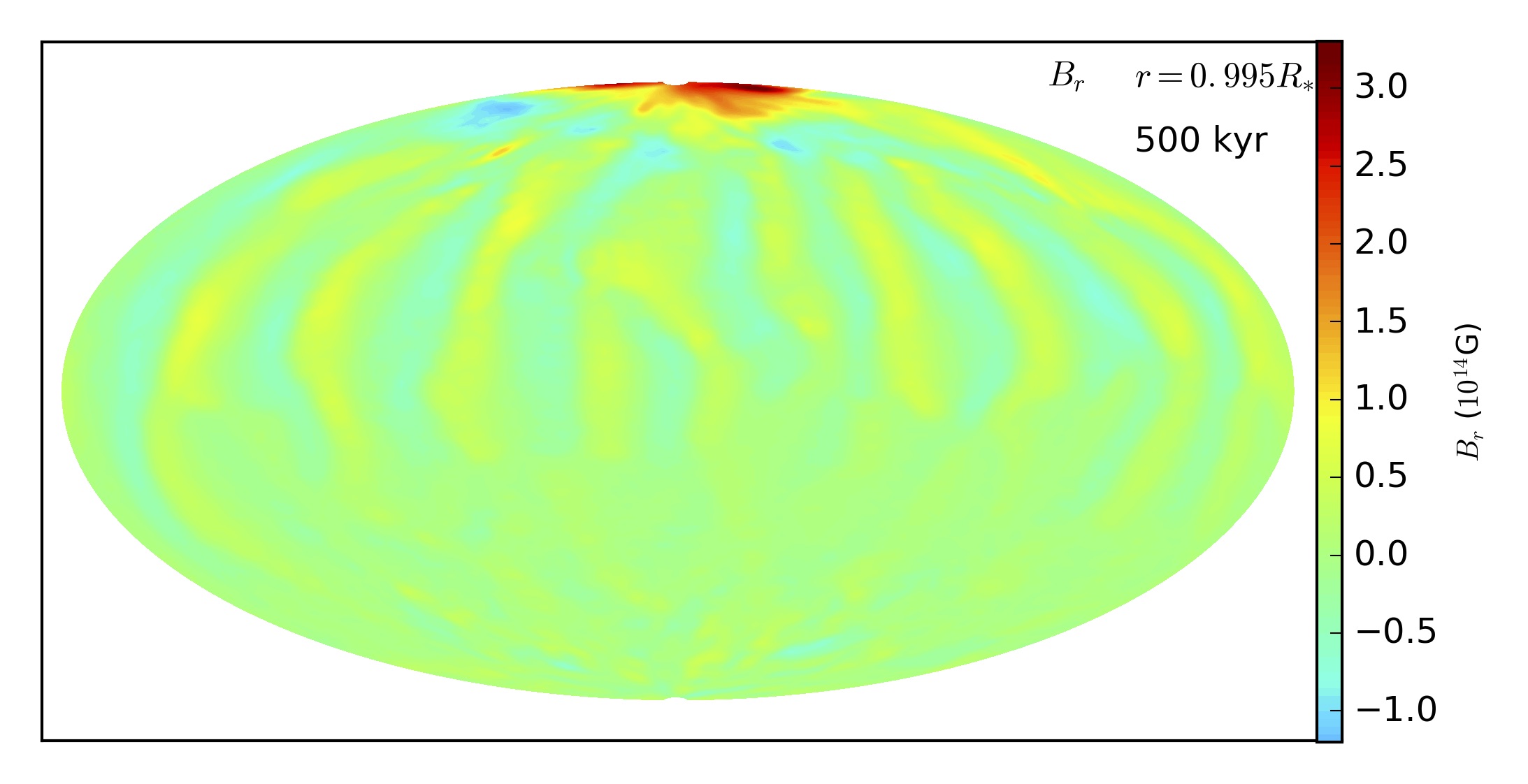}

\includegraphics[width=0.68\columnwidth]{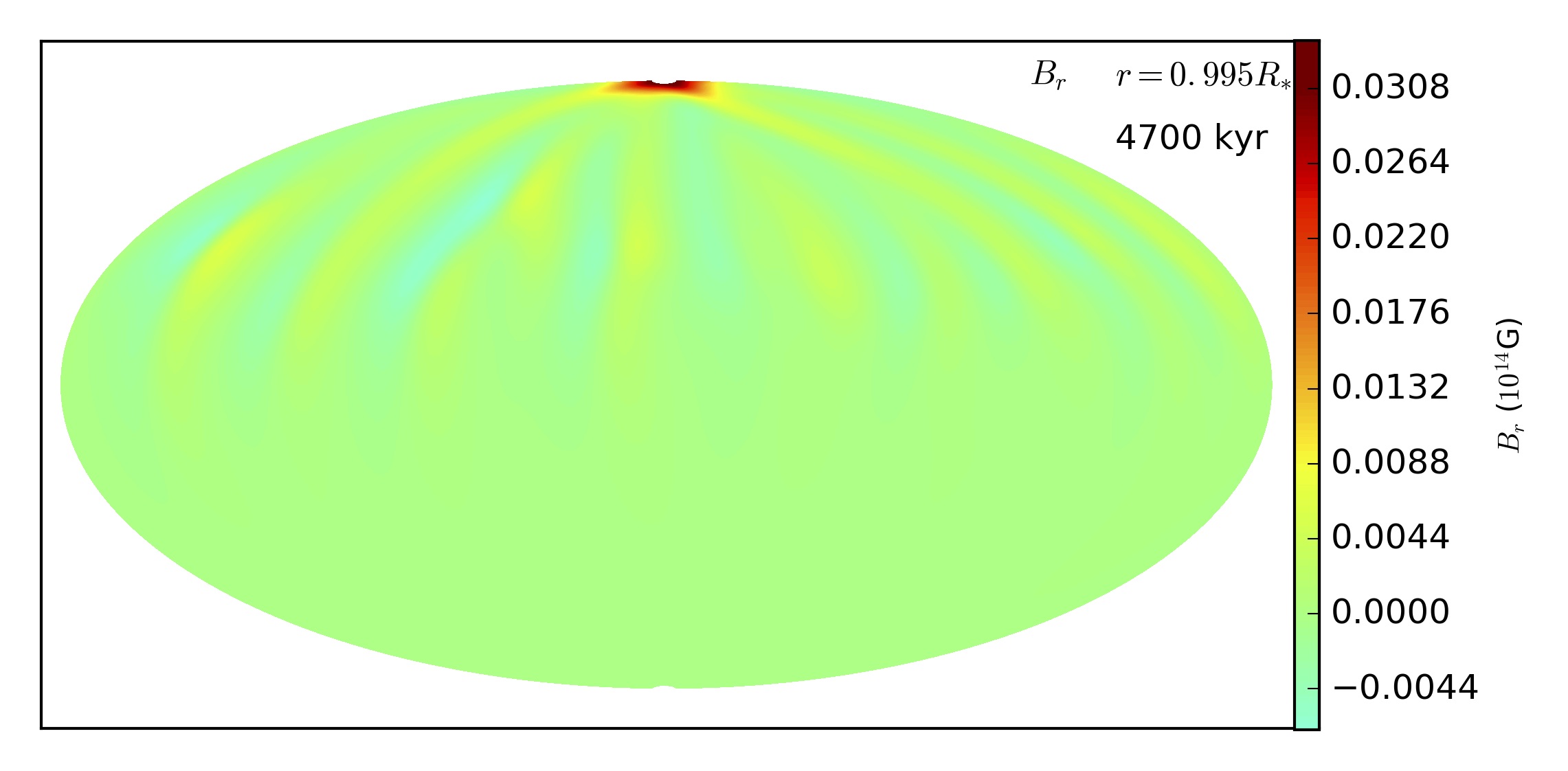}
\includegraphics[width=0.68\columnwidth]{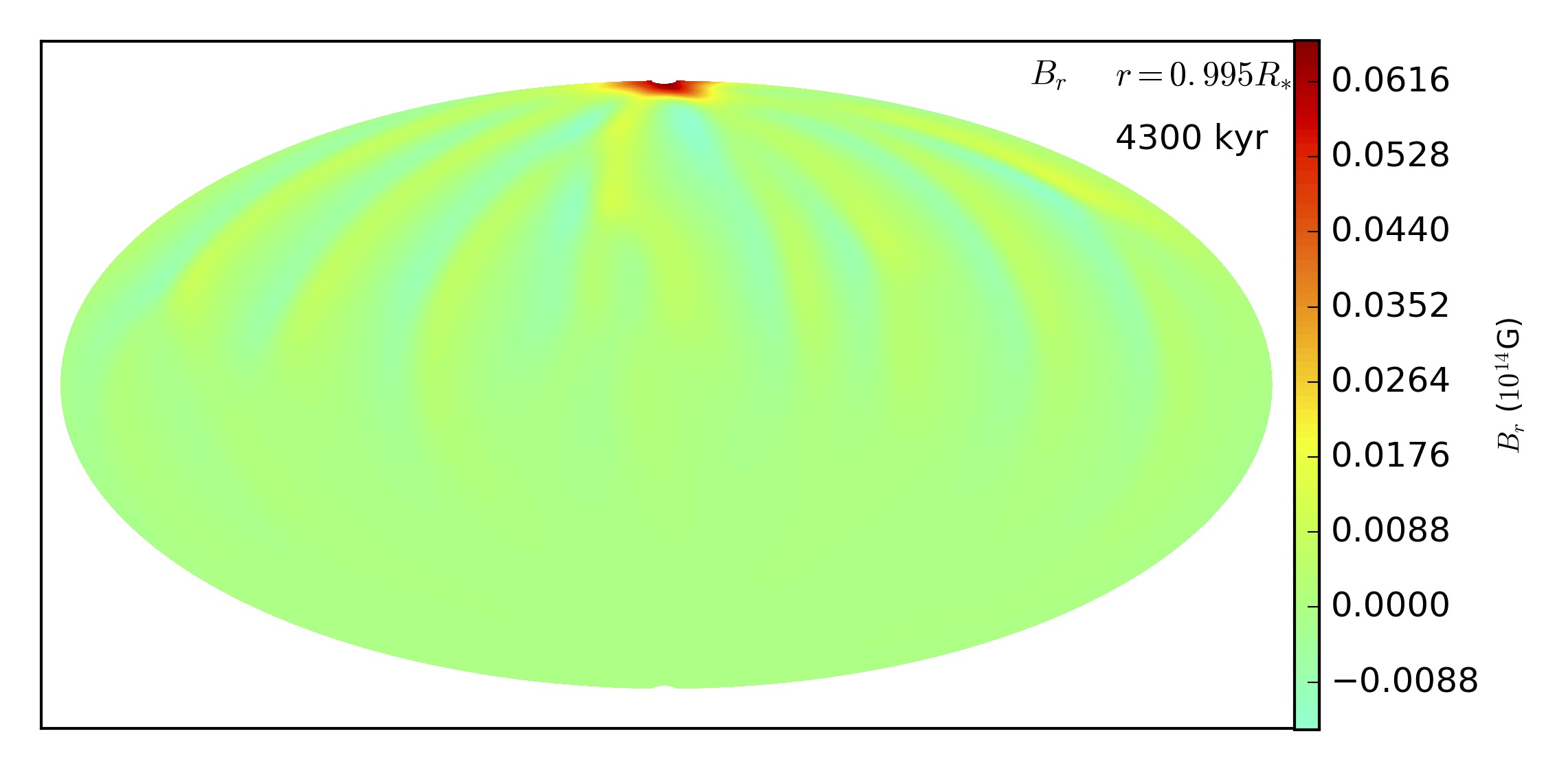}
\includegraphics[width=0.68\columnwidth]{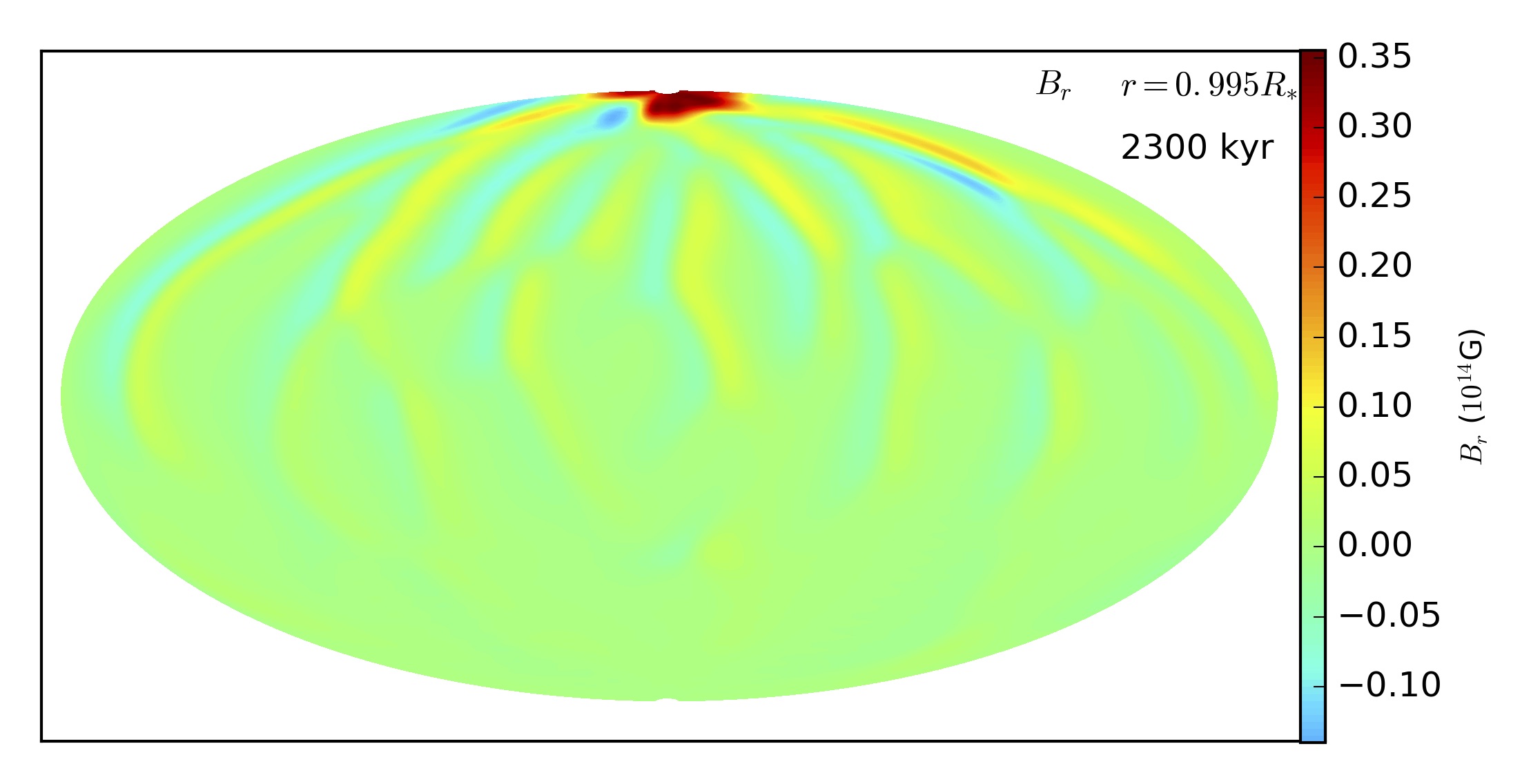}
\caption{Same as Figure \ref{Figure:1}, but for models where $99\%$ of the energy is in the toroidal field. The first column shows model A99-1, the second one A99-2 and the third one  A99-4.}
\label{Figure:3}
\end{figure*} 
\begin{figure*}
\includegraphics[width=0.68\columnwidth]{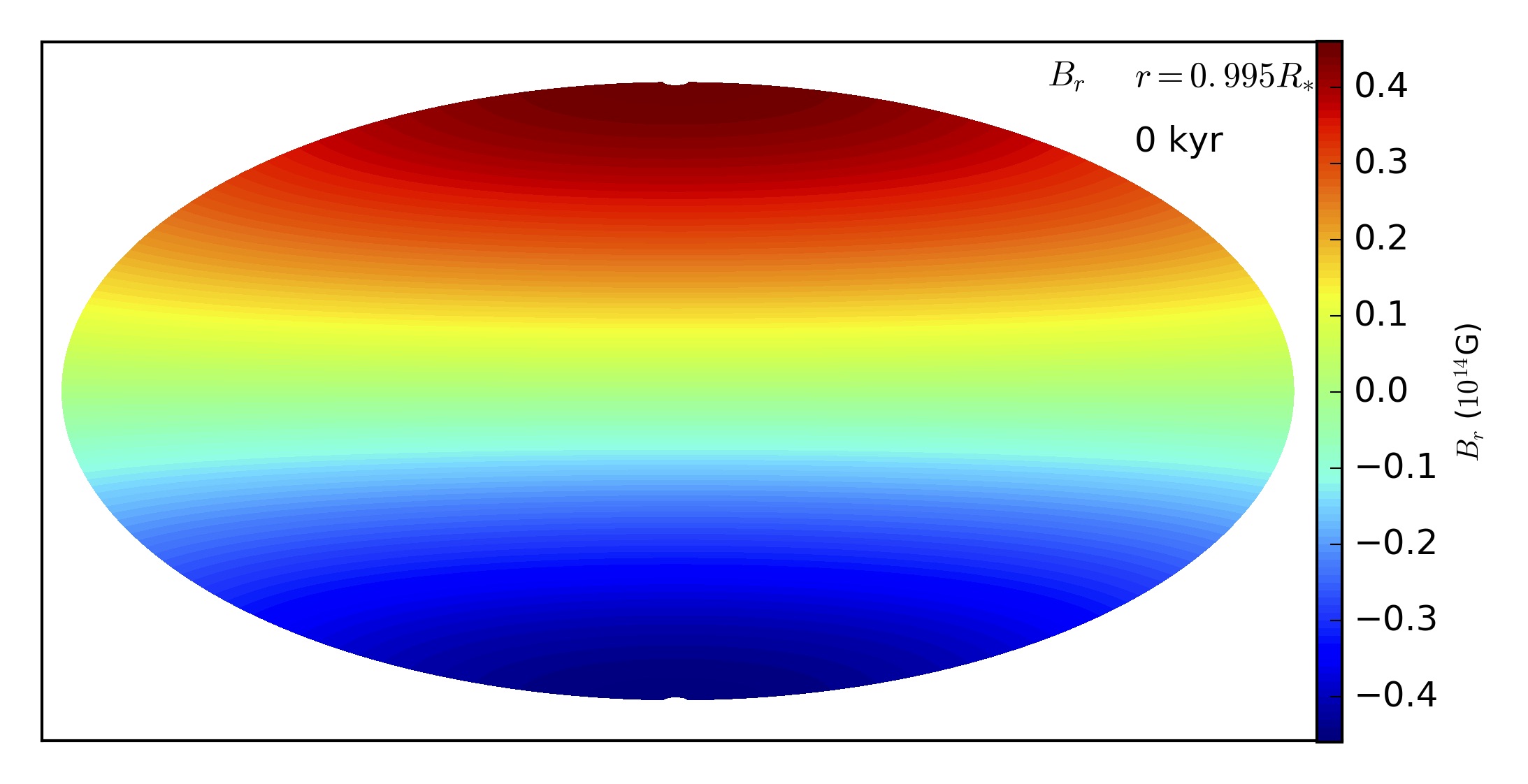}
\includegraphics[width=0.68\columnwidth]{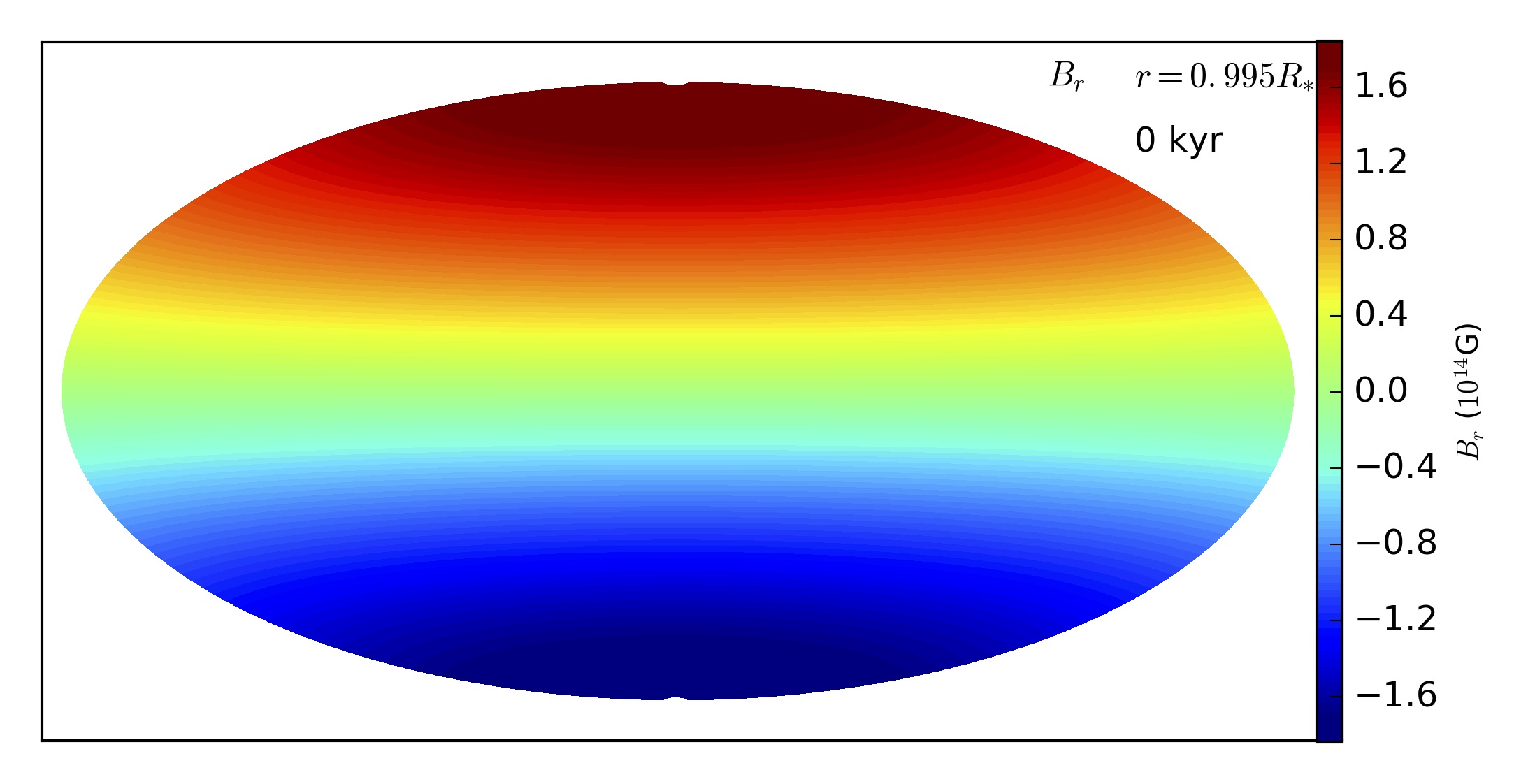}
\includegraphics[width=0.68\columnwidth]{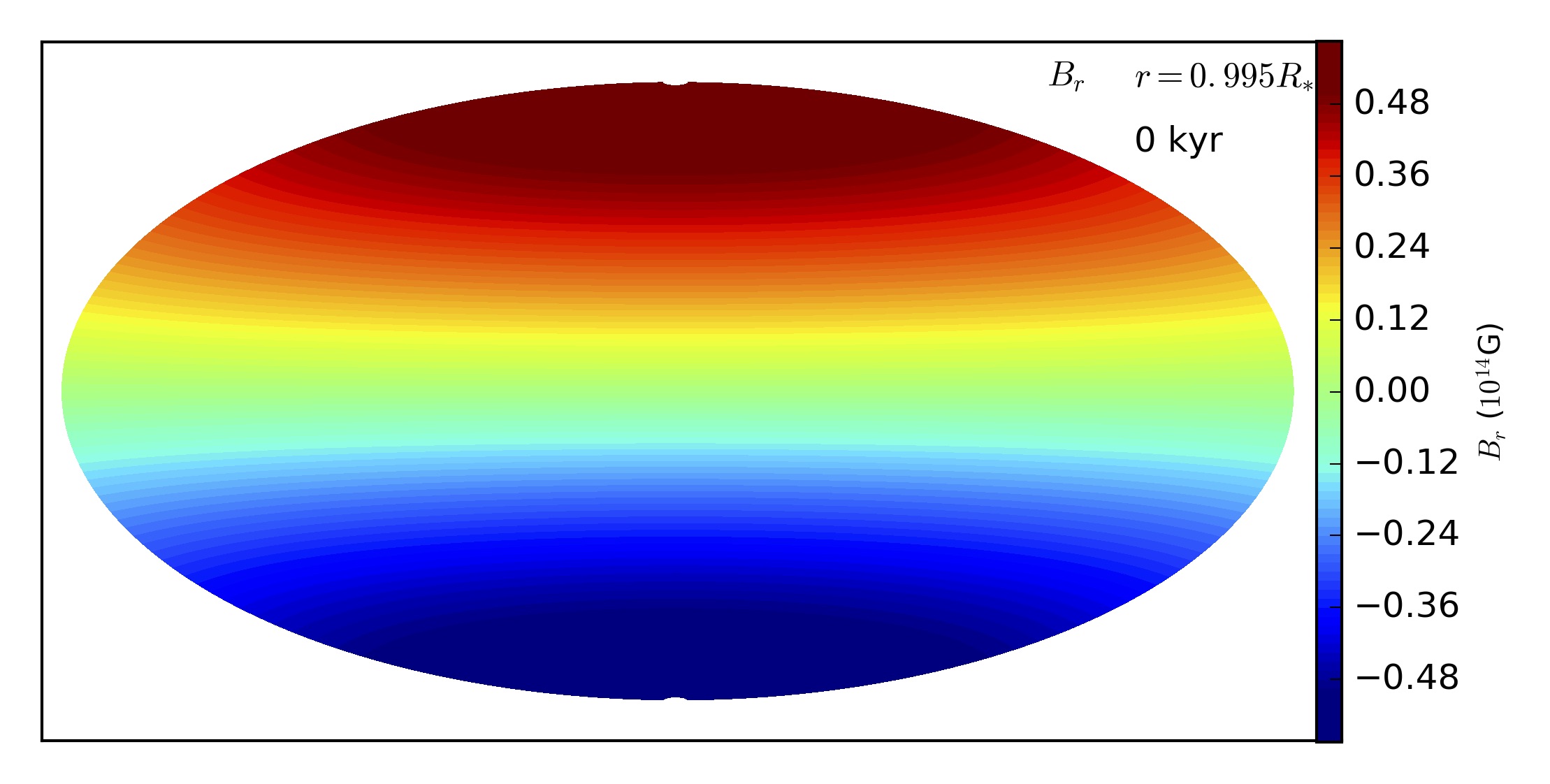}

\includegraphics[width=0.68\columnwidth]{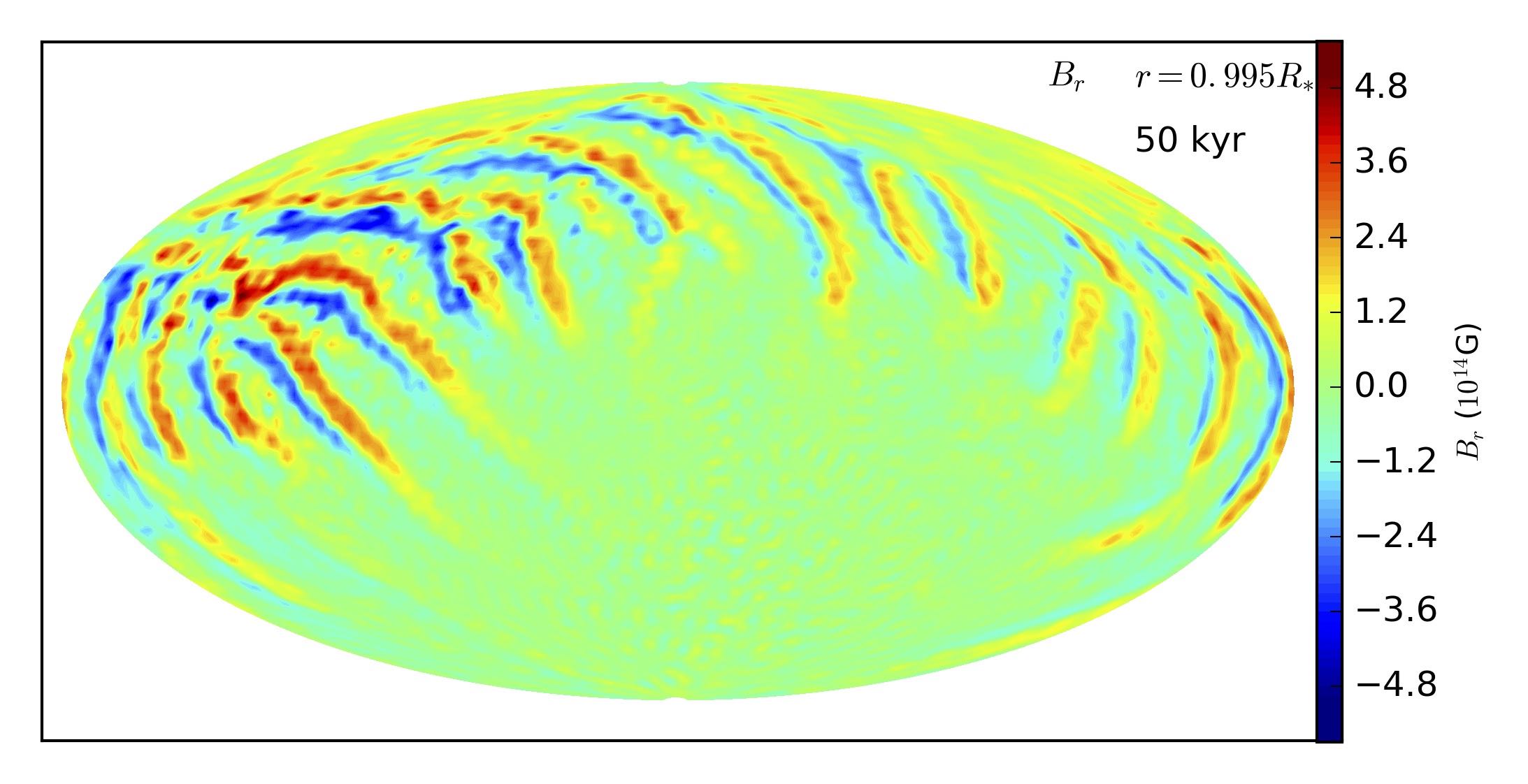}
\includegraphics[width=0.68\columnwidth]{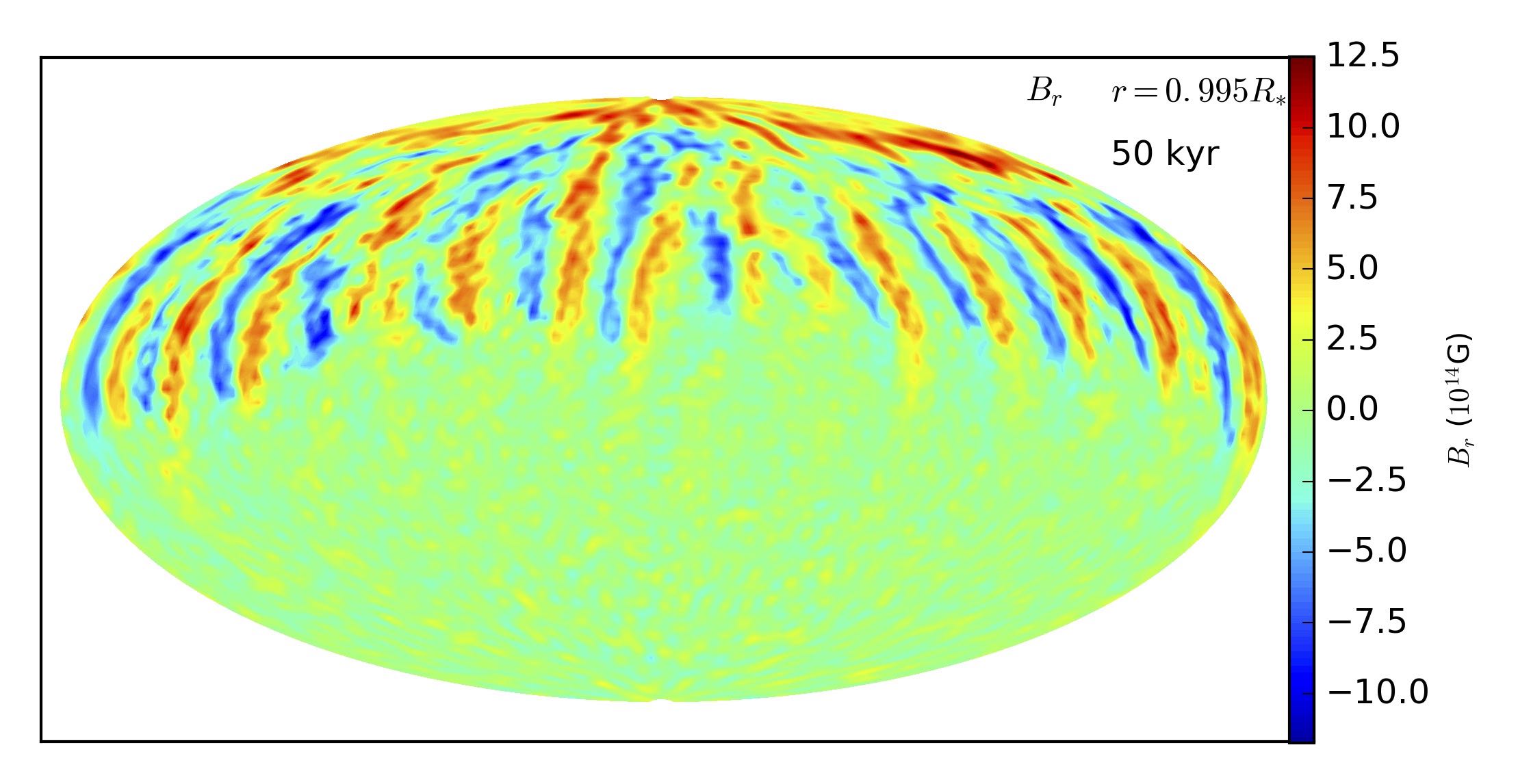}
\includegraphics[width=0.68\columnwidth]{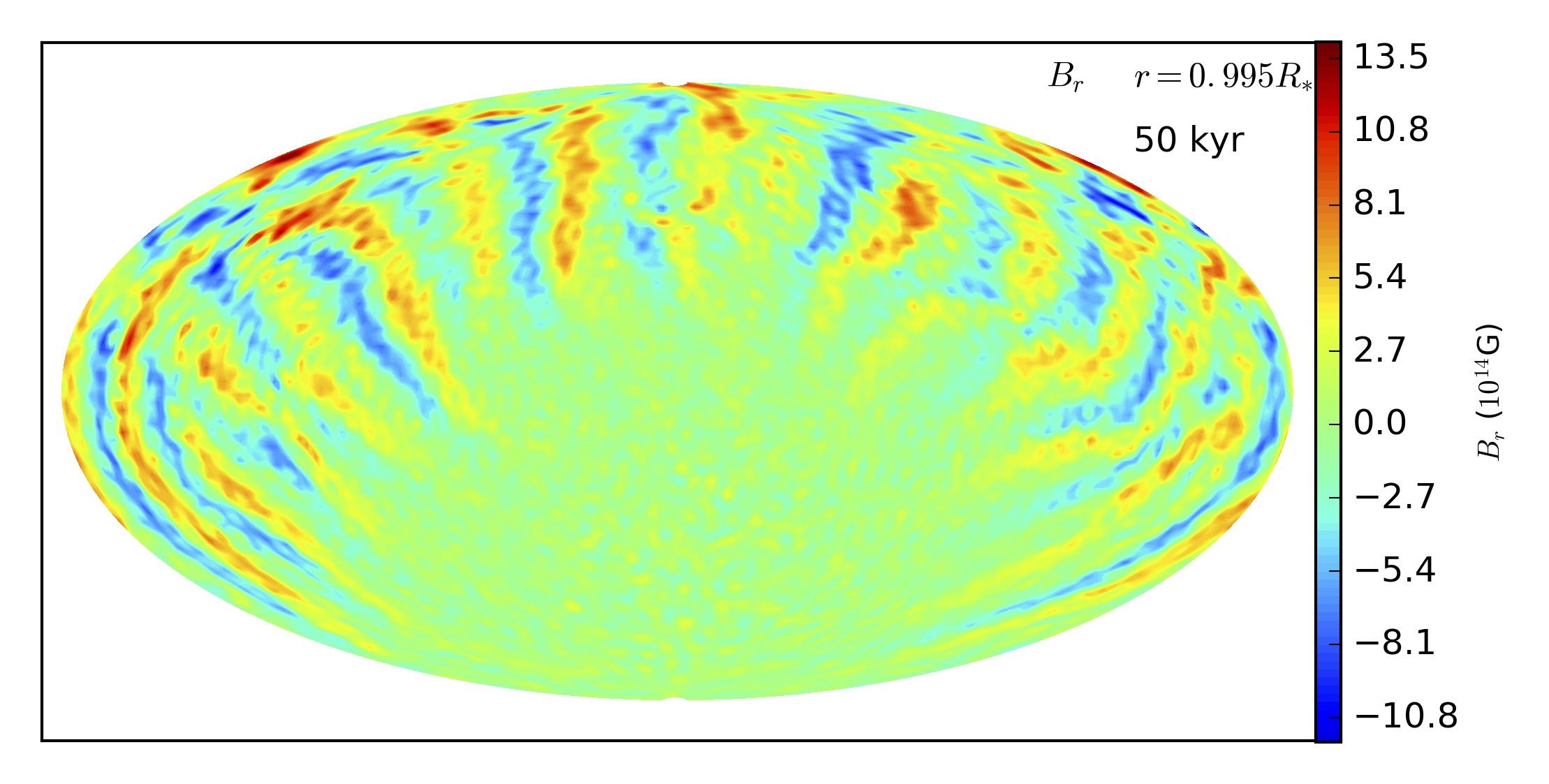}

\includegraphics[width=0.68\columnwidth]{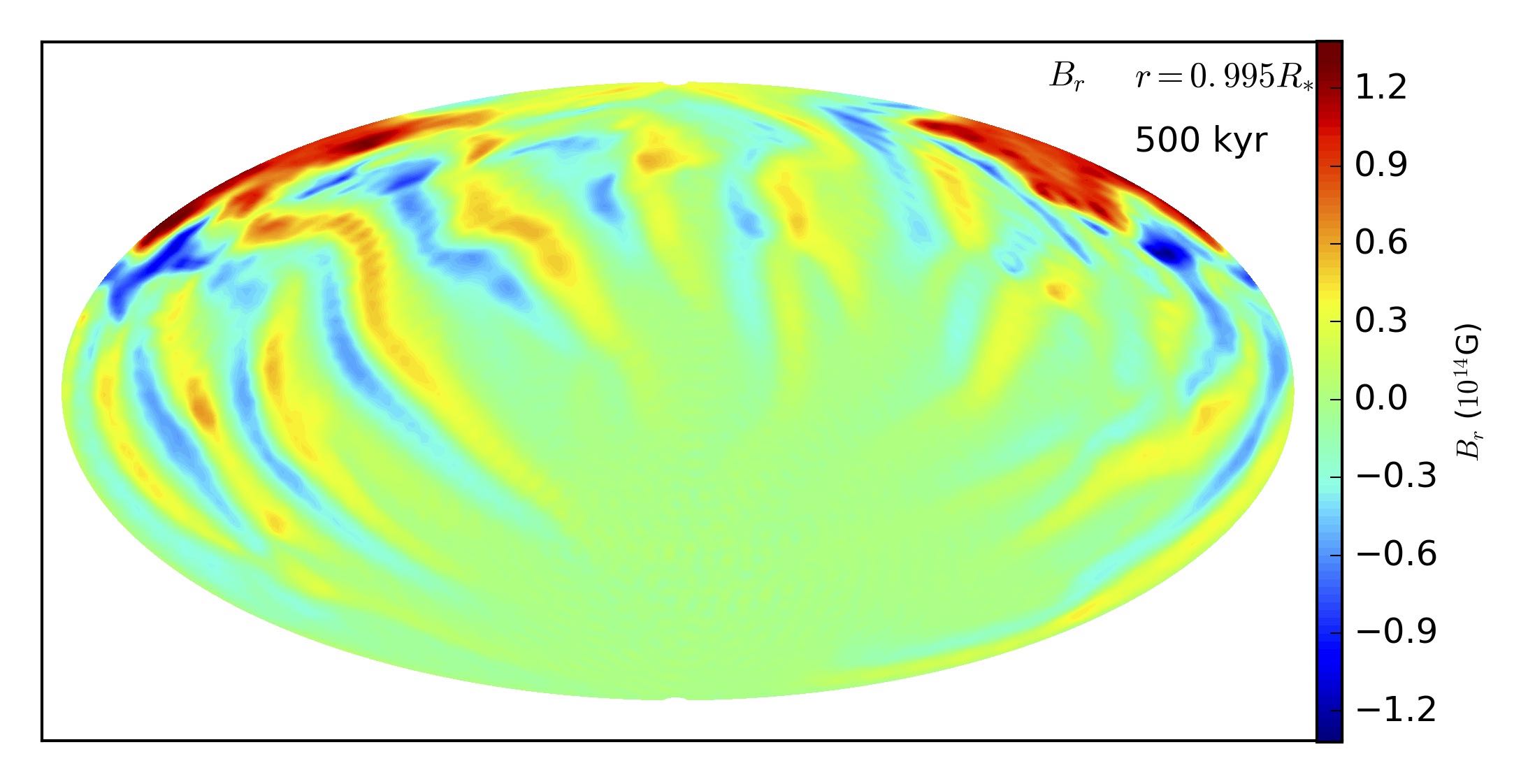}
\includegraphics[width=0.68\columnwidth]{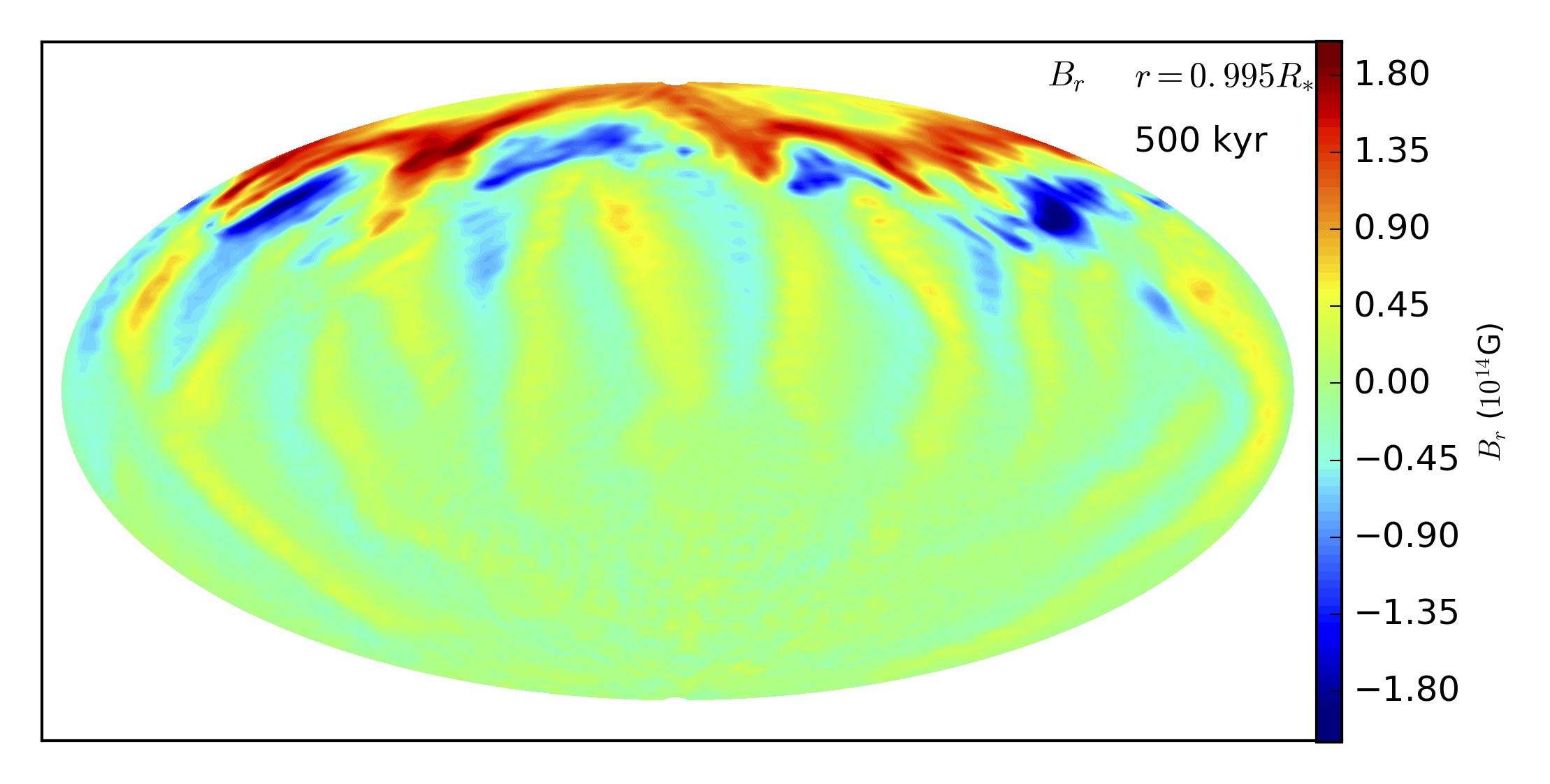}
\includegraphics[width=0.68\columnwidth]{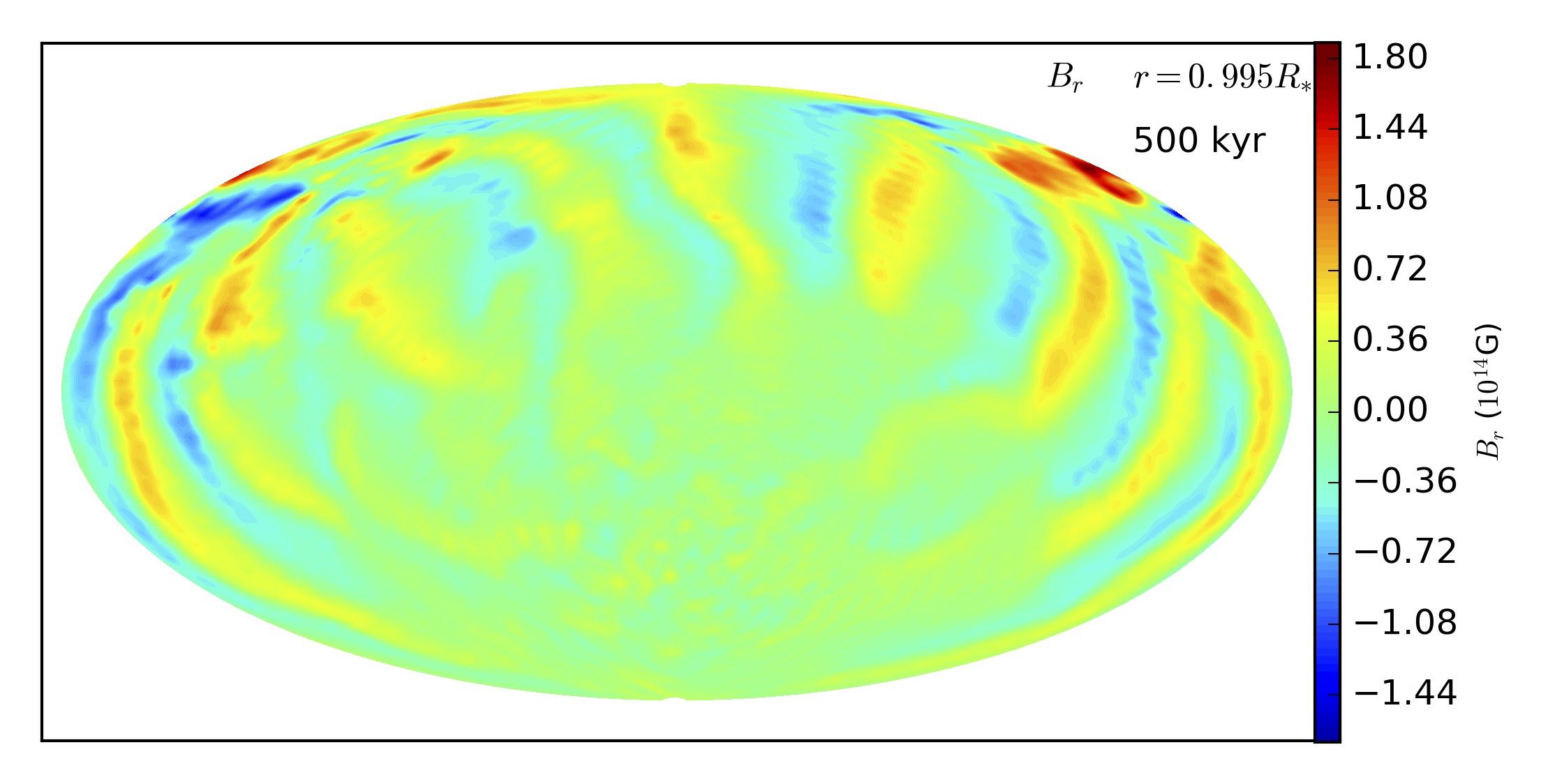}

\includegraphics[width=0.68\columnwidth]{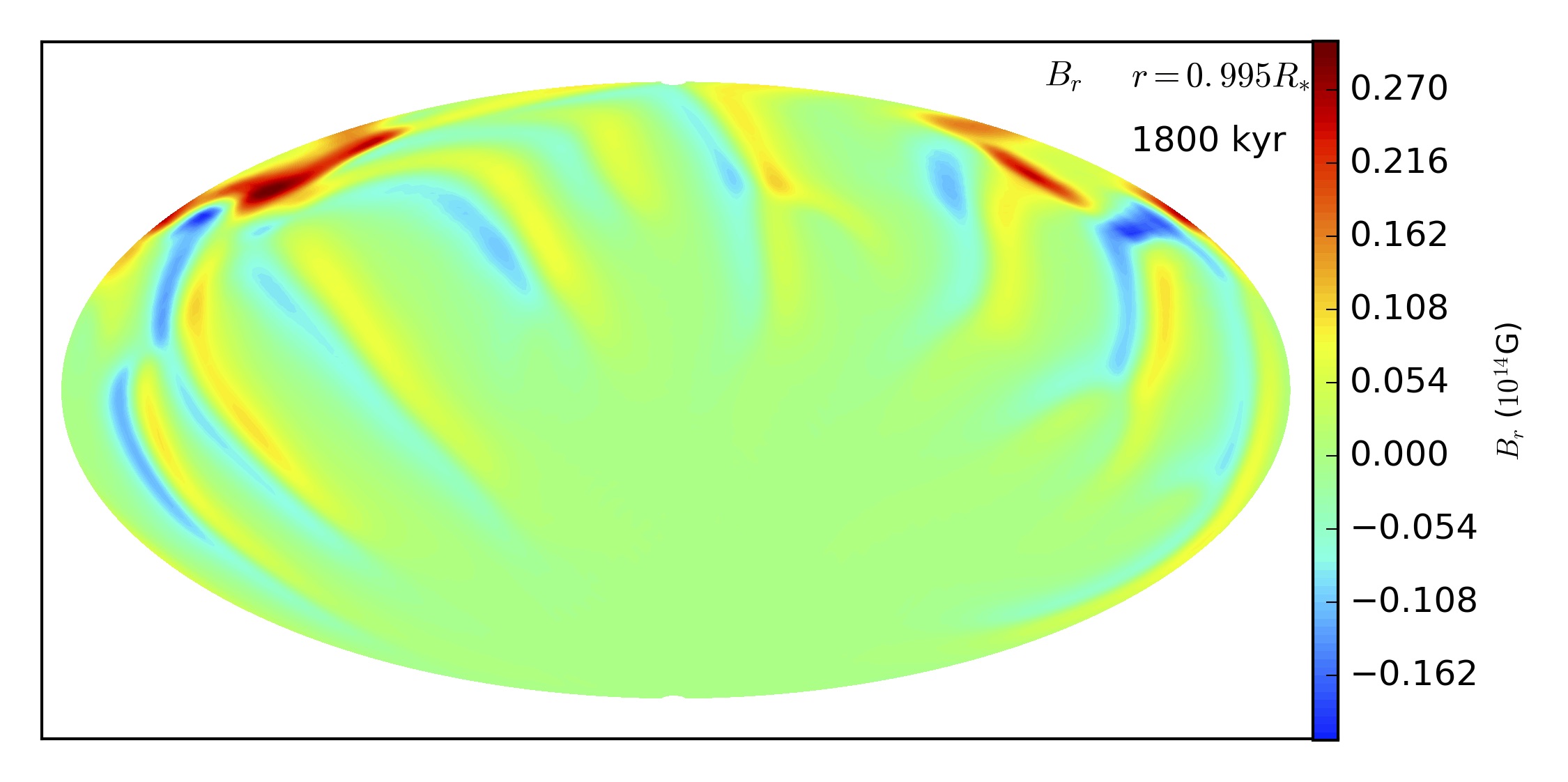}
\includegraphics[width=0.68\columnwidth]{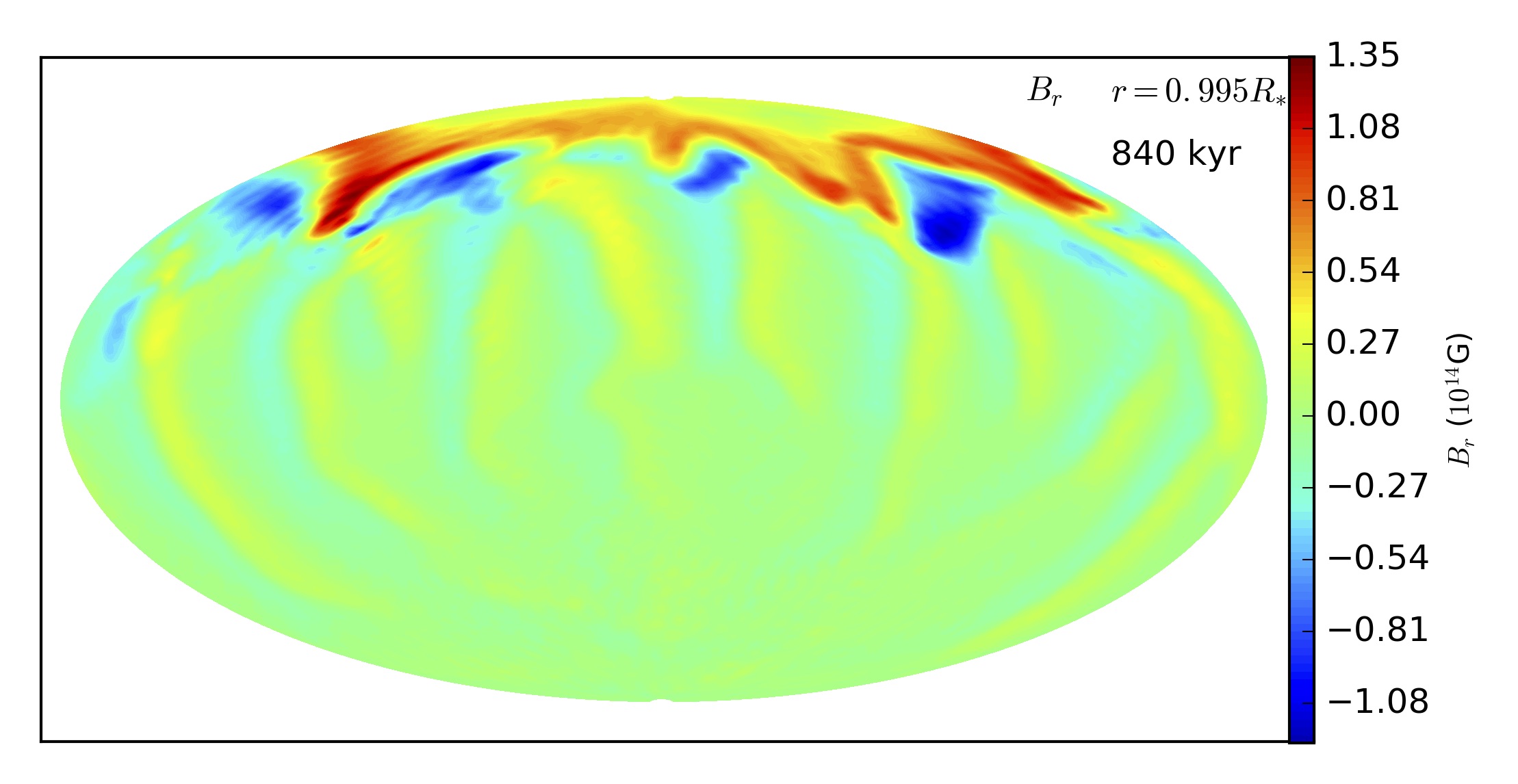}
\includegraphics[width=0.68\columnwidth]{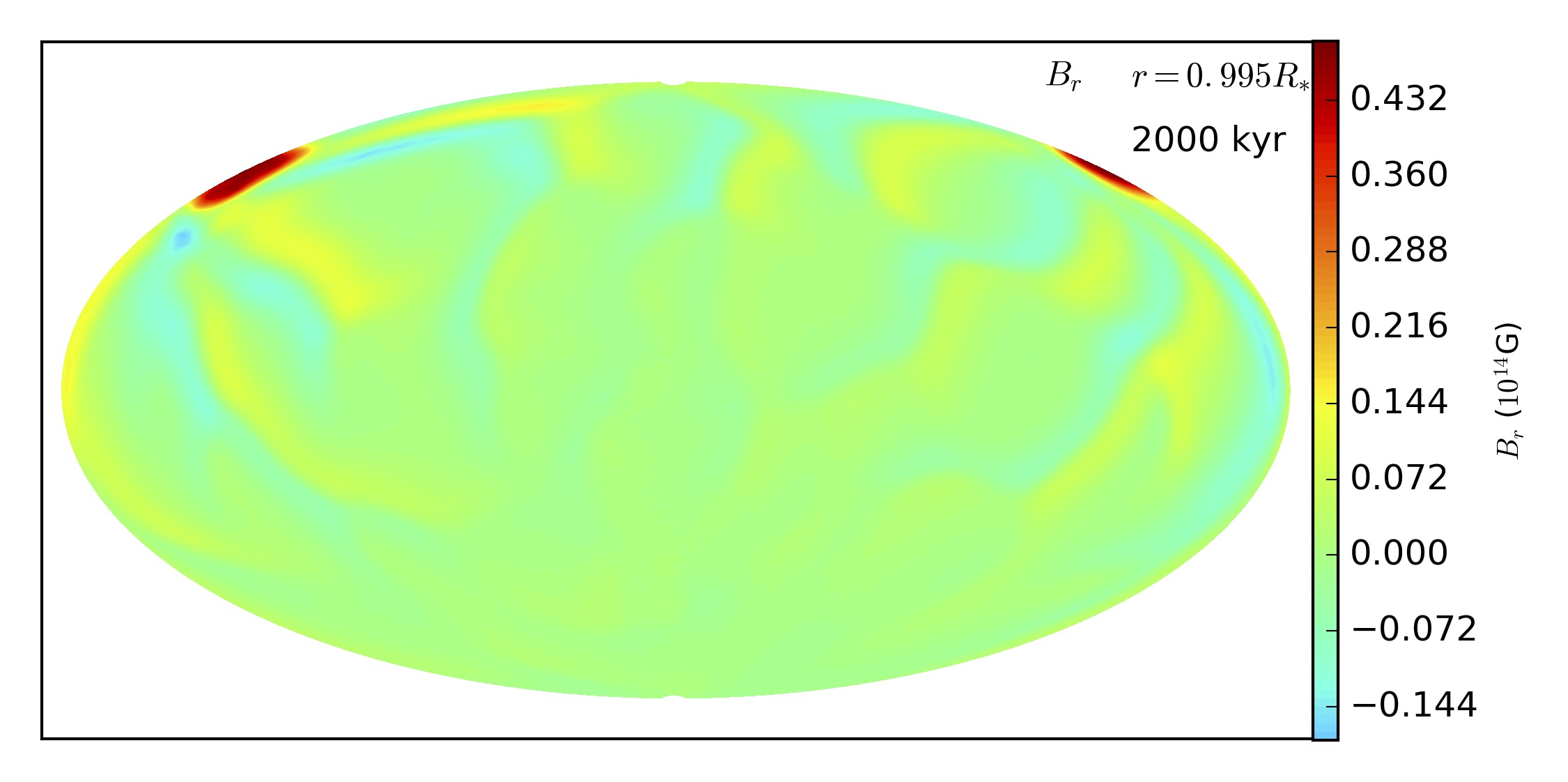}
\caption{Same as Figure \ref{Figure:1}, but for models where the poloidal and toroidal fields are misaligned. The first column shows model B90-2, the second one B90-4 and the third one  B99-4.}
\label{Figure:4}
\end{figure*} 
\begin{figure*}
\includegraphics[width=1.1\columnwidth]{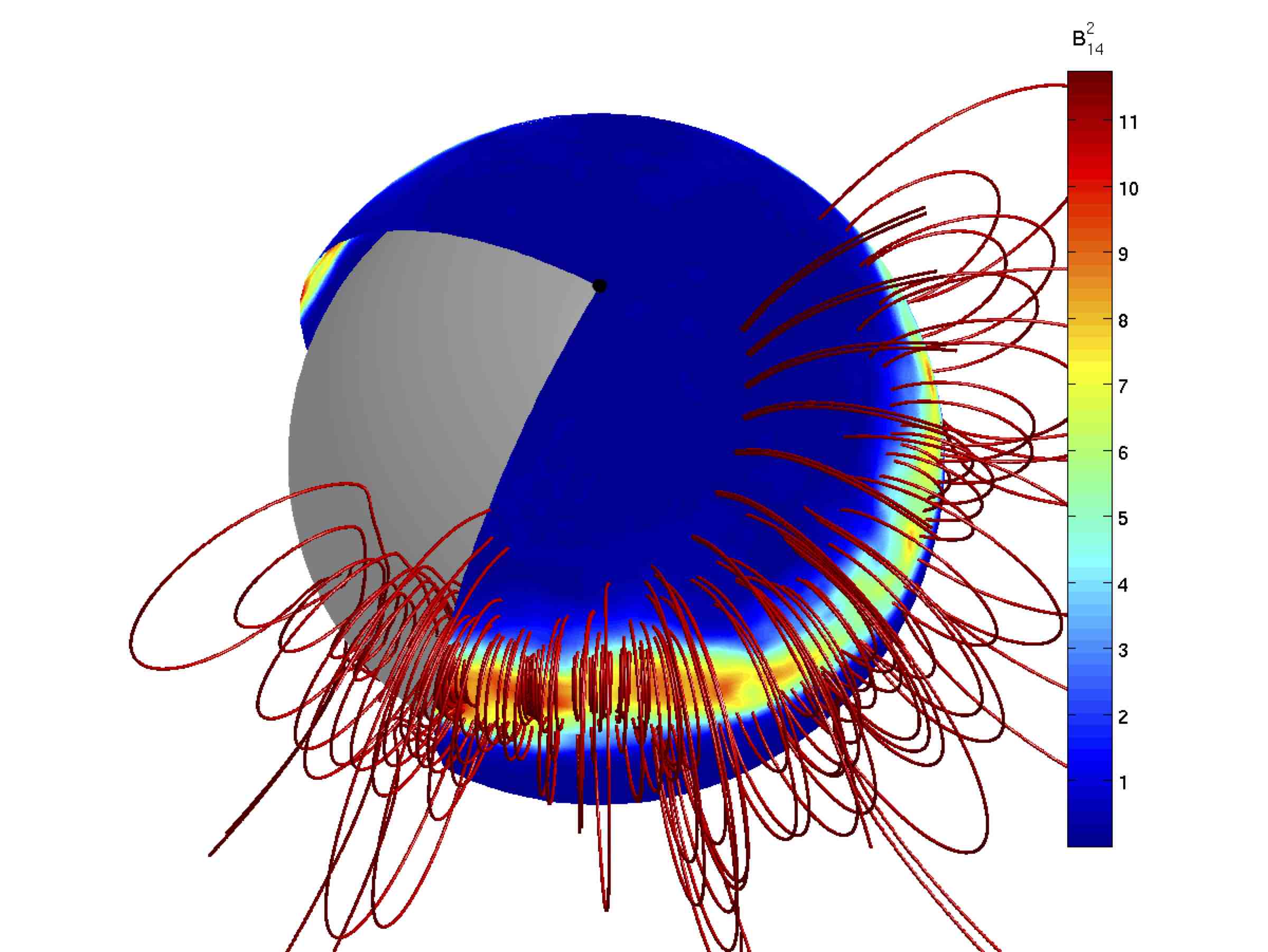}
\includegraphics[width=1.1\columnwidth]{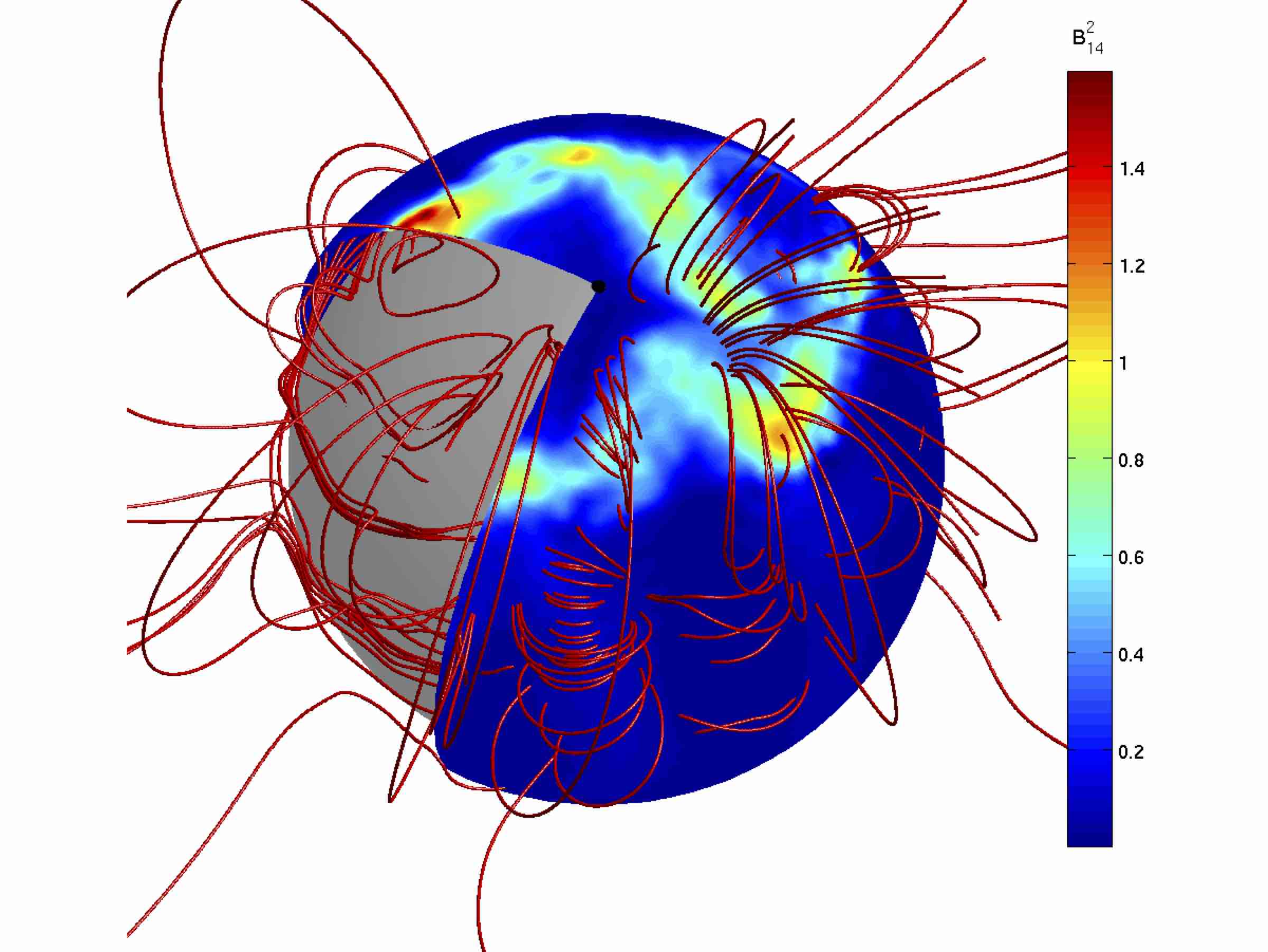}
\includegraphics[width=1.1\columnwidth]{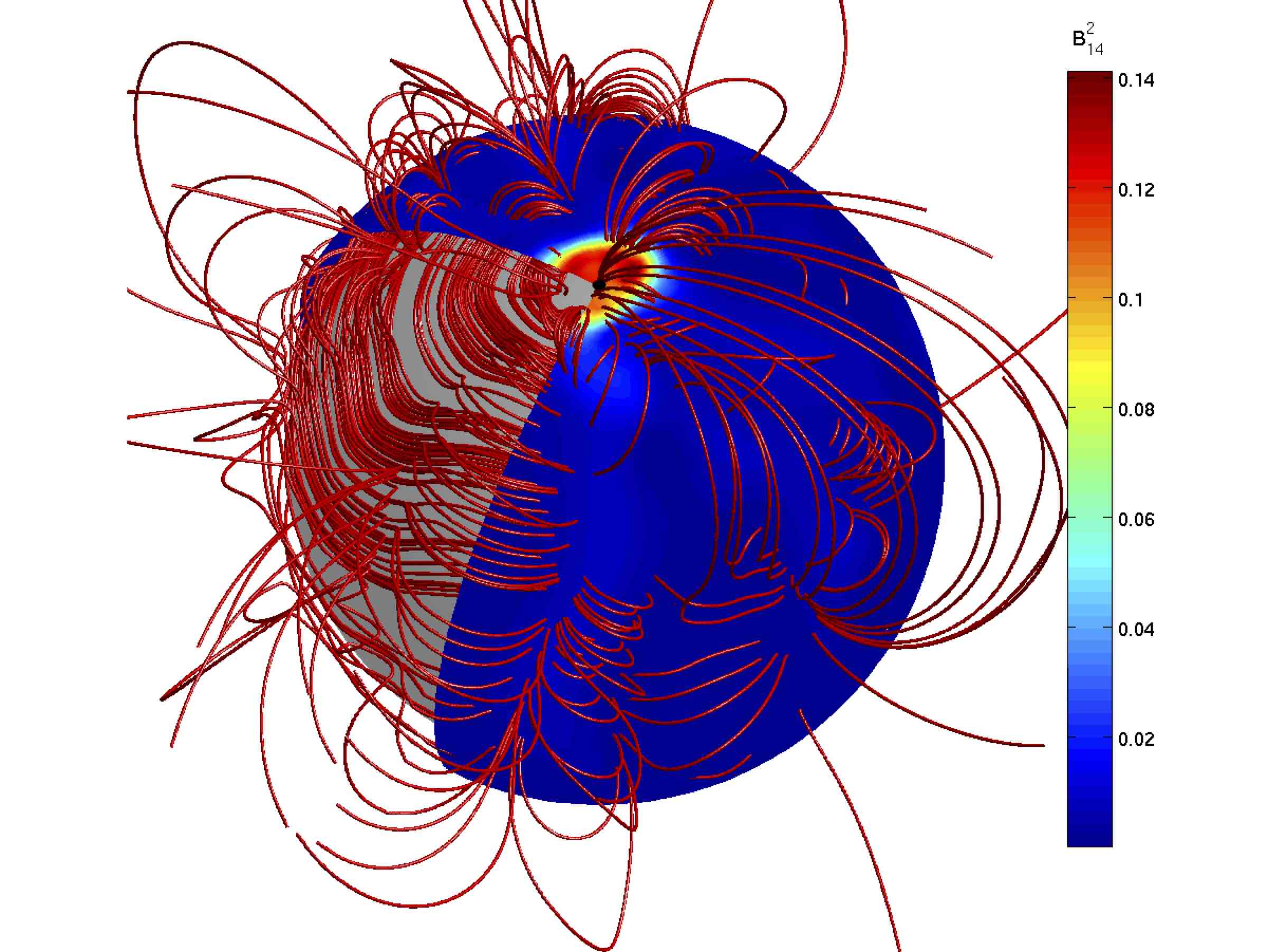}
\includegraphics[width=1.1\columnwidth]{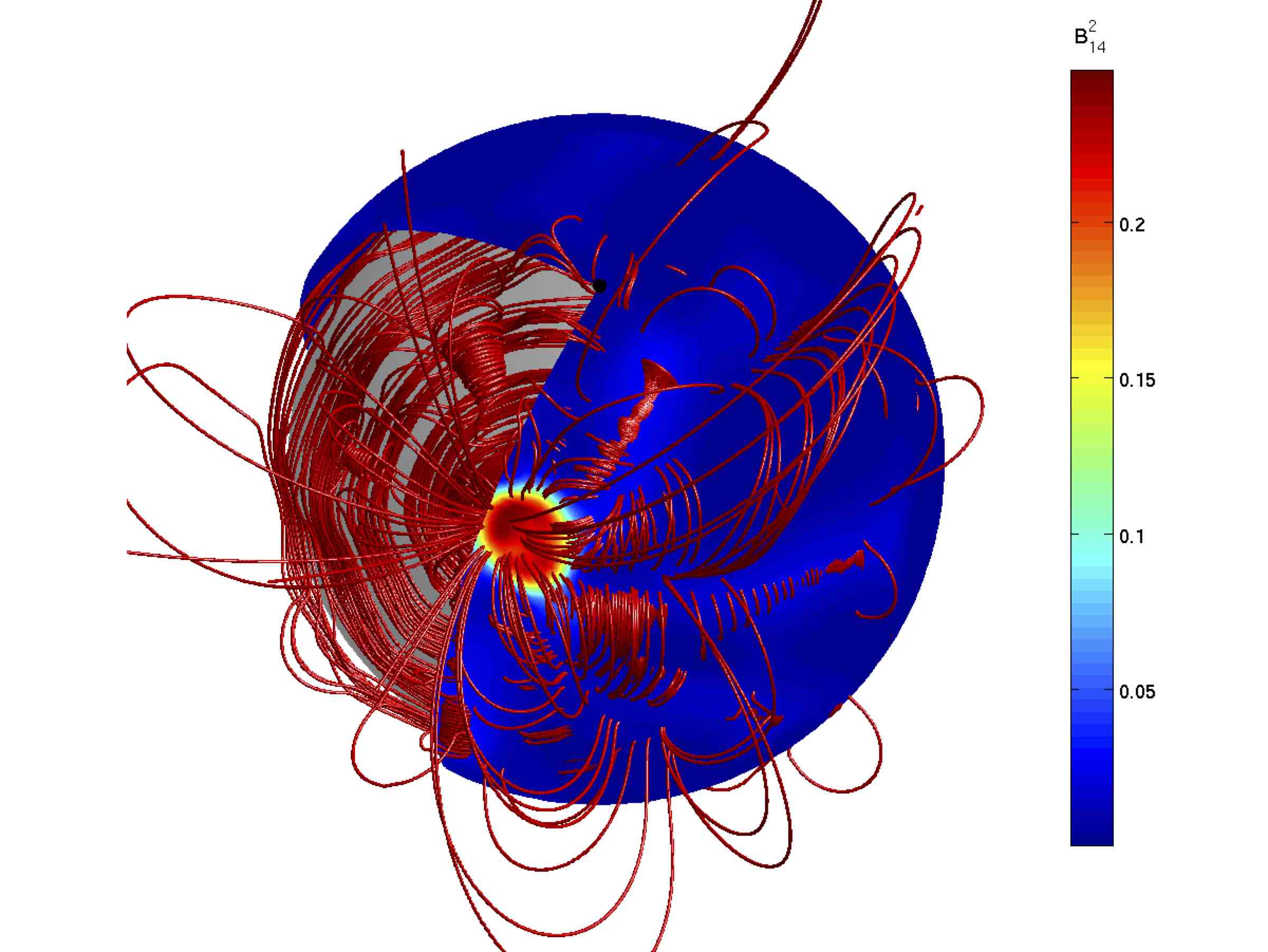}
\caption{Magnetic field structure for models A50-4 (top left), A90-4 (top right), A99-4 (bottom left) and B99-4 (bottom right), at the time when their crossing the Death Line ($t_{d}$). The magnetic field lines are plotted in red, the color-scale corresponds to $(B/10^{14}{\rm G})^{2}$ on the surface of the star. The initial location of the magnetic dipole axis is indicated with a black spot.}
\label{Figure:4a}
\end{figure*} 
Below we present the results of the simulations. Before discussing the simulations results in detail we notice three common patterns in the evolution of the magnetic field. First, the $\ell=1$ toroidal field concentrates towards one hemisphere of the neutron star. In the examples shown it is the northern hemisphere, resulting from the chosen polarity of the toroidal field. This effect is well understood in plane-parallel and axially symmetric studies \citep{Vainshtein:2000, Reisenegger:2007} and survives in 3-D. Second, the field generates non-axisymmetric structures, related to the density shear instability \citep{Wood:2014, Gourgouliatos:2015b}. Third, after the initial drastic change of the magnetic field structure, the evolution saturates and complex features, that formed during the initial stage, survive for a long time.  

In the following three subsections we will consider initial conditions with toroidal fields containing 50\%, 90\% and 99\% of the total energy, and with the toroidal and poloidal fields aligned. A fourth subsection presents initial conditions with toroidal field containing 90\% of the total energy, and with the toroidal and poloidal fields mis-aligned. How the spatial structures for these four configurations evolve in time is shown in Figures 1-5, respectively. Figure 6 then compares how the angle of the dipole axis evolves in time for these four cases, Figure 7 shows how the dipole field strength evolves, and finally Figure 8 shows how the four cases evolve on $P-\dot P$ diagrams.	

\subsection{Models with $50\%$ of the energy in toroidal field (A50-1/2/3/4)}

In this set of runs the energy in the initial poloidal and toroidal fields are equal. During the first 50kyr of evolution the field is displaced towards the northern hemisphere, forming zones of alternating inward and outward radial field, Figure \ref{Figure:1}. These zones are wider when the field is stronger, in particular they extend to polar angles from 44$^{\circ}$ to 61$^{\circ}$ for  A50-2,  41$^{\circ}$ to 69$^{\circ}$ for  A50-3, and 27$^{\circ}$ to 74$^{\circ}$ for  A50-4, where 0$^{\circ}$ and 90$^{\circ}$ is the north pole and the equator respectively. Mild non-axisymmetric features appear in the form of waves in the zones of magnetic field. The zones afterwards move to lower latitudes, with the system with strongest magnetic field (A50-4) rebounding closer to the equator ($\theta \approx 80^{\circ}$). Throughout the evolution, a strong radial magnetic field develops in the location of the zones. This magnetic field is about one order or magnitude higher than the dipole field, indicative of the strong multipolar structure of the field. However, it is located away from the magnetic dipole axis, and it is only connected to closed field lines, Figure \ref{Figure:4a} top left panel. Therefore, it can have little impact on the radiation mechanisms, which initiate at open (or the interface between open and closed) magnetic field lines. There is a small drift of the location of the axis of the magnetic dipole, less than $2^{\circ}$ throughout the evolution of the system, Figure \ref{Figure:5} top left panel. The intensity of the dipole component drops monotonically, by about one order of magnitude over the course of a million years, Figure \ref{Figure:6} top left panel. The tracks of the period-period derivative diagram for this run are shown Figure \ref{Figure:7} top left panel. They follow straight lines for the first few kyr; beyond this point, the decrease of their dipole component leads to downwards trend. All models cross the 1Myr characteristic age when their real age is less than 0.5Myr, suggesting that this estimate is already off by a factor of 2.

\subsection{Models with $90\%$ of the energy in toroidal field (A90-1/2/4)}

In this set of runs, we observe the concentration of the magnetic field near the pole, Figure \ref{Figure:2}. During the first 50kyr of the magnetic field evolution we notice the formation of stripes of alternating inward and outward field caused by the density shear instability. These stripes appear  at polar angles of 35$^{\circ}$, 25$^{\circ}$ and 20$^{\circ}$ for runs A90-1, A90-2 and A90-4 respectively. The field structure above this point depends on the intensity of the initial magnetic field. If the field is relatively weak (A90-1), a ring of outward field forms, where the strongest field appears and eventually a polar field extends from the pole to 28$^{\circ}$. In the A90-2 run, a strong radial field appears at 23$^{\circ}$, followed by spots of inward magnetic field and eventually at the polar cap a radial outward, yet not uniform field, appears. In the A90-4 run, where the strongest field was simulated, there is no evident ring or polar cap, but an overall positive radial field with varying intensity. The strongest field appears at $\approx$20$^{\circ}$ from the north pole. After 500kyr, a ring of strong radial magnetic field at 25$^{\circ}$ appears in the A90-1 case. At the same time, a polar cap extending to $20^{\circ}$ of a strong magnetic field with small variations in its intensity appears in the case A90-2. Finally in the A90-4 run, a radial field of irregular shape around the north pole of the star appears, and the maximum magnetic field at a polar angle of $45^{\circ}$. The magnetic field structure inside the star maintains a substantial tangential field which emerges from the star in the form of arcade-like structures, Figure \ref{Figure:4a} top right panel. There is a drift of the magnetic dipole axis by up to a 8$^{\rm o}$, Figure \ref{Figure:5} top right panel, with rapid changes in the A90-4 run. The intensity of the dipole component of the magnetic field drops monotonically with time, Figure \ref{Figure:6} top right panel, decreasing by about one order of magnitude in one Myr. Finally in the $P-\dot{P}$ diagram, Figure \ref{Figure:7} top right panel, the stars follow tracks of constant magnetic field for the first few kyr of their lives; later on the strong decay of their dipole component leads to a downwards trend. 

\subsection{Models with $99\%$ of the energy in toroidal field (A99-1/2/4)}

During the first 50 kyr of evolution, we notice, in all three runs, the formation of alternating zones of inward and outward field in the northern hemisphere, Figure \ref{Figure:3}. The state of the field near the pole, however, depends on the strength of the initial field.  In runs A99-1 and A99-2 the field near the pole is weaker compared to that of the stripes, whereas in A99-4 a magnetic spot appears, extending up to $10^{\circ}$ from the pole. After 500kyr, in all runs we notice that the strongest magnetic field is at the pole. In the A99-2 run we notice that inward and outward field lines appear very close to each other. At later times, all three runs have a single magnetic spot at the north pole. The drift of the magnetic dipole axis exceeds 20$^{\circ}$ from the geometric pole, Figure \ref{Figure:5} bottom left panel. Therefore, even though the magnetic spot is located very close to the initial location of the dipole component of the field, it does not coincide with the dipole axis later. However, the magnetic field lines that emerge from the magnetic spot close at various latitudes and longitudes, Figure \ref{Figure:4a} bottom left panel.  The intensity of the dipole field has a decreasing trend, however there are phases during which the field slightly increases, Figure \ref{Figure:6} bottom left panel. The strongest field on the surface of the star can be up to two orders of magnitude higher than the intensity of the dipole field; in the run A99-4, at 50 kyr the dipole component is $10^{13}$G while the field at the magnetic spot on the pole is $3\times 10^{15}$G. The evolutionary tracks of the neutron star in the period-period derivative diagram are affected by the increases of the intensity of the dipole component and have locations where they move to higher $\dot{P}$. Moreover, they pass through the densely populated part of the diagram and they cross the Death Line when their ages are between 2.3 and 4.7 Myrs, while their characteristic age exceeds 100 Myr.

\subsection{Models with misaligned toroidal fields (B90-2, B90-4, B99-4)}

In these runs the dipole field is misaligned with the toroidal field. The field has been initiated so that the initial poloidal field is aligned with the geometric pole of the sphere while the toroidal field is offset by 10$^{\circ}$ or $45^{\circ}$, Figure \ref{Figure:4}.  In the early stage of evolution we notice the toroidal field concentrating towards the axis of symmetry of the toroidal field, something that is clearly evident in run B90-2 and B99-4, where the stripes of inward and outward field converge at a location with polar angle approximately  $60^{\circ}$, whereas in the B90-4 run the stripes converge at a point closer to the pole. At 500 kyrs there is an appearance of strong radial magnetic fields at an extended area  $\approx 60^{\circ}$ for the B90-2 run. In the B90-4 run a ring of outward magnetic field forms, centred at 20$^{\circ}$, into which the strongest radial field appears. In the B99-4 a spot of strong radial outward field is located at $40^{\circ}$ from the pole. At later times, we notice the survival of a magnetic spot complex in B90-2  at $45^{\circ}$, an extended zone of radial field near the geometric pole of the star in B90-4 and a single spot at $55^{\circ}$ in the B99-4 run. The magnetic field develops a complex topology with the magnetic field lines that emerge from the magnetic spot returning to the star a distant locations, Figure \ref{Figure:4a} bottom right panel. We notice that in runs B90-2 and B90-4 the displacement of the magnetic dipole axis is about $10^{\circ}$ during the first 1Myr and reaches 20$^{\circ}$ later on in run B90-2. In contrast, run B99-4 shows a much more drastic displacement of the magnetic dipole axis, exceeding $50^{\circ}$, Figure \ref{Figure:5} bottom right panel. The magnetic dipole intensity drops monotonically with time in B90-2 and B90-4 run. In B99-4, while the overall trend is towards a decaying dipole, the evolution is more noisy and there are intervals during which the field's dipole strength increases. The drop in the initial dipole component of the field in the B99-4 run is so fast, that it eventually becomes weaker than in B90-2, even though it was initially higher. Finally in the $P-\dot{P}$ diagram for these runs we notice an overall behaviour that is similar to corresponding runs A90-2, A90-4 and A99-4 . In general comparing the cases where the initial dipole and toroidal field are misaligned we notice that the drift of the dipole axis, the intensity of the dipole field and the tracks on the period-period derivative diagrams look qualitatively similar to the respective runs where the field axes were aligned. There is a clear distinction on the formation and location of the magnetic spots, however, which are sensitive to the toroidal field.

\section{Discussion}

Below we discuss the implications of these simulations to the overall neutron star magnetic field evolution. We focus on the following aspects: magnetic spots, magnetic dipole axis drift, magnetic dipole evolution and evolution on the $P-\dot{P}$ diagram.

\subsection{Magnetic dipole axis drift}
\begin{figure*}
\includegraphics[width=1.1\columnwidth]{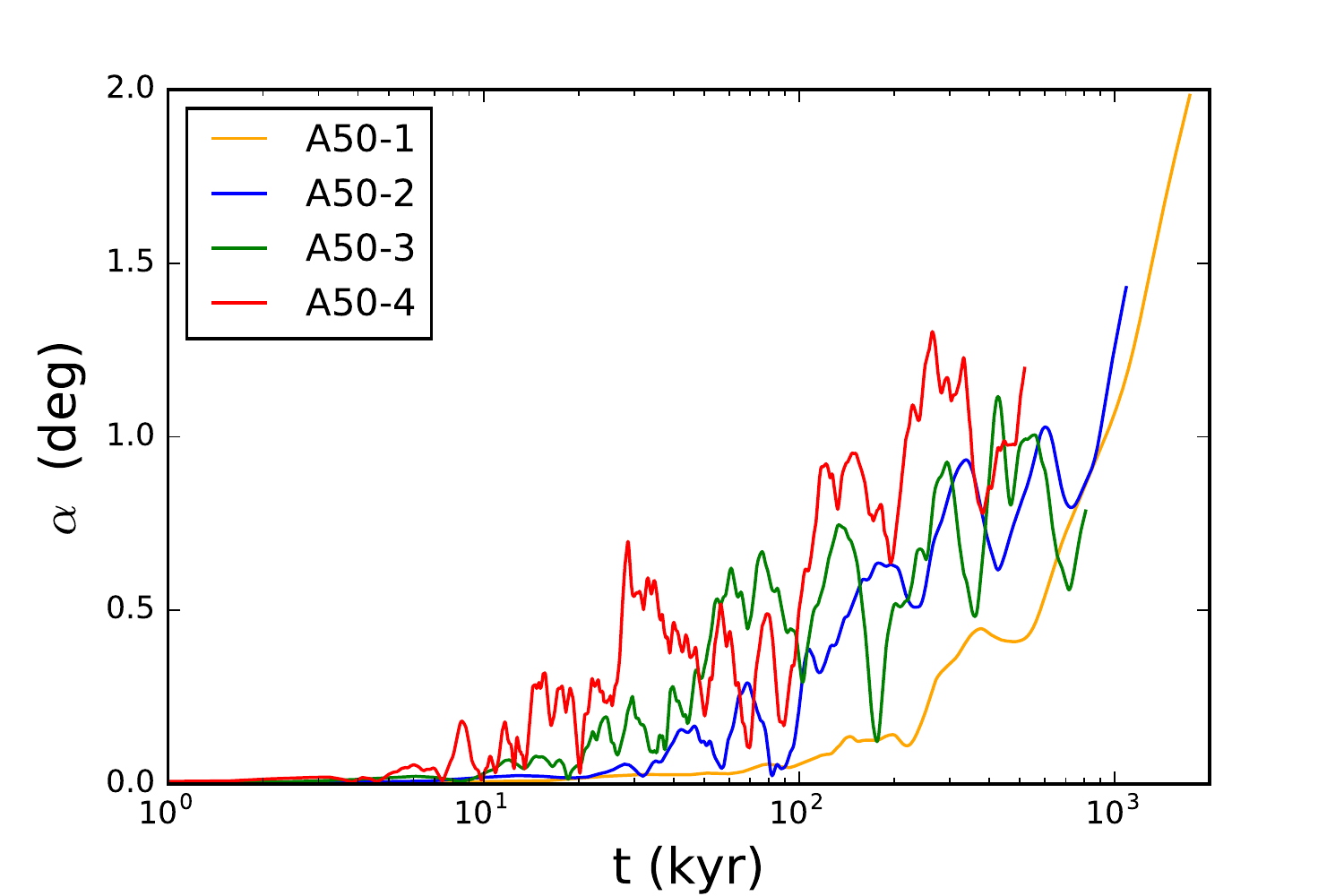}
\includegraphics[width=1.1\columnwidth]{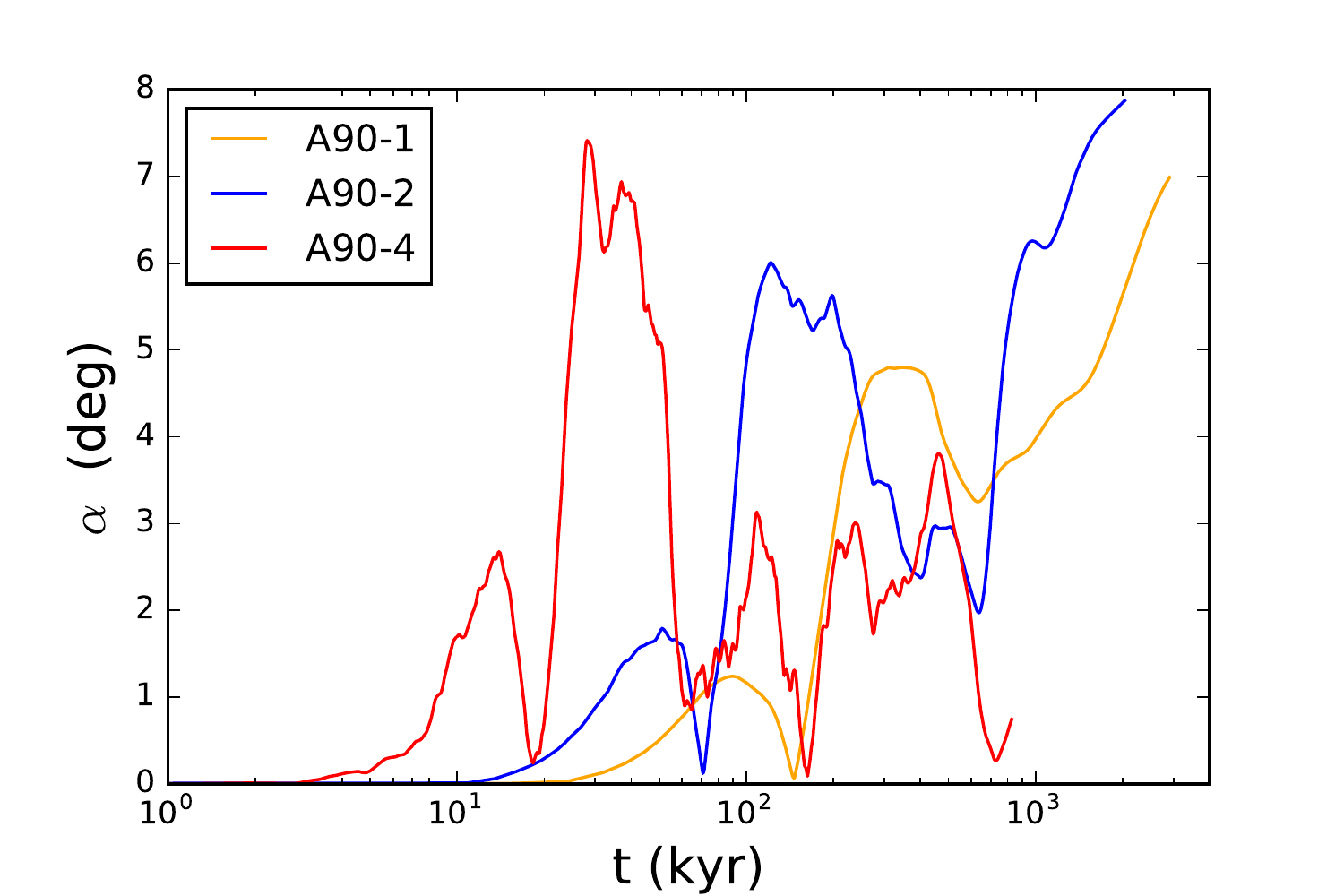}
\includegraphics[width=1.1\columnwidth]{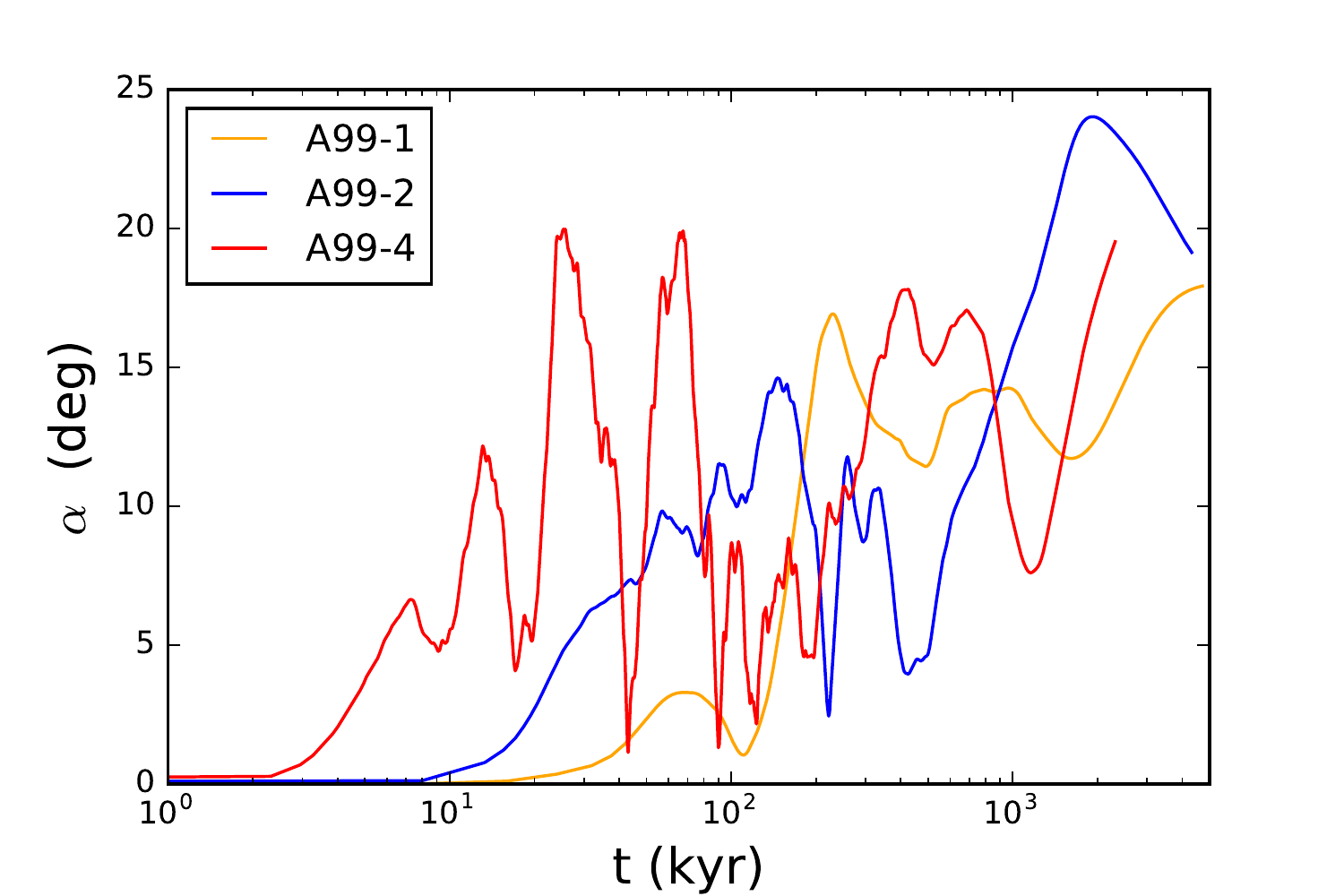}
\includegraphics[width=1.1\columnwidth]{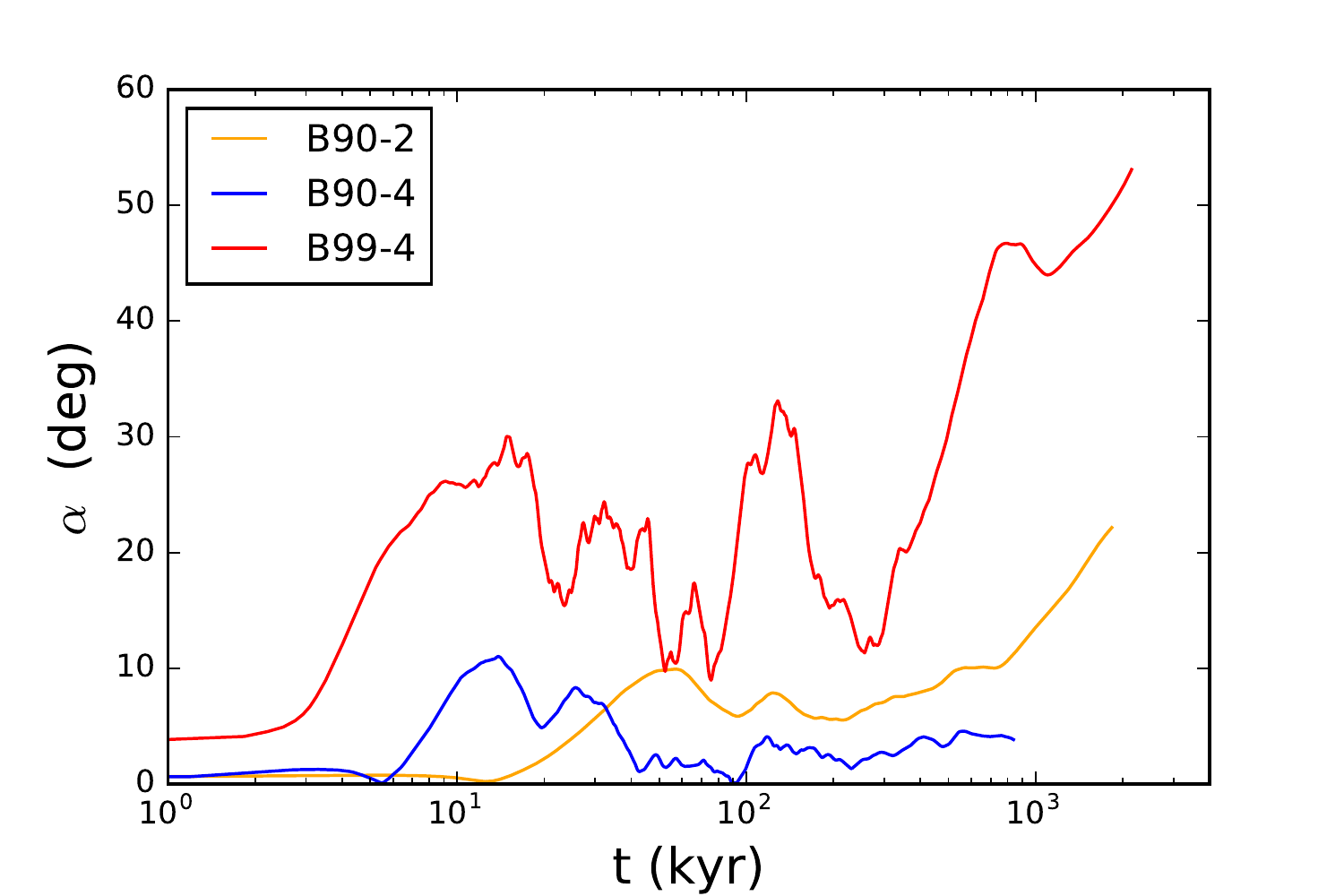}
\caption{Evolution of $\alpha$, the angle between the magnetic dipole axis and its initial position for all the simulations performed.}
\label{Figure:5}
\end{figure*} 
By construction, axially symmetric simulations do not allow dipole and higher multipole axes to drift on the surface of the star. Once the assumption of axial symmetry is relaxed,  we observe a noticeable drift of the dipole axis, Figure \ref{Figure:5}. In particular, the drift is  higher and faster in simulations where most of the energy is in the toroidal component. In the simulations where the initial condition is predominantly axially symmetric, it takes a few thousand years to displace the magnetic dipole axis from its initial position, given the time needed for the non-axisymmetric instability to grow. In contrast, if the toroidal field axis and the dipole axis are not aligned with each other, the drift happens immediately from the beginning (see bottom right pane of Figure \ref{Figure:5}). For a given ratio of poloidal to toroidal initial energy, however, the overall deviation of the magnetic dipole axis does not depend on the strength of the field; it only takes a longer time for this drift to develop when the field strength is lower. The fastest motion of the dipole axis is noticed in the B99-4 simulation where between 3 and 8 kyrs the axis drifts from 5$^{\rm o}$ to 25$^{\rm o}$, at an average rate of $0.4^{\rm o}$ per century. Because in these models the rotation axis of the neutron star is not taken into account, it is not granted whether a displacement will lead to higher or lower obliquity (angle between the magnetic moment and the rotational axis). This depends on the initial angle between the two vectors: if the dipole axis is initially aligned with the rotational axis, the random walk will lead to an increase of the obliquity  \citep{Roberts:1960}.

\subsection{Surface magnetic field and spot formation}

We find in all simulations that the radial magnetic field on the surface of the neutron star strengthens. At $50$ kyrs the strongest field on the surface becomes about $10$ times higher than the dipole field at this time for the run where $50\%$ of the energy is the poloidal field. For instance in A50-4 the strongest field at that time is $1.2\times 10^{15}$G (Figure \ref{Figure:1}, second row last column) and the dipole component is $10^{14}$G (Figure \ref{Figure:6}, top left panel, red curve), with similar ratios for the other A50 runs. In the runs where most of the energy is in the toroidal field the strongest radial magnetic field becomes about $200$ higher than the dipole magnetic field at the same time; for instance in run A99-4 the strongest field is $1.5\times 10^{15}$G (Figure \ref{Figure:3}, second row last column) while the dipole component is $8\times 10^{12}$G (Figure \ref{Figure:5}, bottom left panel red curve). A ratio of about $25$ between the strongest and the dipole magnetic field is deduced for the A-90 runs. These ratios persist throughout the evolution. Thus the presence of a toroidal magnetic field leads to a strong surface field. 

Whereas strong magnetic fields on the surface appear in all our simulations, the formation of a $\sim 1$ km sized magnetic spot is only feasible for the runs where $99\%$ of the energy is in the toroidal field. Even if $90\%$ of the energy is in the toroidal field, strong magnetic field will cover a more extended area whose effective radius is several kilometers. The spot will form near the toroidal field initial axis of symmetry, which is not necessarily the magnetic dipole axis, i.e. in run B99-4 ( Figure \ref{Figure:4}, third column) the location of the spot is at mid-latitudes whereas the initial dipole was located near the axis. Given that the magnetic dipole axis drifts significantly in the A99 simulations, while the spot stays at the geometric pole, the open field lines and consequently the polar cap are not located at the same area of the star where the spot lies. 

\subsection{Dipole magnetic field strength}
\begin{figure*}
\includegraphics[width=1.1\columnwidth]{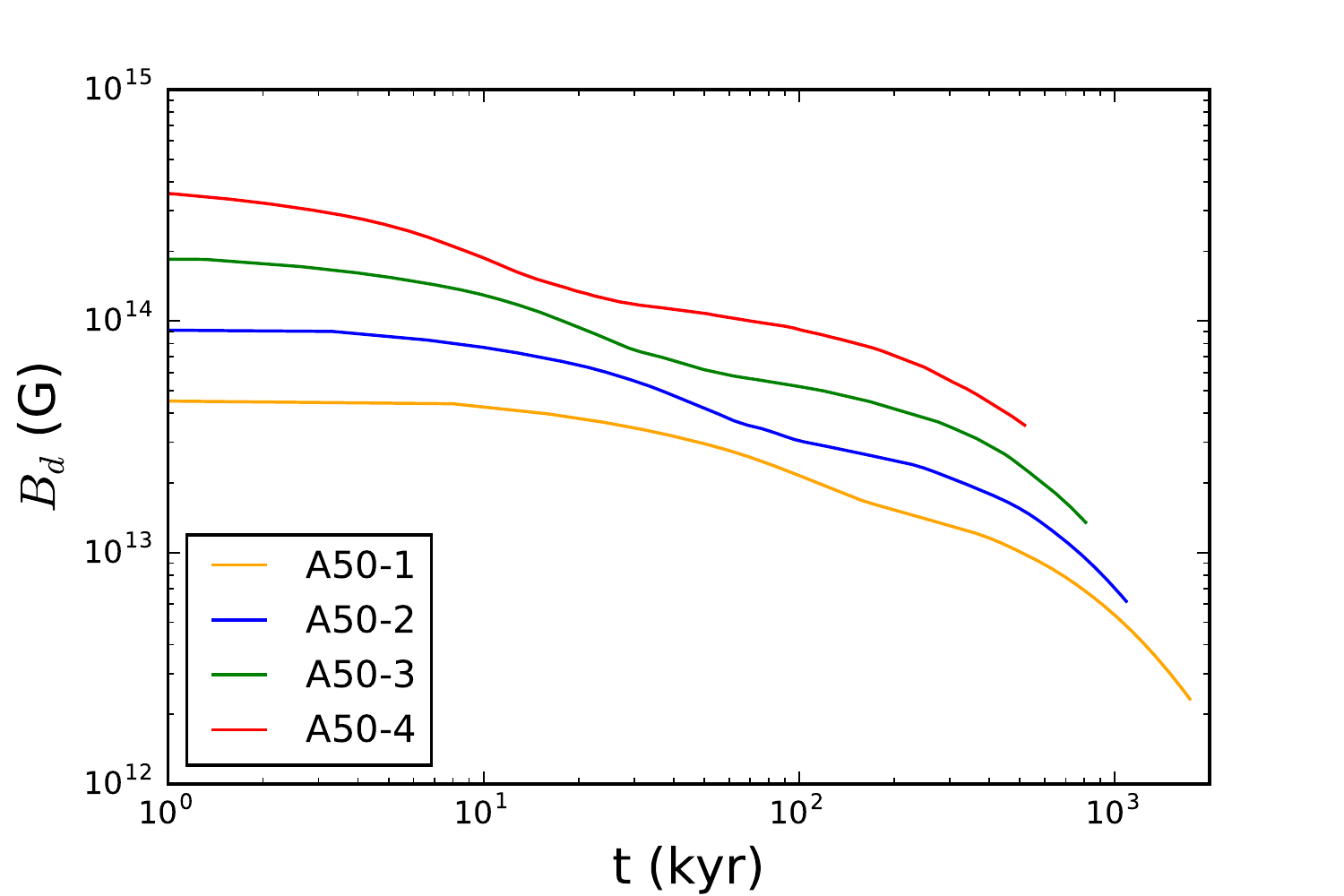}
\includegraphics[width=1.1\columnwidth]{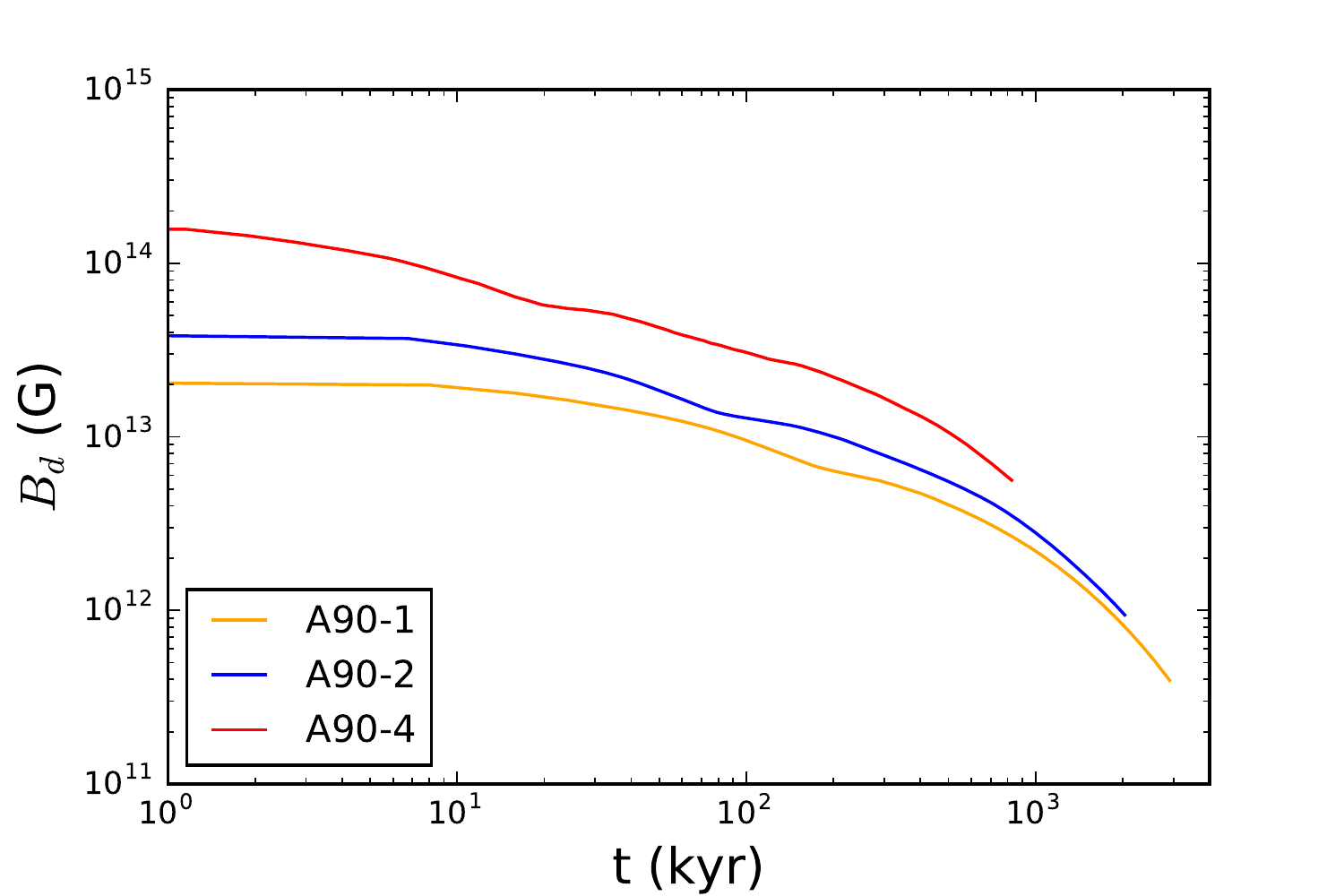}
\includegraphics[width=1.1\columnwidth]{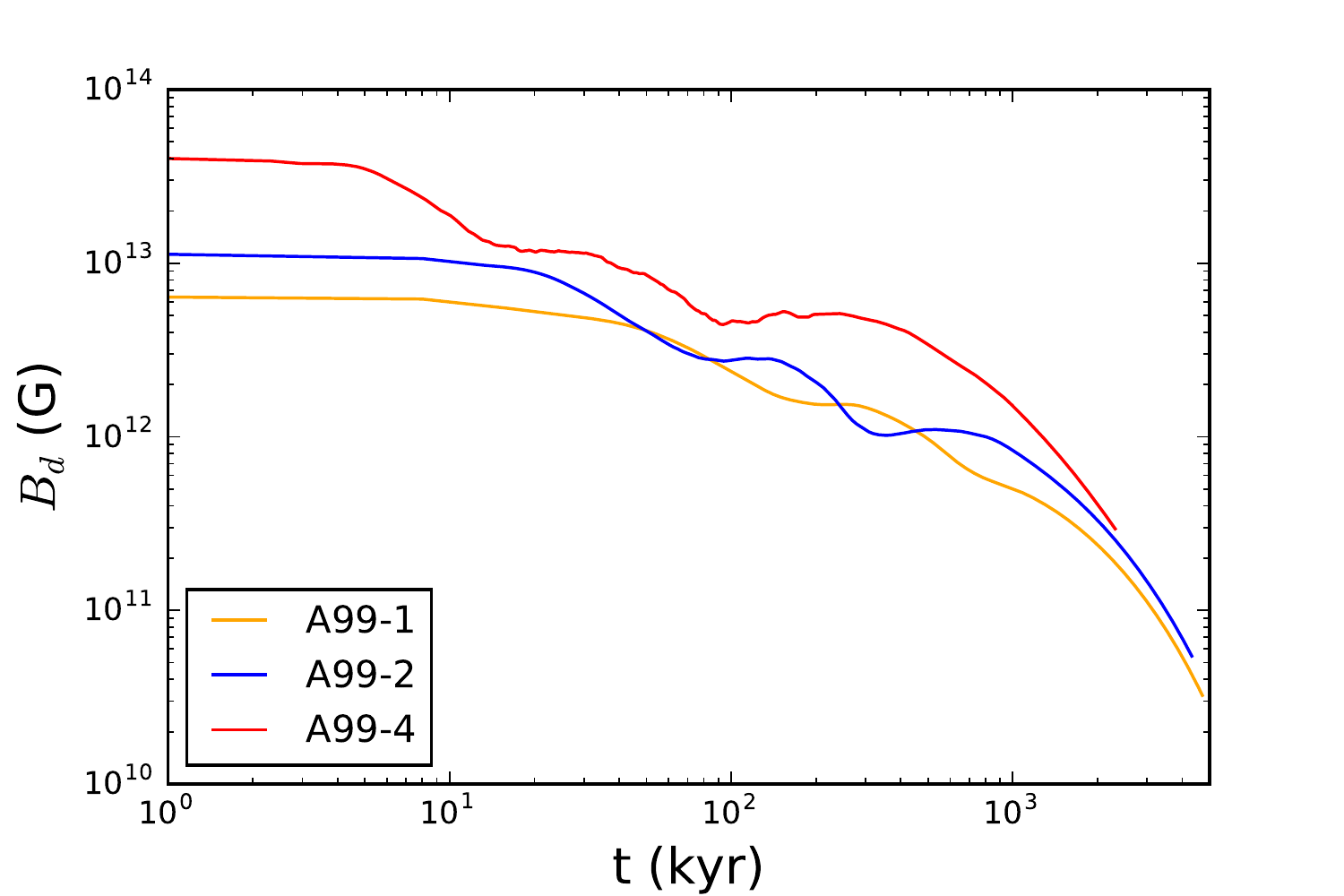}
\includegraphics[width=1.1\columnwidth]{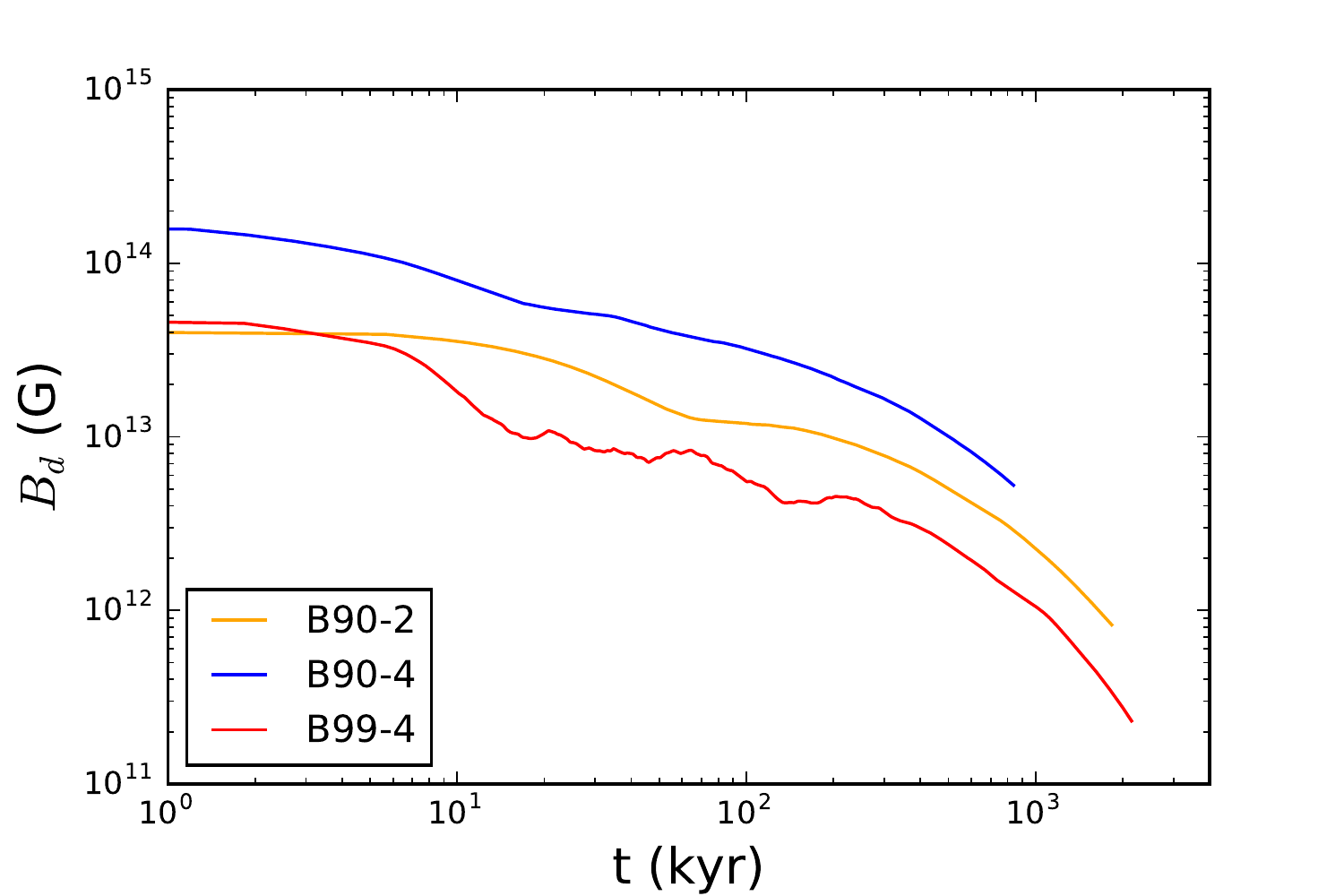}
\caption{Evolution of the dipole magnetic field strength for all the simulations performed.}
\label{Figure:6}
\end{figure*} 
There is a clear trend in all simulations for a decrease of the dipole component of the magnetic field, Figure \ref{Figure:6}. This reflects the decay of the magnetic field due to Ohmic dissipation, and the inevitable loss of magnetic energy. However, there are some temporary increases of the dipole strength, in the form of a superimposed oscillatory term which are evident in A99 and B99-4 runs. The effect of this variation can provide an up to $10\%$ increase of the magnetic dipole strength between a consecutive local minimum and maximum. In the A50 and A90 runs while some fluctuation is still present, the dipole component monotonically decreases.

\subsection{Evolution on $P-\dot{P}$ diagram}
\begin{figure*}
\includegraphics[width=1.1\columnwidth]{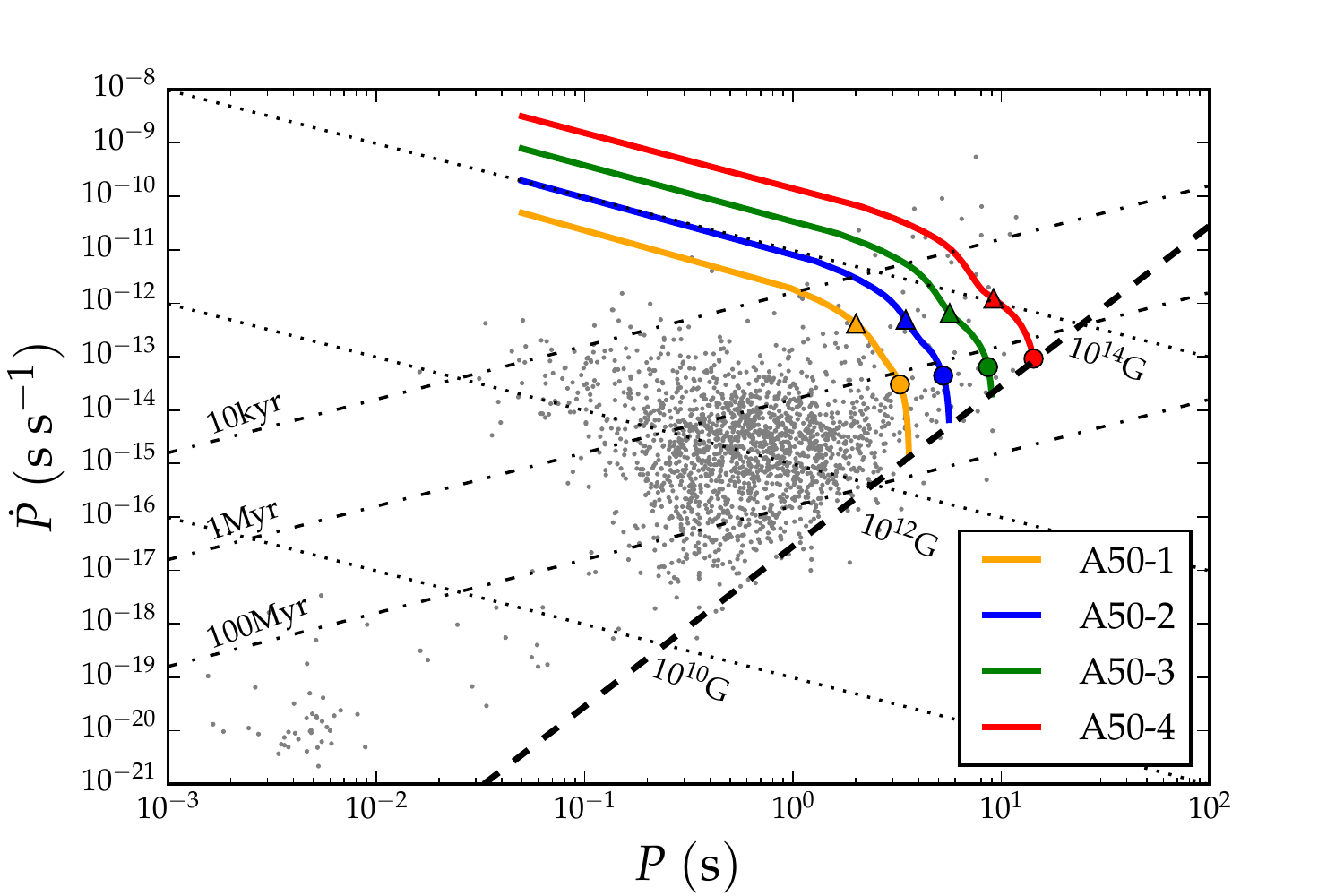}
\includegraphics[width=1.1\columnwidth]{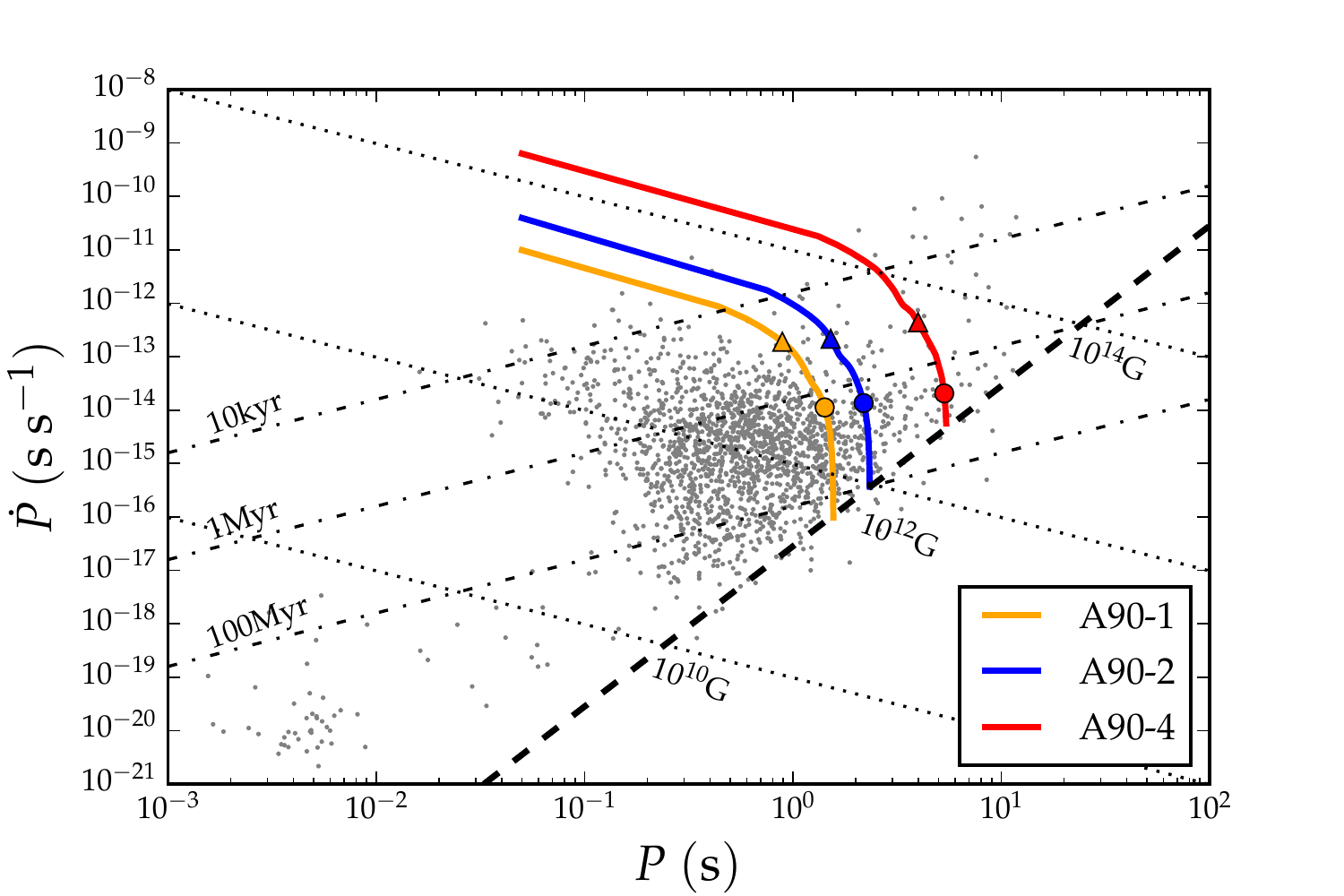}
\includegraphics[width=1.1\columnwidth]{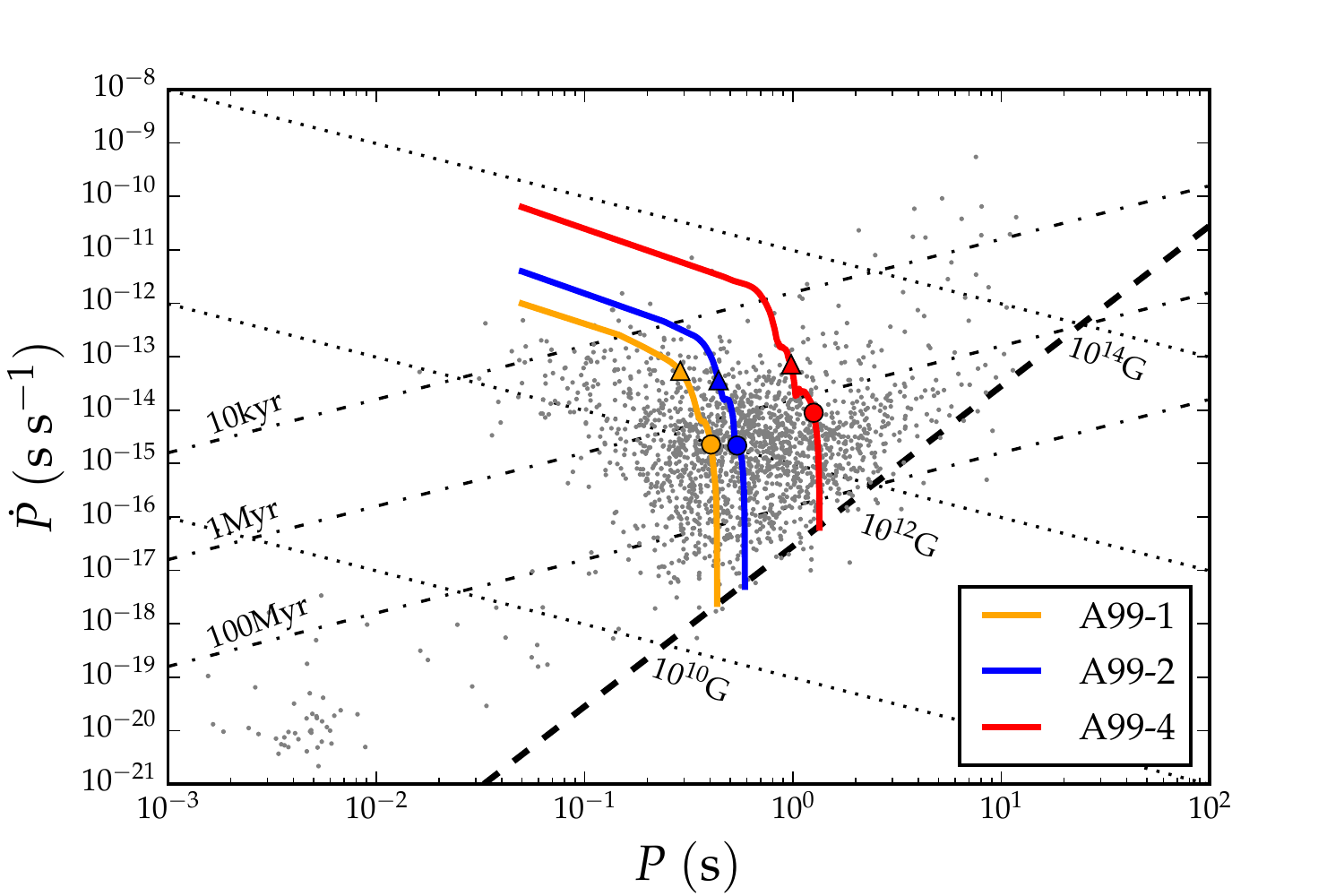}
\includegraphics[width=1.1\columnwidth]{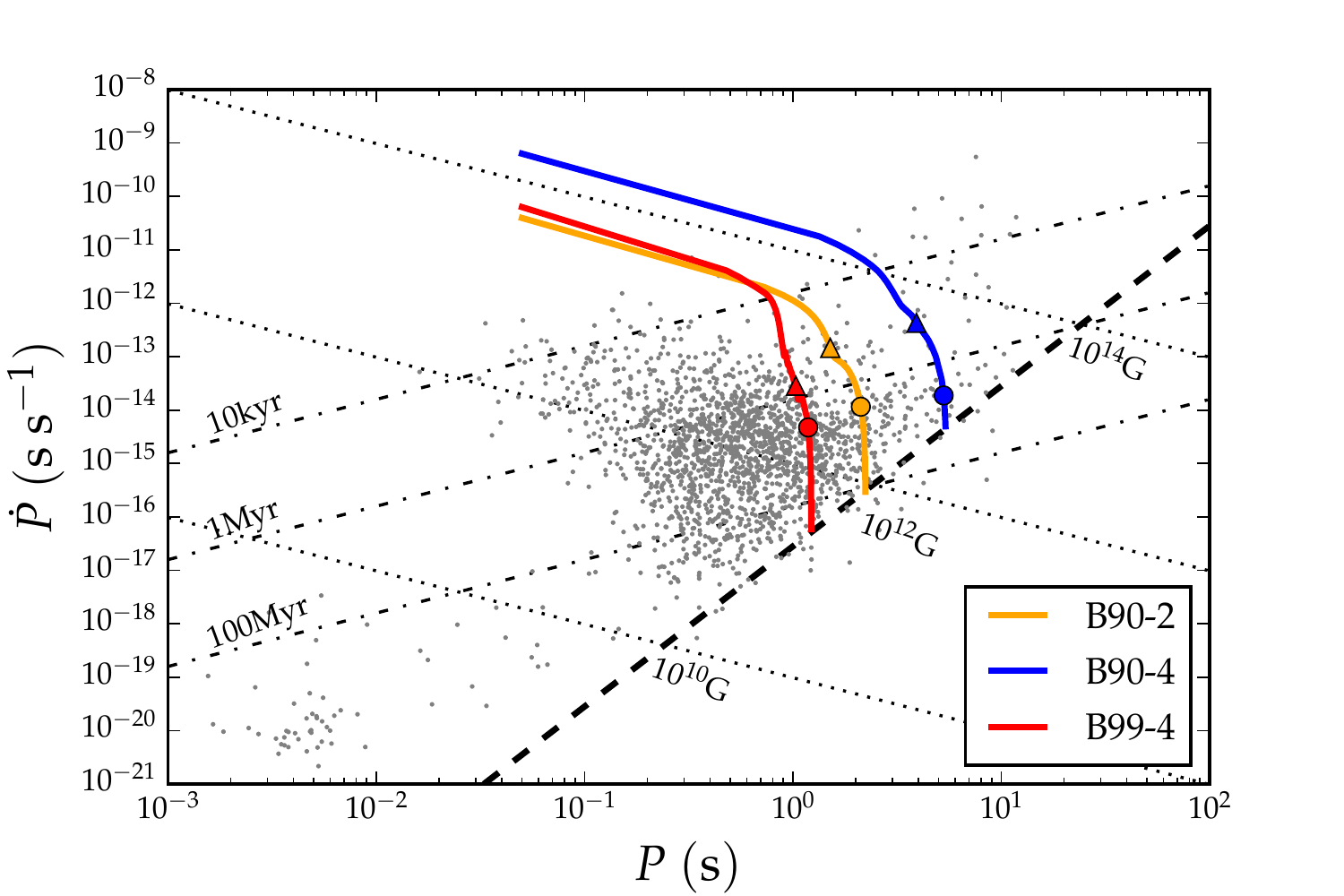}
\caption{Period-period derivative diagram ($P-\dot{P}$) for all the simulations performed. A triangle denotes the location of the pulsar when its real age is $50$kyr, and a circle when its age is $500$kyr. The thick dashed line denotes the Death-Line, equation 4. Grey points are the locations of the non-binary pulsars taken from the ATNF pulsar catalogue \citep{Manchester:2005}. }
\label{Figure:7}
\end{figure*} 
Assuming that a pulsar spins down entirely due to its magnetic field, and using canonical values for its radius and the moment of inertia \citep{Lorimer:2012}, we can relate its dipole magnetic field its period and period derivative
\begin{eqnarray}
B_{d}=3.2\times 10^{19}{\rm G} \left(P \dot{P}/s\right)^{1/2}\,.
\end{eqnarray}
We then calculate its trajectory on the $P-\dot{P}$ diagram by integrating in time 
\begin{eqnarray}
\frac{d P^{2}}{dt} = 2\left(\frac{B_{d} }{3.2\times 10^{19} {\rm G}}\right)^{2}~s\,. 
\end{eqnarray}
Assuming an initial spin period of $50$ms we deduce the trajectories shown in Figure \ref{Figure:7} for the models simulated. We find that all pulsars follow straight line trajectories, parallel to the lines of constant magnetic field $P\dot{P}=$const.~for their first 10 kyr. Later on, as the dipole magnetic field decays they deviate from these trajectories, and eventually follow vertical lines moving towards lower $\dot{P}$'s, without any significant change on $P$. We notice large discrepancies between characteristic age of the pulsar $\tau=P/(2\dot{P})$, and the real age, especially once the real age exceeds $500$kyrs. In all cases the circles drawn on the trajectory denote the position at which the age of the pulsar is 500 kyrs, and lie well beyond the 1Myr line, Figure \ref{Figure:7}.  We continue our integration until the pulsar crosses the Death Line described by the equation \citep{Bhattacharya:1992}
\begin{eqnarray}
\log B_{d} -2 \log P =25.86\,,
\end{eqnarray}
where their radio emission mechanism is expected to turn off.

We notice in a few runs with strong toroidal fields (A99-2, A99-4, B99-4) that the trajectories move temporarily towards higher $\dot{P}$. This occurs because of the temporary increase in the dipole component of the magnetic field and the faster spin-down rate.

\section{Implications for neutron stars}

\subsection{Plasma generation and acceleration}

One of the main challenges of pulsar theory is the generation of abundant plasma to initiate the pair cascade that is essential for radio emission. The magnetic field at the polar cap region for this process to start is $5\times 10^{13}$ G, or even higher than that, depending on the curvature radius of the magnetic field in this area \citep{Ruderman:1975, Gil:2006, Medin:2007, Geppert:2012}. It is evident that this condition does hold for the bulk of the population of neutron stars if their dipole magnetic field is considered, being typically in the range of $10^{12}$G or lower \citep{Manchester:2005}. From our exploration of simulations we found that it is feasible to create magnetic spots whose magnetic field strength exceeds $10^{13}$G and survive for most of its life, provided a strong toroidal field is included in the initial setup. For instance model A99-2 that passes through  the middle of the main clustering of rotation radio pulsars, Figure \ref{Figure:7} bottom left panel, crosses the Death Line with a magnetic field of $6\times 10^{12}$G in its magnetic spot and a radius of curvature of a few kilometers. This is below the limit set for the slot-gap model to work, but it is two orders of magnitude higher than the $5\times 10^{10}$G dipole field that corresponds to spin-down. The dipole axis in this instance is about $20^{\rm o}$ off the magnetic spot. The corresponding opening angle of the polar cap for a rotation period $0.6$s is $\theta_{pc}=(R_{*}/R_{LC})^{1/2}\approx 1^{\rm o}$ \citep{Sturrock:1971}, where $R_{*}=10^{6}$cm is the radius of the neutron star and $R_{LC}=cP/(2\pi)$ the light cylinder radius, suggesting that they are well separated from each other. Thus, while magnetic spots do form, they are not located at the polar cap and are unlikely to contribute to the slot-gap mechanism.  A possible resolution to this puzzle is a  complex topology of the magnetosphere where open field lines are connected to a magnetic spot a few kilometers above the stellar surface \citep{Szary:2015, Gralla:2016}. Indeed we notice in Figure \ref{Figure:4a} that the field lines emerging from the magnetic spots are connected to distant locations on the surface of the star. It is likely that the end points of the open magnetic field lines may not even coincide with the locating of the magnetic dipole axis. In addition to the above effect, frame dragging due to general relativity, decreases the limit on the essential magnetic field for the activation of the radiation process \citep{Philippov:2015, Philippov:2017}.

\subsection{Pulsar braking indices}

In a few pulsars it has been feasible to perform accurate timing and determine the second period derivative. Using this information it is possible to distinguish between various models of pulsar spin-down and constrain changes in the magnetic field, moment of inertia and obliquity \citep{Blandford:1988}. This second derivative is usually quantified through the braking index \citep{Pacini:1968, Ostriker:1969}
\begin{eqnarray} 
n=\frac{\nu \ddot{\nu}}{\dot{\nu}^{2}}
\end{eqnarray}
where the $\nu=1/P$ is the frequency of the pulsar. Considering spin-down due to a dipole in vacuum \citep{Deutsch:1955}, the torque on the neutron star is 
\begin{eqnarray}
\frac{d}{dt}\left(I \nu\right)= -\frac{2}{3c^{3}}R_{*}^{6}B_{d}^{2}\left(2 \pi \nu\right)^{3}\sin^{2}\psi
\label{VAC}
\end{eqnarray}
where $I$ is the moment of inertia and $\psi$ the angle between the rotational axis and the magnetic dipole. Alternatively, if a force-free magnetosphere  is considered \citep{Spitkovsky:2006}, the torque becomes  
\begin{eqnarray}
\frac{d}{dt}\left(I \nu\right)=-\frac{R_{*}^{6} B_{d}^{2}\left(2 \pi \nu \right)^{3}}{c^{3}}\left(1+\sin^{2}\psi\right)\,.
\label{FF}
\end{eqnarray}
Assuming that $I$ and $R_{*}$ remain unchanged, we can quantify the impact that magnetic field intensity and obliquity variations have on the spin-down of a neutron star. We find that the corresponding braking index for the vacuum dipole case, is:
\begin{eqnarray}
n=3-4\tau\left(\frac{\dot{B}_{d}}{B_{d}}+\dot{\psi} \cot\psi\right)\,,
\end{eqnarray}
where a dot denotes the time derivative of a quantity. In the force-free magnetosphere, the expression for the braking index becomes:
\begin{eqnarray}
n=3-4\tau\left(\frac{\dot{B}_{d}}{B_{d}}+\frac{\sin\psi \cos\psi }{1 +\sin^{2}\psi} \dot{\psi}\right)\,,
\end{eqnarray}
Changes of the magnetic field strength or obliquity angle lead to deviations of the braking index from the canonical value of $3$. We further remark that the two models differ very drastically if $\psi\approx 0$, with dramatic effects in the vacuum case and mild in the force-free. 

There are 11 pulsars with a known braking index, and the vast majority has $n<3$ \citep{Manchester:1985, Middleditch:2006, Livingstone:2007, Espinoza:2011, Lasky:2017, Ferdman:2017,Antonopoulou:2018, Archibald:2016}, with all the systems having characteristic ages less than 11kyrs. Braking indices smaller than 3 can be accommodated by a positive net combination of magnetic field variation and obliquity change. A temporary increase of the dipole magnetic field is feasible if a quadrupolar toroidal field is included \citep{Gourgouliatos:2015a}, or a buried field remerges from the surface of the star \citep{Muslimov:1996, Ho:2011}. In the simulations we performed in this work, with a toroidal field of $\ell=1$ multipole order, we notice that the change of the magnetic dipole axis is dominant, and any change of the magnetic dipole component happens at much later times in the life of the star rather than the first few kyr. A displacement of the dipole axis with respect to a fixed point, here being its initial position, should affect also $\psi$.  In our simulations we have found that the magnetic dipole axis may drift as fast as $\dot{\alpha}=0.5^{\rm o}$ per century and assuming that $\psi=30^{\rm o}$ with $\dot{\psi} \sim \dot{\alpha}$  in a $\tau=1$kyr pulsar, we obtain a braking index  $n=2.5$ in the vacuum model and $n=2.9$ in the force-free. Further support of the role of obliquity variation on the braking index comes from the Crab pulsar. In this system, apart from a measurement of its braking index, pulse profile evolution has been observed \citep{Lyne:2013, Lyne:2015}. The pulse profile change suggests that the beam direction moves at a rate of $0.6^{\rm o}$ per century. We remark that the torque on the neutron star can also force the rotation axis to move. Once the torque is considered, however, the trend is towards an aligned rotator both for the vacuum dipole and the force-free magnetosphere \citep{Philippov:2014}, whereas the observed braking indices imply in general an increase on $\psi$. In the magnetic axis drift model, the drift of the magnetic axis is random. If however, the axes are initially approximately aligned, the effect of a random walk, at least early in the evolution, will more likely lead to misalignment with the approximation of $\dot{\psi}\approx \dot{\alpha}$ being valid. Given the rapid differential rotation during the birth of a neutron star, the magnetic and the rotational axis are likely to be aligned, making this scenario a viable option \citep{Mosta:2015}.

\subsection{Low magnetic field magnetars}

X-ray timing observations of SGR 0418+5729 \citep{Rea:2010}, 3XMM J185246.6+003317 \citep{Rea:2014} and SGR 1822-1606 \citep{Rea:2012, Scholz:2014}, reveal dipole magnetic fields lower than $4\times 10^{13}$G, while all of them have magnetar-type behaviour. Such magnetic fields are typical of radio pulsars, and too weak to produce the observed outbursts, so most likely a strong local magnetic field is superimposed on the weak dipole. This has been strongly supported by the observation of a variable feature in the spectrum of SGR 0418+5729, which is caused by a local magnetic field between $2\times 10^{14}$ to $10^{15}$G \citep{Tiengo:2013}. In this system thermal radiation originates from a $4^{\rm o}$ spot and a further  absorption line caused by the superimposed magnetic field has been observed \citep{Guillot:2015}. In the A99 models simulated here, formation of a single spot on the axis is evident, which for the case of A99-4 has a local magnetic field of $1.5\times 10^{15}$G within 50kyrs from the birth of the star. At this instance the dipole component of the magnetic field is $8\times 10^{12}$G. The reservoir of magnetic energy in the crust, $1.8\times 10^{47}$erg, can provide sufficient energy to power the thermal part of the spectrum via Ohmic dissipation; however the details of this process will depend on the heat transfer properties of the crust which are not simulated in this work.

\section{Conclusions}

In this work we have simulated the evolution of the magnetic field in neutron stars starting from poloidal dipoles and $\ell=1$ toroidal fields. We examined systems where the toroidal field contains $50\%$, $90\%$ and $99\%$ of the total energy, and systems where the toroidal field is misaligned with the poloidal field. We find that a toroidal field containing $99\%$ of the total energy leads to the generation of a single spot on the axis of symmetry of the initial field. Even though the spot forms, the dipole axis at the same time drifts and does not necessarily intersect the surface of the neutron star at the location of the spot. Milder toroidal fields do not lead to the generation of spots: a configuration where $50\%$ of the energy is initially in the toroidal field eventually relaxes to a state where the field is almost equatorially symmetric. 

In our approach we have considered the evolution of the magnetic field due to the Hall effect and Ohmic decay, solely. In particular, we have assumed that the magnetic field evolves only inside the crust which extends to the outer $0.1 R_{*}$, without penetrating the core and being coupled to an external potential field. While these assumptions make the numerical computation feasible, further processes may affect the formation of magnetic spots. In principle the magnetic field evolution is coupled to its thermal evolution, as conductivity is a function of temperature \citep{Chugunov:2012}. Moreover, thermoelectric effects can in principle increase the magnetic field and consequently enhance its strength in regions where high temperature gradients appear \citep{Urpin:1980, Geppert:2017}. In addition to that, high temperatures near the surface of the crust, could cause thermoplastic effects which in turn may activate $\alpha-\omega$ dynamos. The effect of thermal feedback has been quantified by \cite{Vigano:2013} and does not drastically alter the generation of smaller scale structure, whereas the thermoelectric effects are still to be explored in realistic neutron star configurations. Ambipolar diffusion operating at the base of the crust and the core where neutrons are highly abundant affects the evolution of the magnetic field but the typical timescales are in the order of several million years \citep{Passamonti:2017, Castillo:2017}. The superconducting core may also steadily force magnetic flux to the crust affecting the overall evolution by slowing-down the decay of the magnetic field \citep{Elfritz:2016}. The crust is treated here as a rigid solid, while in principle it may deform, elastically or plastically and even fail \citep{Beloborodov:2014, Thompson:2017}. These effects have been explored in one dimensional and axially symmetric calculations: the elastic response of the crust is not sufficient to impede Hall evolution \citep{Bransgrove:2017}, whereas the plastic flows are indeed important close to the surface of the neutron star and lead to Hall avalanches \citep{Li:2016}. We remark that the result of the above mentioned studies, however, suggest that the dominant driving mechanism of the magnetic field evolution in the bulk of the crust, in the range of strengths explored in this work, remains the Hall drift and the other effects may be important locally or lead to migration of the magnetic spots towards cooler regions.

We have applied the results of our simulations in the context of pulsar and magnetar behaviour. Rotation powered pulsars require a strong magnetic field on their dipole axis for the slot-gap acceleration to be activated, and it is usually speculated that this condition is satisfied by multipole components. In this work, we find that strong magnetic fields in the form of spots indeed appear for some initial configurations, but they are not  located at the magnetic dipole axes, where the open field lines lie, and cannot contribute to particle acceleration. Thus, alternatives need to be sought either regarding the formation of magnetic spots at the exact position of the dipole pole, or the overall particle acceleration mechanism needs to be revisited. This drift of the dipole axis could be applicable to the question of braking indices on pulsars. We find that the magnetic dipole axis may drift up to $0.4^{\rm o}$ per century, which is in accordance with the value need  to explain the braking indices of pulsars and the change in the pulse shape of the Crab pulsar. Finally, magnetic spots are explored in the context of magnetars with weak dipole magnetic fields. In this case, we find that the single spot formed and the intensity of the local magnetic field are in accordance with the expectations from thermal emission and its spectral features. Given that thermal emission does not need to originate from the location of open magnetic field lines, which is typically the magnetic dipole axis, the drift of the dipole axis is no longer an issue.

\section*{Acknowledgements}

KNG and RH were supported by STFC Grant No. ST/K000853/1. The numerical simulations were carried out on the STFC-funded DiRAC I UKMHD Science Consortia machine, hosted as part of and enabled through the ARC HPC resources and support team at the University of Leeds. KNG acknowledges support from ``NewCompStar'', COST Action MP1304. We also thank an anonymous referee whose insightful comments and suggestions improved this paper. We acknowledge the use of the ATNF Pulsar Catalogue http://www.atnf.csiro.au/research/pulsar/psrcat. 

\bibliographystyle{apj}
\bibliography{BibTex.bib}

\end{document}